\documentclass[twocolumn]{aastex63}

\usepackage{graphicx}
\usepackage{amsmath}
\usepackage{amssymb}
\usepackage{txfonts}
\usepackage{ulem}

\usepackage{color}
\usepackage{colortbl} 
\definecolor{pinegreen}{RGB}{1, 121, 111}
\definecolor{salmon}{RGB}{255,160,122}

\usepackage{natbib}

\usepackage{rotating}

\definecolor{YC}{RGB}{115,80,185}

\newcommand{\Msun}{\ensuremath{\,M_\odot}}
\newcommand{\Rsun}{\ensuremath{\,R_\odot}}
\newcommand{\Lsun}{\ensuremath{\,L_\odot}}
\newcommand{\kms}{\ensuremath{\,\rm{km}\,\rm{s}^{-1}}}
\newcommand{\Msunyr}{\ensuremath{\,M_\odot\,\rm{yr}^{-1}}}

\newcommand{\HeII}{He~\textsc{ii}\ }
\newcommand{\HeI}{He~\textsc{i}\ }
\newcommand{\NIII}{N~\textsc{iii}\ }
\newcommand{\NIV}{N~\textsc{iv}\ }
\newcommand{\NV}{N~\textsc{v}\ }

\newcommand{\HeIIl}{He~\textsc{ii}\ $\lambda$}
\newcommand{\HeIl}{He~\textsc{i}\ $\lambda$}

\defcitealias{DroutGotberg23}{Drout \& G\"{o}tberg et al., under review}

\newcommand{\rev}[1]{{{{#1}}}}
\newcommand{\revv}[1]{{{{#1}}}}
\newcommand{\revvv}[1]{{{{#1}}}}

\usepackage{refs}   
\usepackage{booktabs}    
\usepackage{morefloats}

\bibpunct{(}{)}{;}{a}{}{,}

\submitjournal{ApJ}

\begin{document}

\shorttitle{Stellar properties of stripped stars}
\shortauthors{Y.~G\"otberg et al.}

\title{Stellar properties of observed stars stripped in binaries in the Magellanic Clouds}

\correspondingauthor{Y.~G\"{o}tberg, M.~R.~Drout}
\email{ygoetberg{@}carnegiescience.edu} \email{maria.drout{@}utoronto.ca}

\author[0000-0002-6960-6911]{Y.~G\"{o}tberg}\thanks{Hubble Fellow}
\affiliation{The observatories of the Carnegie institution for science, 813 Santa Barbara Street, Pasadena, CA 91101, USA}

\author[0000-0001-7081-0082]{M.~R.~Drout}
\affiliation{David A.\ Dunlap Department of Astronomy and Astrophysics, University of Toronto, 50 St.\ George Street, Toronto, Ontario, M5S 3H4 Canada}

\author[0000-0002-4863-8842]{A.~P.~Ji}
\affiliation{Department of Astronomy \& Astrophysics, University of Chicago, 5640 S Ellis Avenue, Chicago, IL 60637, USA}

\author{J.~H.~Groh}
\affiliation{Independent researcher}

\author{B.~A.~Ludwig}
\affiliation{David A.\ Dunlap Department of Astronomy and Astrophysics, University of Toronto, 50 St.\ George Street, Toronto, Ontario, M5S 3H4 Canada}

\author{P.~A.~Crowther}
\affiliation{Department of Physics and Astronomy, University of Sheffield, Hicks Building, Hounsfield Road, Sheffield S3 7RH, UK}

\author{N.~Smith}
\affiliation{Steward Observatory, University of Arizona 933 N. Cherry Ave., Tucson, AZ 85721, USA}

\author{A.~de~Koter}
\affiliation{Anton Pannekoek Institute for Astronomy, University of Amsterdam, Science Park 904, NL-1098 XH Amsterdam, the Netherlands}
\affiliation{Institute of Astronomy, KU Leuven, Celestijnenlaan 200D, B-3001 Leuven, Belgium}

\author[0000-0001-9336-2825]{S.~E.~de~Mink}
\affiliation{Max-Planck-Institut f\"{u}r Astrophysik, Karl-Schwarzschild-Stra\ss e 1, D-85741 Garching, Germany}
\affiliation{Anton Pannekoek Institute for Astronomy, University of Amsterdam, Science Park 904, NL-1098 XH Amsterdam, the Netherlands}

\begin{abstract}
Massive stars ($\sim$8-25\Msun) stripped of their hydrogen-rich envelopes via binary interaction are thought to be the main progenitors for merging neutron stars and stripped-envelope supernovae. We recently presented the discovery of the first set of such stripped stars in a companion paper. Here, we fit the spectra of ten stars with new atmosphere models in order to constrain their stellar properties precisely. We find that the stellar properties align well with the theoretical expectations from binary evolution models for helium-core burning envelope-stripped stars. The fits confirm that the stars have high effective temperatures ($T_{\rm eff} \sim$ 50-100kK), high surface gravities ($\log g \sim$ 5), and hydrogen-poor/helium-rich surfaces ($X_{\rm H, surf} \sim$ 0-0.4) while showing for the first time a range of bolometric luminosities ($10^3$-$10^5$\Lsun), small radii ($\sim$ 0.5-1\Rsun), and low Eddington factors ($\Gamma_e\sim0.006$-$0.4$). Using these properties, we derive intermediate current masses ($\sim$1-8\Msun), which suggest that their progenitors were massive stars ($\sim$5-25\Msun) and that a subset will reach core-collapse, leaving behind neutron stars \rev{or} black holes. Using the model fits, we also estimate the emission rates of ionizing photons for these stars, which agree well with previous model expectations. Further, by computing models for a range of mass-loss rates, we find that the stellar winds are weaker than predicted by any existing scheme ($\dot{M}_{\rm wind}\lesssim 10^{-9}$\Msunyr). The properties of this first sample of intermediate mass helium stars suggest they both contain progenitors of type Ib and IIb supernovae, and provide important benchmarks for binary evolution and population synthesis models.
\end{abstract}



\section{Introduction}

Helium stars with masses intermediate between subdwarfs and Wolf-Rayet (WR) stars ($\sim 2$-$8\Msun$) have been predicted to be created through mass transfer or common envelope ejection in binary stars with initial primary star masses of $\sim 8$-$25\Msun$ \citep[e.g.,][]{1967ZA.....65..251K, 1967AcA....17..355P, 2011ApJ...730...76I}. 
These envelope-stripped stars should be common \citep{2019A&A...629A.134G, 2021ApJ...908...67S}, because a large fraction of massive binaries go through envelope-stripping \citep[$\sim$30\%,][]{2012Sci...337..444S}, and the long-lasting helium-core burning phase usually remains after envelope-stripping \citep[e.g.,][see however also \citealt{2022A&A...662A..56K}]{2002ApJ...573..283P, 2008AIPC..990..230D}. 
Because of their ubiquity, stripped stars have been proposed as the main progenitors of stripped-envelope supernovae \citep{2011MNRAS.412.1522S, 2017ApJ...840...10Y, 2019ApJ...885..130S}, which also matches with their low ejecta masses \citep{2011ApJ...741...97D, 2016MNRAS.457..328L}. 
Envelope-stripping is also considered necessary for the creation of merging compact objects \citep{2007PhR...442...75K}. For example, the evolutionary channel to merging binary neutron stars includes two stripped stars \citep{2017ApJ...846..170T, 2020PASA...37...38V,2020ApJ...888L..10Y}. In addition, stripped stars are also so small that they can emit low-frequency gravitational waves detectable with the Laser Interferometer Space Antenna (LISA), when stripped by a compact object \citep{2004MNRAS.349..181N, 2018A&A...618A..14W, 2020A&A...634A.126W, 2020ApJ...904...56G, 2020ApJ...891...45K, 2022arXiv221008434L}
Furthermore, with their high effective temperatures ($T_{\rm eff}\sim$50-100kK), stripped stars should emit most of their radiation in the ionizing regime, thus providing a boost of ionizing emission several tens of millions of years after a starburst \citep{2016MNRAS.456..485S,2019A&A...629A.134G, 2020A&A...634A.134G}. 
However, although ``intermediate mass'' stripped stars have many interesting implications, an observed sample of them was missing until recently.

Previous efforts have been made in the search for stripped helium stars, resulting in discoveries on the low- and high-mass ends. 
In an impressive search for hot companions orbiting Galactic Be stars using ultraviolet (UV) spectroscopy, a set of hot subdwarf companions have been revealed \citep{2017ApJ...843...60W, 2018ApJ...853..156W, 2021AJ....161..248W}. With flux contributions of only up to $\sim$10\% in the UV, the subdwarfs likely have low masses of $\sim 0.5$-$1.5\Msun$ \citep{2022ApJ...926..213K,2022arXiv221003090K}, which suggests that the bright and early type Be-star companions became more massive and more luminous after they gained significant mass from the donor star during conservative mass transfer. 
Subdwarfs that instead orbit faint companions have been studied for example by \citet{2022A&A...666A.182S}.  
Also, during the recent searches for black holes, a number of inflated, low-mass ($\sim 0.5\Msun$) stripped stars were unveiled instead \citep[e.g.,][]{2020A&A...633L...5I, 2020A&A...641A..43B, 2022arXiv220306348E}.  
In addition, the star $\upsilon$~Sag, which was thought to be a $\sim$3\Msun\ intermediate mass helium giant \citep{1990MNRAS.247..400D}, has recently been determined to have $<$1\Msun\ \citep{2022arXiv220914315G}.
In the higher mass range, searches for companions to WR stars that may have been responsible for the envelope-stripping \citep{1998NewA....3..443V} has been done \citep{2017MNRAS.464.2066S, 2020MNRAS.492.4430S, 2019A&A...627A.151S}. In particular, the WR X-ray binary Cyg X-3 likely evolved via binary interaction, indicated from its short orbital period \citep{1973A&A....25..387V, 1992Natur.355..703V}. 

While the above described studies are important for our understanding of interacting binaries, none of them included helium stars of intermediate mass. 
In fact, the only previously known intermediate mass stripped star is the $\sim$4\Msun\ quasi Wolf-Rayet (WR) star in the binary system HD~45166, however, even this star has recently been observed to have lower mass than previously thought ($\sim$2\Msun, T.~Shenar, private communication). 
However, in \citetalias{DroutGotberg23}, we presented a new sample of 25 stars in the Magellanic Clouds. Originally identified has having excess UV radiation in comparison to the main-sequence  \citep{2018A&A...615A..78G}, we demonstrate that they have colors, brightnesses, and optical spectra consistent with expectations for binary systems containing intermediate mass helium stars. In particular, their spectral morphologies fall into three broad categories, as expected for systems with a range of mass ratios: (1) those consistent with a stripped helium star dominating the optical flux of the system, (ii) those consistent with both a stripped star and a main-sequence companion contributing to the optical flux, and (iii) those consistent with a main sequence companion dominating the optical flux of the system. By comparing the measured equivalent widths of several diagnostic lines for the stars in Class 1, 
we were able to obtain rough estimates for their physical properties, demonstrating that they have hot temperatures ($T_{\rm{eff}}$$\gtrsim$70kK), 
high surface gravities ($\log(g)\sim5$), 
and depleted surface compositions ($X_{\rm{H,surf}}\lesssim0.3$), further solidifying their nature as intermediate mass helium stars. 

Full characterization of the stripped star binary sample of \citetalias{DroutGotberg23} will deepen our understanding of binary interaction significantly, as it would produce direct constraints for binary evolution and population models. While the approximate effective temperatures, surface gravities and surface compositions presented in \citetalias{DroutGotberg23} were sufficient to establish their nature as intermediate mass stripped helium stars, more precise measurements and additional properties are needed to serve as benchmarks for detailed evolutionary models.  In particular, obtaining bolometric luminosities would allow placement on the Hertzsprung-Russell diagram, stellar radii can inform their current evolutionary stage, and constraints on the stellar winds of stripped stars are important for understanding both the evolutionary past and future. 
Historically, envelope-stripping of massive stars were predominantly considered via strong stellar winds, but recent measurements of the mass-loss rates the suggested previous evolutionary stage, the red supergiants, are surprisingly low \citep{2020MNRAS.492.5994B}. Low mass-loss rates of helium stars would further strengthen the binary-stripping scenario \citep{2022ApJ...933...41B}. 
For the future evolution, the stripped star winds directly affect the amount of hydrogen leftover from interaction and thus the supernova type \citep{2019MNRAS.486.4451G}. They also determine the orbital widening of short-period stripped star + compact object binaries and therefore also their ability to merge in gravitational wave events \citep{2022MNRAS.516.5737B, 2022MNRAS.517.4034S}. 

While full characterization of these stripped star binaries will ultimately require orbital solutions and ultraviolet spectroscopy, here we initiate the effort. We present a detailed analysis of the stellar properties of ten stripped stars that dominate over their companion stars even in their optical spectra using atmosphere modelling and spectral fitting.  We provide precise measurements of their surface hydrogen and helium content, effective temperatures, surface gravities, stellar radii bolometric luminosities.
We further estimate their stellar masses, emission rates of hydrogen- and helium-ionizing photons, calculate their Eddington parameters, and estimate rough mass-loss rates via stellar winds. 
The paper is structured as follows. 
In \secref{sec:sample}, we describe the specific sample of stars that we perform spectral fitting of in close detail, while in \secref{sec:observations}, we describe how the spectra and photometry for this sample were obtained. \secref{sec:spectral_fitting} is dedicated to describing a newly computed spectral model grid and the methodology we use to fit the spectra and obtain stellar parameters for the observed stars. We summarize the best-fit properties with associated for the stellar parameters of the stars in \secref{sec:stellar_properties}, while the full spectral fits for the individual stars are presented in \appref{app:details_spectral_fitting}. In \secref{sec:evol_stage}, we motivate what evolutionary stage we believe the stars to be in. In \secref{sec:wind}, we present a rough analysis for obtaining stellar wind mass-loss rate estimates, and in \secref{sec:ionizing_flux} we present estimates for the emission rates of ionizing photons. In \secref{sec:implications}, we discuss implications of the derived stellar parameters for massive binary evolution, and in \secref{sec:summary_conclusion} we summarize and conclude our findings.

\section{Stellar sample}\label{sec:sample}

The full sample of 25 stars presented in \citetalias{DroutGotberg23} was divided into three spectral groups. Specifically, they were divided based on a comparison of the equivalent widths of \HeIIl 5411 and\ H$\eta$/\HeIIl 3835 lines (chosen to probe the presence of a hot helium star and a B-type MS star, respectively) for the observed stars to a model grid of helium star plus MS star binaries. We found that (i) 8 stars have significant \HeII absorption and minimal short-wavelength Balmer lines, consistent with models where the stripped star contributes 80--100\% of the optical flux (ii) 8 stars exhibit both \HeII absorption and non-negligible short-wavelength Balmer lines, consistent with models where the stripped star contributes 20--80\% of the optical flux, and (iii) 9 stars have strong Balmer lines an lack an detectable \HeII absorption, only possible in the model grid if any stripped star component contributes $<20$\% of the optical flux.

In \citetalias{DroutGotberg23} these were designated Class 1: ``Helium-star type'', Class 2: ``Composite type'', and Class 3: ``B-type'', respectively. Members of all three classes with multiple epochs of spectroscopy showed evidence of radial velocity shifts, indicative of binary motion. While orbital solutions/spectral disentangling will ultimately allow for characterization of the spectral properties of both binary components in the full sample, here we describe the motivation for the subset of 10 objects that we present detailed spectral fits for in this manuscript (\secref{sec:samples}) and review the basic spectral features present in these stars (\secref{sec:samplep}).

\begin{figure*}
\centering
\includegraphics[width=\textwidth]{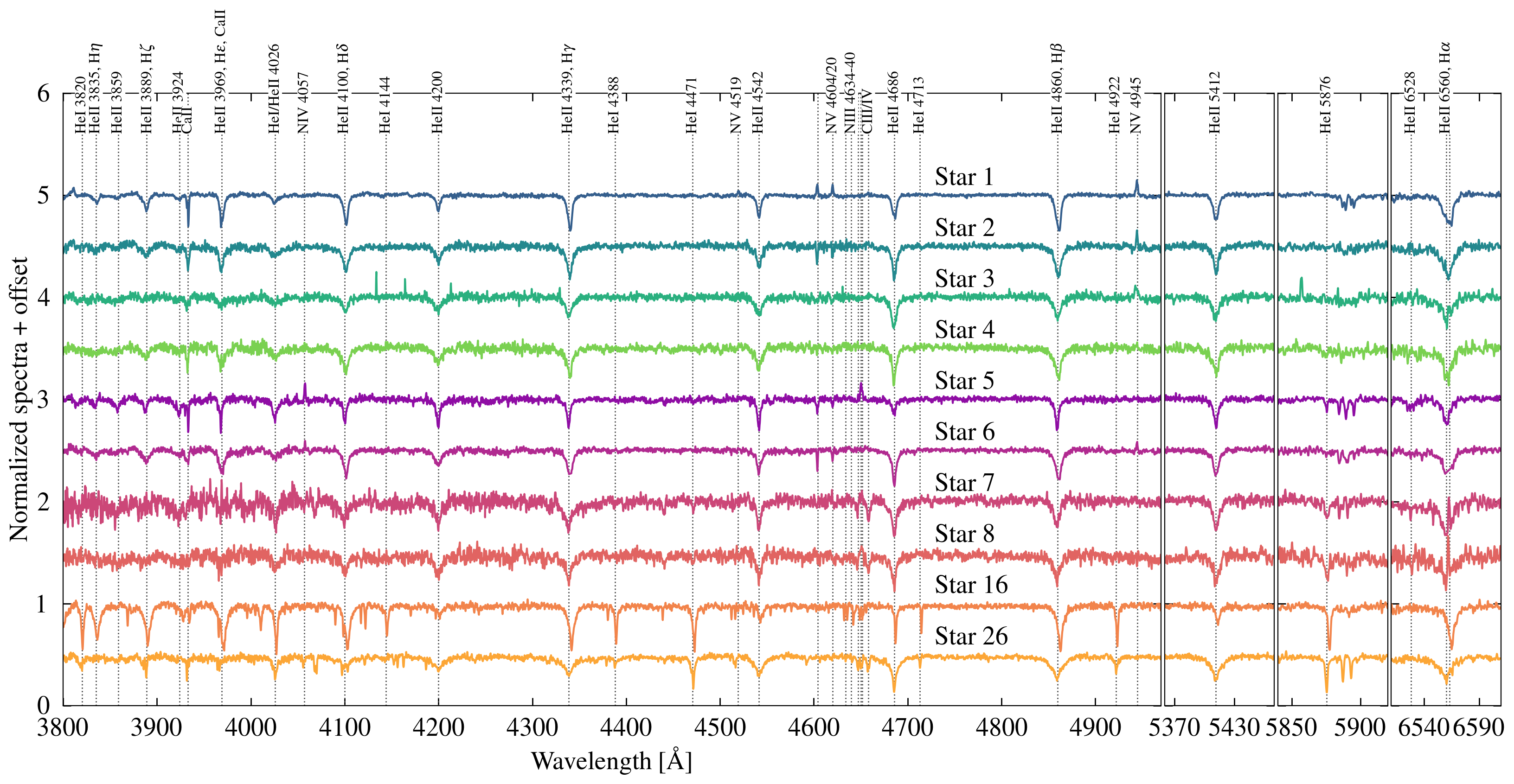}
\caption{The normalized spectra of the stars in the observed spectroscopic sample, described in \secreftwo{sec:sample}{sec:observations}. These stars are selected from the sample of \citetalias{DroutGotberg23} and thought to be stars stripped of their hydrogen-rich envelopes via binary interaction. }
\label{fig:optical_spectra}
\end{figure*}

\subsection{Sample Selection}\label{sec:samples}

Our goal in this first follow-up manuscript is to provide detailed stellar properties for a set of intermediate mass helium stars. We therefore begin by selecting a set of 10 stars where we believe that the stripped star dominates the optical flux and the companion contributes minimally. For this sample, we can therefore adopt a simplified analysis and model the optical spectrum as a single star. 
Specifically, in this manuscript we will analyze:
\begin{itemize}
    \item The 8 stars of Class 1 from \citetalias{DroutGotberg23} (stars 1-8). Of these, stars 1-4 are located in the SMC and 5-8 in the LMC.
    \item A single object from Class 2 (star 16; located in the LMC).
    \item An additional star that was originally identified in search for stripped helium stars described by \citetalias{DroutGotberg23}, but rejected from their final sample based on its kinematics (star 26; likely a foreground halo object).
\end{itemize}
The optical spectra of these ten stars are displayed in \figref{fig:optical_spectra}. We have used the full set of information available to us in assessing that the optical spectrum of a specific star is likely dominated by the flux of a single object. Here we elaborate on each item above.

The Class 1 stars from \citetalias{DroutGotberg23} all had spectral morphologies consistent with models for ``isolated'' helium stars, we are able to achieve a good spectral fit assuming contributions from a single star (see \secref{sec:stellar_properties}). In addition, while they all show radial velocity shifts, they appear as \emph{single-lined} spectroscopic binaries. This requires that the companion stars are optically faint: either compact objects or low mass main-sequence stars (M$\lesssim$3\Msun). However, in \citetalias{DroutGotberg23} we found that a MS companion star could potentially contribute up to 20\% of the optical ($V$-band) flux and still be classified as a ``helium-star type'' spectrum. Therefore, in \appref{app:impact_companion} we present a set of tests on how the presence of a MS companion may impact the results of our spectral fitting, concluding only minor effects could arise.

While star 16 was placed in the ``Composite-type'' class by \citetalias{DroutGotberg23} due to a combination of short-wavelength Balmer lines and \HeII absorption, it is most likely an inflated stripped star. When inflated, the surface temperature and surface gravity of a stripped star will decrease, leading to stronger Balmer absorption if any hydrogen remains on the surface. This interpretation is strengthened by the good spectral fit (see \secref{sec:stellar_properties} and \figref{fig:fit_star16}) and the analysis of its evolutionary stage in \secref{sec:evol_stage}. It also exhibits radial velocity shifts indicative of a single-lined spectroscopic binary. This is in stark contrast to the other Class 2 objects from \citetalias{DroutGotberg23}, which (i) we were unable to achieve a reasonable spectral fit for assuming contributions from a single star and (ii) show indications of anti-correlated motion in their \HeII and Balmer absorption lines, suggestive of double-lined spectroscopic binaries.

Finally, we address star 26. This object shows a significant UV excess in it spectral energy distribution and has an optical spectrum that would be grouped with the ``Helium-star type'' class from \citetalias{DroutGotberg23} due to strong \HeII absorption and weak short-wavelength Balmer lines. However, it has a mean radial velocity and proper motions from \emph{Gaia} DR3 that are sufficiently in-consistent with the bulk of stars in the LMC that we consider it a likely foreground, halo star (see \appref{app:kinematic} for a detailed kinematic assessment). \emph{Gaia} does not detect a parallax at the 3$\sigma$ level and we place a lower limit on its distance of $\sim$3.5 kpc (approximated by taking three times the parallax error provided by \emph{Gaia}). Analysis presented in \secreftwo{sec:stellar_properties}{sec:evol_stage} suggest that at a distance of 10 kpc, the properties of star 26 would be consistent with a subdwarf nature. In the rest of the paper, we therefore predominantly adopt the 10 kpc distance for star 26, but also present the stellar properties for the star assuming it is located in the LMC (completeness and for comparison).

\subsection{Spectral Morphology}\label{sec:samplep}

The optical spectra for the 10 stars are shown in \figref{fig:optical_spectra}. All objects show strong \HeII absorption, indicative of high temperatures. Stars 1-8 and 26 all show weak short-wavelength Balmer/\HeII-blends while star 16 shows stronger features in this regime, consistent with their classifications in Class 1 and 2, respectively, in \citetalias{DroutGotberg23}. 

\HeI lines are present in the spectra of stars 5, 7, 8, 16, and 26, while they are not present in the spectra of stars 1-4 and 6. 
Stars 1-4 and 6 all show \NV lines in emission and/or absorption. Stars 5 and 6 display \NIV $\lambda 4057$ in emission, while in star 26 it appears in absorption. In the case of star 16, \NIII lines are visible. Stars 7 and 8 have too poor signal-to-noise ratio spectra for these weak N-features to be detectable.
In the case of stars 2, 5, 7, 8, and 26, carbon lines are visible. We will not discuss these further here, but address it in a future study on the CNO abundances of stripped stars.
Finally, the Ca~\textsc{ii} H \& K doublet visible in several of the spectra at 3935 and 3970 \AA\ is interstellar.

\section{Observations}\label{sec:observations}

In order to derive detailed stellar properties for the stars described above, we utilize both the moderate-resolution optical spectra and UV-optical photometry. Data acquisition and reduction are described in detail in \citetalias{DroutGotberg23}. Here we briefly review key details of our methods. 

\subsection{Spectroscopy}\label{sec:obs_spectroscopy}


\begin{table*}
\centering
\caption{Observations to obtain optical spectra.}
\label{tab:observations}
\begin{tabular}{lccccccc}
\toprule\midrule
Star & Location & RA & DEC & Dates of observation & \# spectra & \rev{Approximate} exposure times & \rev{SNR$^*$} \\ 
\midrule
1 & SMC & 01:00:59.70 & -72:37:13.7 & 2018-2022 & 18 & 2$\times$600s  & $120$ \\ 
2 & SMC & 00:57:01.56 & -72:36:03.3 & 2019-2022 & 11 & 2$\times$1200s & $60$ \\ 
3 & SMC & 00:57:40.09 & -71:59:16.5 & 2019-2022 & 30 & 1$\times$1200s & $50$ \\ 
4 & SMC & 01:04:00.48 & -72:16:42.7 & 2019-2022 & 7  & 3$\times$1200s & $50$ \\
\midrule
5 & LMC & 05:08:49.38 & -69:05:29.8 & 2019-2022 & 14 & 2$\times$800s & $70$ \\
6 & LMC & 05:04:46.68 & -69:02:25.3 & 2018-2022 & 18 & 2$\times$850s & $80$ \\
7 & LMC & 05:28:01.15 & -69:59:48.7 & 2019-2022 & 8 & 2$\times$1200s & $30$ \\
8 & LMC & 05:47:28.01 & -69:06:07.6 & 2019-2022 & 6 & 3$\times$1200s & $40$ \\
16 & LMC & 05:35:33.63 & -70:19:06.1 & 2019-2022 & 12 & 2$\times$900s & $60$\\
\midrule
26 & Foreground & 05:46:56.80 & -70:05:36.4 & 2019-2022 & 10 & 3$\times$600s & $70$ \\ 
\bottomrule
\end{tabular}

\rev{$^*$ The signal-to-noise ratios are calculated \revv{per pixel} for the stacked spectra, rounded to the nearest \revv{multiple of ten} and \revv{then averaged} over the wavelength ranges 4230-4300, 4400-4430, 4730-4820, 5030-5250\AA.}
\end{table*}

We obtained multiple epochs of medium-resolution ($\mathcal{R}\sim 4100$) optical spectra ($\lambda \sim 3700-7000\AA$) for the stars detailed in \tabref{tab:observations} using the The Magellan Echellette (MagE) spectrograph on the Magellan/Baade 6.5m telescope at Las Campanas Observatory \citep{2008SPIE.7014E..54M}. Spectra were taken during 22 dark/grey nights between December 2018 and February 2022 (PI: G\"{o}tberg \& Drout). Observations were typically taken at the parallactic angle, but on some occasions a rotation was applied to exclude other nearby stars from the slit. This can result in slightly lower signal-to-noise in the blue portion of the spectra (e.g., Star 7; \figref{fig:optical_spectra}).


Initial data reduction was performed using the \texttt{CarPy} python-based pipeline\footnote{\url{https://code.obs.carnegiescience.edu/mage-pipeline}} \citep{2000ApJ...531..159K,2003PASP..115..688K}. The pipeline performs bias/flatfield correction, sky subtraction, 1D spectral extraction, and wavelength calibration. Individual echelle orders were normalized by fitting low order polynomials to the continuum after performing 2.5$\sigma$ clipping to reject contributions from absorption lines. Orders were then stitched together after normalization. 
We manually clip artifacts caused by both cosmic rays and by imperfect sky subtraction in cases where stars are located in bright/clumpy H~\textsc{ii} regions (e.g., Star 6). 
\rev{Finding the true continuum is challenging, especially for the upper Balmer series ($\lambda \lesssim$3900\AA), and we therefore carefully flatten each spectrum manually and exclude members of the Balmer series above H$\delta$ in our analysis. We note that artifacts could be present in our final spectra that relate to slight variation of the continuum in the wings of broad lines or averaged spectra where the orders overlap. However, we do not consider that these artifacts are sufficiently large to significantly impact our results.}

Finally, to produce the highest signal-to-noise ratio (SNR) spectra of each stars, we stack together observations taken on different occasions. However, all the stars considered here display radial velocity shifts and appear as single-lined spectroscopic binaries. We therefore must correct for binary motion when stacking spectra obtained days to months apart. This process is discussed in detail in \citetalias{DroutGotberg23}. \revv{The SNR is then calculated per pixel within the wavelength ranges, 4230-4300, 4400-4430, 4730-4830, and 5030-5250\AA, and then averaged, resulting in final SNRs of our combined spectra ranging} from $\sim$30-120 (see \tabref{tab:observations}). These combined spectra are show in in \figref{fig:optical_spectra} and  \revvv{will be made publicly available upon publication of this manuscript.}
Stars 1--8 and 16 were originally published in \citetalias{DroutGotberg23}, and we have now made Star 26 available as well.

\subsection{Photometry}\label{sec:obs_photometry}

\begin{table*}
\centering
\caption{Photometric data with 1$\sigma$ errors obtained from the UBVI survey at the Swope \rev{telescope \citep{Zaritsky2002,Zaritsky2004}} and photometry perfomed on the Swift/UVOT images of the Magellanic Clouds \citep{2014AJ....148..131S, 2015arXiv150402369S, 2019AJ....158...35S} (see \citetalias{DroutGotberg23} and Ludwig et al., in preparation). These apparent magnitudes are presented in the AB system\rev{, where we have converted the optical data from Vega magnitudes following the description in \secref{sec:obs_photometry}}.}
\label{tab:photometry}
\begin{tabular}{lccccccc} 
 \toprule \midrule 
Star & UVW2 & UVM2 & UVW1 & U & B & V & I\\ 
\midrule 
1 & $ 16.15 \pm 0.05$ & $ 16.26 \pm 0.05$ & $ 16.34 \pm 0.05$ & $ 16.72 \pm 0.04$ & $ 17.14 \pm 0.03$ & $ 17.45 \pm 0.04$ & $ 18.22 \pm 0.05$\\ 
 2 & $ 17.91 \pm 0.06$ & $ 17.98 \pm 0.07$ & $ 17.98 \pm 0.07$ & $ 18.29 \pm 0.05$ & $ 18.63 \pm 0.04$ & $ 18.93 \pm 0.06$ & $ 19.63 \pm 0.07$\\ 
 3 & $ 17.83 \pm 0.06$ & $ 17.99 \pm 0.08$ & $ 18.10 \pm 0.08$ & $ 18.44 \pm 0.04$ & $ 18.86 \pm 0.03$ & $ 19.22 \pm 0.04$ & $ 19.77 \pm 0.08$\\ 
 4 & $ 17.79 \pm 0.06$ & $ 17.87 \pm 0.08$ & $ 17.95 \pm 0.09$ & $ 18.46 \pm 0.10$ & $ 18.86 \pm 0.03$ & $ 19.22 \pm 0.06$ & $ 19.99 \pm 0.08$\\ 
 \midrule
 5 & $ 17.70 \pm 0.07$ & $ 17.79 \pm 0.08$ & $ 17.74 \pm 0.08$ & $ 17.97 \pm 0.06$ & $ 18.03 \pm 0.04$ & $ 18.33 \pm 0.05$ & $ 19.03 \pm 0.06$\\ 
 6 & $ 17.27 \pm 0.06$ & $ 17.55 \pm 0.08$ & $ 17.55 \pm 0.07$ & $ 18.02 \pm 0.07$ & $ 18.30 \pm 0.04$ & $ 18.57 \pm 0.06$ & $ 19.30 \pm 0.06$\\ 
 7 & $ 17.83 \pm 0.07$ & $ 17.97 \pm 0.08$ & $ 17.99 \pm 0.08$ & $ 18.44 \pm 0.07$ & $ 18.54 \pm 0.08$ & $ 18.68 \pm 0.19$ & -- \\ 
 8 & $ 18.13 \pm 0.06$ & $ 18.27 \pm 0.07$ & $ 18.33 \pm 0.07$ & $ 18.83 \pm 0.07$ & $ 19.10 \pm 0.05$ & $ 19.48 \pm 0.09$ & $ 20.31 \pm 0.14$\\ 
 16 & $ 18.05 \pm 0.08$ & $ 18.13 \pm 0.09$ & $ 18.13 \pm 0.11$ & $ 18.13 \pm 0.07$ & $ 18.50 \pm 0.07$ & $ 18.77 \pm 0.12$ & $ 19.94 \pm 0.15$\\ 
 \midrule
 26 & $ 16.46 \pm 0.05$ & $ 16.57 \pm 0.05$ & $ 16.59 \pm 0.05$ & $ 17.05 \pm 0.06$ & $ 17.37 \pm 0.03$ & $ 17.70 \pm 0.03$ & $ 18.45 \pm 0.04$\\ 
 \bottomrule 
\end{tabular}

\end{table*}

In this manuscript, we utilize photometry of the stars in our sample in 3 UV and 4 optical photometric bands:$UVW2$, $UVM2$, $UVW1$, $U$, $B$, $V$, and $I$.  Specifically, this data is used to estimate the bolometric luminosity and extinction of each star by fitting magnitudes computed for the best-fit spectral models to the observed photometry.  The optical photometry for all sources comes from the Magellanic Cloud Photometric Survey \citep{Zaritsky2002,Zaritsky2004}. \rev{Originally, these data are presented in the Vega magnitude system. We calculate zeropoint offsets to convert these to AB magnitudes by performing synthetic Vega and AB photometry on a subset of the stripped star models in our synthetic grid (described below) in order to minimize systematics due to the underlying spectral shape of the star. For a range of stripped star models, the resulting zeropoints vary by significantly less than the catalog magnitude uncertainties ($<$0.001 mag)}. The UV photometry was performed on images from the \emph{Swift-}UVOT Magellanic Cloud Survey \citep{Siegel2015,Hagen2017} as described in \citetalias{DroutGotberg23} and Ludwig et al.\ in prep. In particular, to mitigate the effects of crowding in the \emph{Swift} images, we performed forced point-spread-function photometry at the positions of the optical sources using the forward-modelling code \texttt{The Tractor} \citep{2016ascl.soft04008L}. Final magnitude calibration was then performed using standard HEASARC routines, and multiple observations the same source were averaged. 

All photometric data that are used in this study are presented in \tabref{tab:photometry}. UV photometry for stars 1--8 and 16 were originally published in \citetalias{DroutGotberg23}, and we have now added magnitudes computed via the same method for star 26.

%
\section{Spectral fitting}\label{sec:spectral_fitting}

To obtain stellar properties for the stars in the spectroscopic sample, we compute a grid of spectral models and adopt a $\chi^2$ minimization technique to identify the best-fit model and associated errors. Below, we describe these steps in detail. 

\subsection{Spectral model grid}\label{sec:spectral_model_grid}

We used the publicly available 1D non-LTE radiative transfer code CMFGEN \citep{1990A&A...231..116H,1998ApJ...496..407H} to compute a grid of stellar atmosphere models that we can use for spectral fitting and obtain properties of the stars in our spectroscopic sample. A subset of the models described here were used in \citetalias{DroutGotberg23} to estimate the effective temperature, surface gravity, and surface hydrogen mass fraction of stars 1--8 via a set of equivalent width diagnostics. We have now expanded this grid to cover a larger parameter space to aid in our spectral fitting. Below we describe the grid and computation method in detail.

These spectral models are based on those presented in \citet{2018A&A...615A..78G}, which in turn stem from \citet{2008A&A...485..245G} and the openly available O-star grid on the CMFGEN website\footnote{\url{http://kookaburra.phyast.pitt.edu/hillier/web/CMFGEN.htm}}. 
For these models, we include the elements H, He, C, N, O, Si, and Fe. 
We compute the model spectra between 50 and 50,000 \AA. Depending on the density of the wind, we adopt a suitable extent of the atmosphere, which is between 6 and 1000 times the surface radius. We use a minimum number of mesh points of 40, but up to more than 100, together with 15 core rays. 

We vary three parameters in the spectral model grid: (1) the temperature (\rev{$T_{\star}=$} 30, 33, 35, 37, 40, 50, 60, 70, 80, 90, 100, 110, 120, 130, 140, 150 kK), (2) the surface gravity (\rev{$\log_{10} g_{\star}/({\rm cm \; s}^{-2})=$} 4.0, 4.3, 4.5, 4.8, 5.0, 5.2, 5.5, 5.7, 6.0), and (3) the surface hydrogen mass fraction ($X_{\rm H, surf}=$ 0.01, 0.1, 0.3, 0.5, 0.7), which also determines the surface helium mass fraction ($X_{\rm He, surf}=$ \rev{0.985, 0.895, 0.695, 0.495, 0.295}). 
We set the metallicity to be that which is expected for helium-core burning stars stripped in binaries, using the $Z=0.006$ evolutionary model grid from \cite{2018A&A...615A..78G}, which was scaled to solar values \citep{1998SSRv...85..161G}. The resulting stripped star metal composition on the surface led to mass fractions of $X_{\rm C, surf} = 3\times 10^{-5}$, $X_{\rm N, surf} = 4\times 10^{-3}$, $X_{\rm O, surf} = 1 \times 10^{-4}$, $X_{\rm Si, surf} = 1.5\times 10^{-4}$, and $X_{\rm Fe, surf} = 2.5 \times 10^{-4}$ after envelope-stripping. Adding the adopted abundances together gives $Z = 0.00453$. 

The CNO abundances originate from layers that once were part of the convective main-sequence core, and thus have experienced complete CNO processing. In the structure models of \citet{2018A&A...615A..78G}, the nitrogen and oxygen abundances have a rough constant level from the surface to the convective helium-burning core, while the carbon abundance increases by roughly a factor of three from the surface in to the hydrogen-free layer. This larger change in carbon is balanced by oxygen. However, because oxygen is more abundant, the fractional abundance change of oxygen is not prominent. Here, we refrain from a detailed analysis of possible variations of CNO abundances, which will be the topic of a future study. 
We note that none of the metal lines are used in our spectral fitting process (see \secref{sec:fitting_routine}).  

To create a spectral model grid that easily can be scaled to the desired radius or luminosity, we fix the stellar radius in the models to 0.5\Rsun. 
While this radius is typical for the expectations of envelope-stripped stars \citep[see Tables 1 and B.1-B.3 of][]{2018A&A...615A..78G}, we note that we scale the spectral models during the fitting procedure so that the radius is a free parameter.
We let the luminosity adapt to the assumed radius and temperature, resulting in $L_{\rm bol}\sim 200-22,000\, L_{\odot}$. 
We set the code to match the input temperature, radius and surface gravity at an optical depth of $\tau = 20$ \rev{(quantities denoted by $\star$)}, following \cite{2008A&A...485..245G}. We note that these properties are \rev{very similar} at the photospheric optical depth of $\tau = 2/3$ \rev{(quantities denoted by $\mathrm{eff}$), but not exactly the same. Differences between the quantities at $\tau=20$ and $\tau=2/3$ are somewhat larger for models closer to the Eddington limit (see below)}.

Because the stars in the spectroscopic sample lack the typical emission lines originating from stellar winds, we adopt weak, fast, and relatively smooth stellar winds for the models in our primary grid. To do this, we assume mass-loss rates of $10^{-9}\Msunyr$, terminal wind speeds of 2500\kms, which corresponds to one to several times the surface escape speed as has been measured for massive stars \citep{1995ApJ...455..269L}, and modest clumping by assuming a volume filling factor, $f_{\rm vol}$, of 0.5. For the wind velocity profile, we assume a $\beta$-law ($v(r) = v_{\infty}(1-R_\star/r)^{\beta}$), setting $\beta = 1$. In section~\ref{sec:wind} we will vary these parameters to obtain rough estimates for the mass-loss rates for the stars in our sample. 
\rev{We adopt a turbulent velocity of 20\kms, in common with Magellanic Cloud O-type stars \citep{2017A&A...600A..81R}. The impact of turbulence and thermal broadening is negligible for the diagnostic \HeII and H\textsc{i} lines, which are dominated by (Stark) pressure-broadening. There is no evidence for rotational broadening contributing significantly to the Pickering-Balmer lines, although we defer an investigation of rotation rates using metal lines to a future investigation. Before using the models for spectral fitting, we also degrade them to the spectral resolution of MagE using a Gaussian kernel.}

\begin{figure}
\centering
\includegraphics[width=0.5\textwidth]{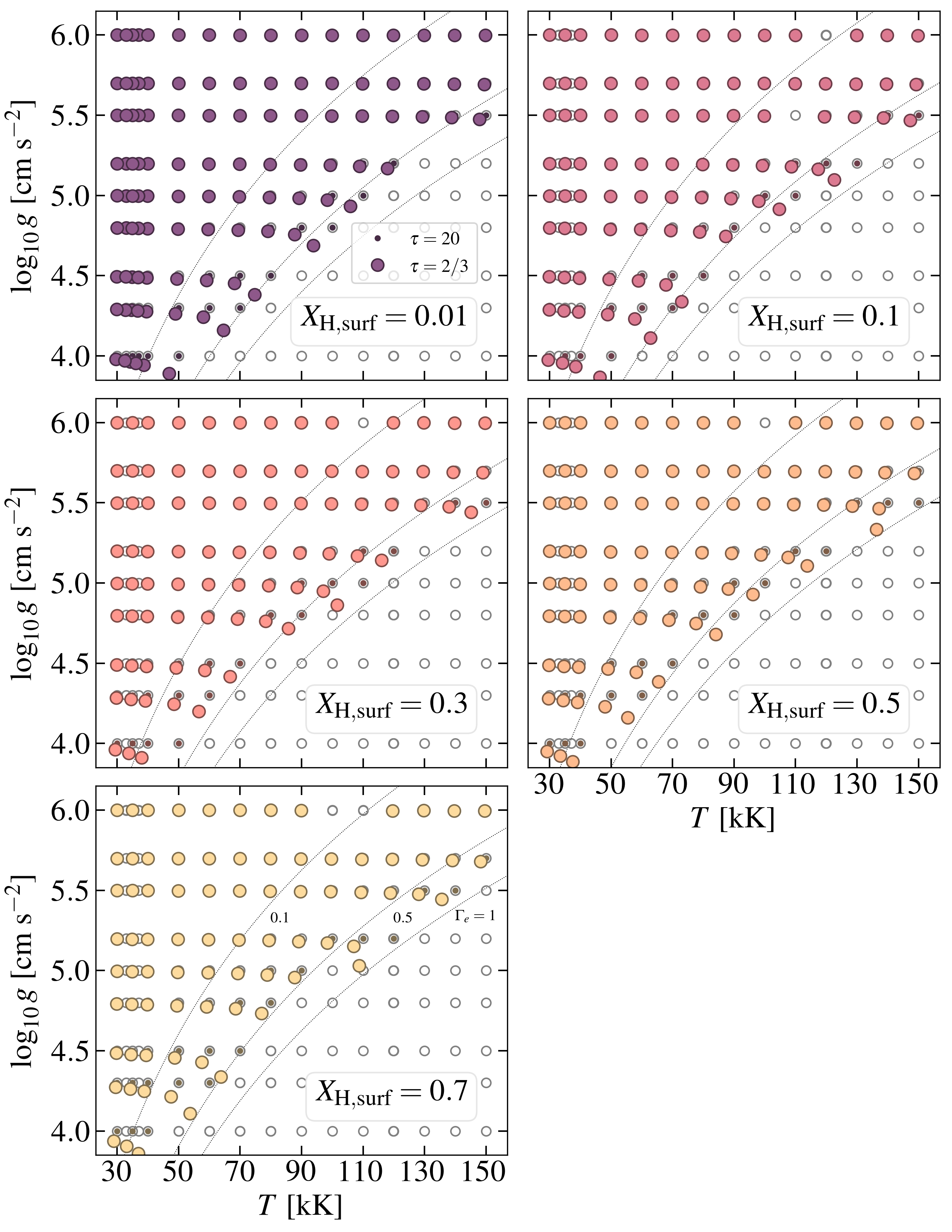}
\caption{Coverage of the spectral model grid used as base for spectral fitting to obtain stellar properties of stripped stars. This visualization shows which models have reached convergence using a colored circle\rev{, where the small and large circles correspond to the temperature and surface gravity at optical depth $\tau=20$ and 2/3, respectively}. Thin, dotted lines indicate where the Eddington factor is 0.1 (leftmost), 0.5 (middle), and 1 (rightmost).}
\label{fig:grid_representation}
\end{figure}

The resulting spectral model grid covers most of the intended parameter space, as shown in \figref{fig:grid_representation}. \rev{The figure shows that the difference between the temperature and surface gravity evaluated at $\tau=20$ and $\tau=2/3$ is negligible for most of the models, and at maximum the temperature ($T_{\star}$ and $T_{\rm eff}$) and surface gravity ($\log_{10} g_{\star}$ and $\log_{10} g_{\rm eff}$) differ by 10\% and 5\%, respectively. }
We encountered numerical convergence issues when high temperatures and low surface gravity are combined, because these combinations approach the Eddington limit ($\Gamma_e = 1$, see \secref{sec:photometric_fit} and the dotted lines in \figref{fig:grid_representation}). We note that the Eddington factor is independent of the assumed radius, mass, and bolometric luminosity.
Because the spectral morphology changes significantly between 30 and 40 kK, we introduce the 35 kK models for all surface hydrogen mass fractions, and also the 33 and 37 kK models for the surface hydrogen mass fraction $X_{\rm H, surf} = 0.01$. 
In total, the grid contains 441 models and \revvv{we make the full grid publicly available on Zenodo under a Creative Commons Public Domain license: \dataset[doi:10.5281/zenodo.7976200]{https://doi.org/10.5281/zenodo.7976200}. Please cite both the present article and the Zenodo dataset when reusing these model grids \citep{prop_zenodo3}.}

\subsection{Fitting routine}\label{sec:fitting_routine}


We employ the $\chi^2$ minimization technique to obtain the best-fit spectral model and models allowed within $1\sigma$ deviation for each star. This gives rise to measurements for their effective temperatures, \rev{effective} surface gravity, hydrogen and helium surface mass fractions, and flux-weighted gravity. We then match the spectral models to the observed photometry to obtain extinction and luminosity, which in turn can be used to calculate the \rev{effective} radius, spectroscopic mass, and Eddington factor. Finally, we use a set of evolutionary models and our derived \rev{bolometric} luminosities to estimate evolutionary masses under the assumption the stars are central helium-burning (this assumption \rev{is} investigated in \secref{sec:evol_stage}).

\rev{Using $\chi^2$ minimization in our rather finely spaced and interpolated grid ensures that the model with the truly smallest $\chi^2$ is found. Because all models within the chosen parameter space are included, the best-fit model will represent the true minimum and not a local minimum.}
\rev{Concerning the errors, artefacts related to the data reduction (\secref{sec:obs_spectroscopy}) and implementation of physical processes in CMFGEN \citep{2013ApJ...768....6M} could mean that the formal 1$\sigma$ errors we obtain in the $\chi^2$ analysis are slightly underestimated.}
Below, we describe the details of the adopted \rev{fitting} procedure\footnote{The spectral fitting routine is made \revvv{publicly available on Zenodo under  a Creative Commons Public Domain license  \citep{prop_zenodo2}.}}

\subsubsection{Treatment of spectral lines}

\begin{figure}
\centering
\includegraphics[width=\columnwidth]{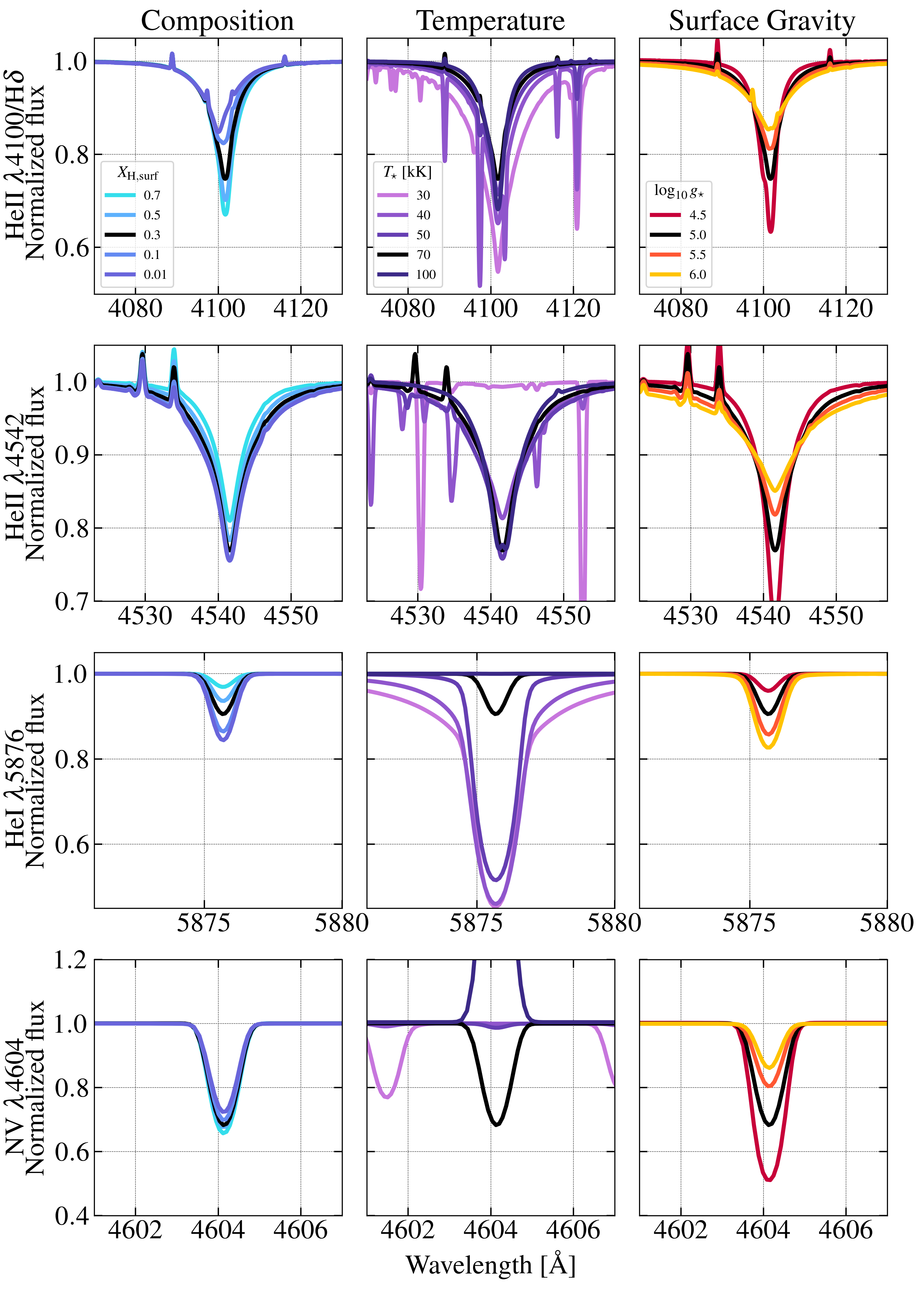}
\caption{Effect of varying surface hydrogen/helium mass fraction (left), \rev{temperature} (center), and surface gravity (right) on the spectral lines H$\delta$/\HeIIl 4100 (first row), \HeIIl 4542 (second row), \HeIl 5876 (third row), and \NV $\lambda$4604 (fourth row). We use the model with $T_{\rm eff} = 70$ kK, $\log_{10} g = 5.0$ and $X_{\rm H, surf} = 0.3$ as a base (black solid line) and vary each parameter according to the legends.
}
\label{fig:grid_exploration}
\end{figure}

When fitting spectral models to the data, we choose to fit only to certain spectral lines \citep[this is, for example, also done in the fitting procedure in the IACOB survey, see][]{2011JPhCS.328a2021S}.
The choice of what lines to fit to is important, because they are affected differently by parameter variations. 
This is demonstrated in \figref{fig:grid_exploration}, where we show the effect of varying the surface hydrogen to helium content, the \rev{temperature}, and the surface gravity, on the four spectral lines \HeIIl 4100/H$\delta$, \HeIIl 4542, \HeIl 5876, and \NV $\lambda$4604. For this figure, we start from the parameters \rev{$T_{\star} = 70$ kK, $\log_{10} g_{\star} = 5.0$} and $X_{\rm H, surf} = 0.3$ and vary each parameter. 

The left panels of \figref{fig:grid_exploration} show that the surface mass fraction of helium and hydrogen affect the central wavelength of the \HeIIl 4100/H$\delta$ line blend along with the strength of \HeIIl 4100/H$\delta$, \HeIIl 4542 and \HeIl 5876. The effect on the nitrogen line is negligible. 
The central panels show that effective temperature significantly affects the strength of \HeIl 5876 and \HeIIl 4542 for \rev{$T_{\star} \lesssim 70$ kK}, but these lines are minimally affected for variations at higher temperature. In fact, \HeIl 5876, the most temperature sensitive \HeI line in the spectral range, disappears for \rev{$T_{\star} >$70-80 kK} (see also \figref{fig:N_He_structure}). The nitrogen ionization balance is also sensitive to temperature variations for \rev{$T_{\star} > 70$ kK}. In \figref{fig:grid_exploration} \NV $\lambda$4604 is not present for \rev{$T_{\star} \leq 50$ kK}, appears in absorption for \rev{$T_{\star} = 70$ kK}, and in emission for \rev{$T_{\star}=100$ kK}. However, to fully trace these variations of the nitrogen features we would require both higher signal-to-noise spectra and to expand the model grid to vary the surface nitrogen mass fraction.
Finally, the right panels of \figref{fig:grid_exploration} show that variations in surface gravity affect both the strength and shape of the hydrogenic line transitions of \HeIIl 4100/H$\delta$ and \HeIIl 4542. The effect of surface gravity on \HeIl 5876 and \NV $\lambda$4604 is moderate. 

Summarizing, to probe the \rev{parameters} of the model grid when fitting the observed spectra, it is important to include (1) both pure \HeII and \HeI lines when possible, since it gives the most accurate temperature determination and (2) a combination of pure \HeII and H/\HeII blended lines to trace surface hydrogen to helium content. This set will thus also include lines that are affected by Stark broadening and trace surface gravity. 
\rev{In choosing the final set of lines to fit, we avoid fitting to the $\alpha$ lines H$\alpha$/\HeIIl 6560 and \HeIIl 4686 because of their sensitivity to stellar wind and nebular contamination.} This choice differs from analysis of the more luminous Wolf-Rayet and WN3/O3 stars where the $\alpha$-lines often are used as primary diagnostic lines \citep{1995A&A...302..457C, 2017ApJ...841...20N}. 
The final set of lines used to fit the spectrum for each star are listed in \tabref{tab:fit_lines} in Appendix~\ref{app:details_spectral_fitting}. 

We renormalize the continuum for each spectral line individually before fitting with models. This is done by fitting a horizontal line to the $\sim$10 \AA\ regions on both sides of each line. We then hold the continuum fixed in our $\chi^2$ minimization. We select wavelength range that will be fit for each line by finding where the wings of the observed line first increase above the continuum level of 1 (due to noise fluctuations) on both sides of the central wavelength.

When computing $\chi^2$ for one model, we compute the $\chi^2$ for each line individually and then sum these together, meaning that all lines are weighted equally. Because some lines are narrower than others, this means that these will carry somewhat less importance to the fit compared to broader lines, which are composed of more data points. However, in tests with higher weighted narrow lines, we did not find significant improvements of the fits and therefore choose to not include different line weights.


\subsubsection{Interpolating and constraining the spectral model grid}

To obtain better fits and finer resolution in the measured parameters, we interpolate the spectral model grid. 
The interpolation is linear in \rev{$T_{\star}$, $\log_{10} g_{\star}$} and $X_{\rm H, surf}$. We choose to sample \rev{$T_{\star}$} every 2 kK between 30 and 150 kK, \rev{$\log_{10} g_{\star}$} every 0.1 steps between 4.0 and 6.0, and $X_{\rm H, surf}$ in steps of 0.05 between 0.05 and 0.7 (in addition to the computed models at 0.01). We do not extrapolate the grid, meaning that the high temperature and low surface gravity corner still is not populated with models (cf.\ \figref{fig:grid_representation}). 

\begin{figure}
\centering
\includegraphics[width=0.35\textwidth]{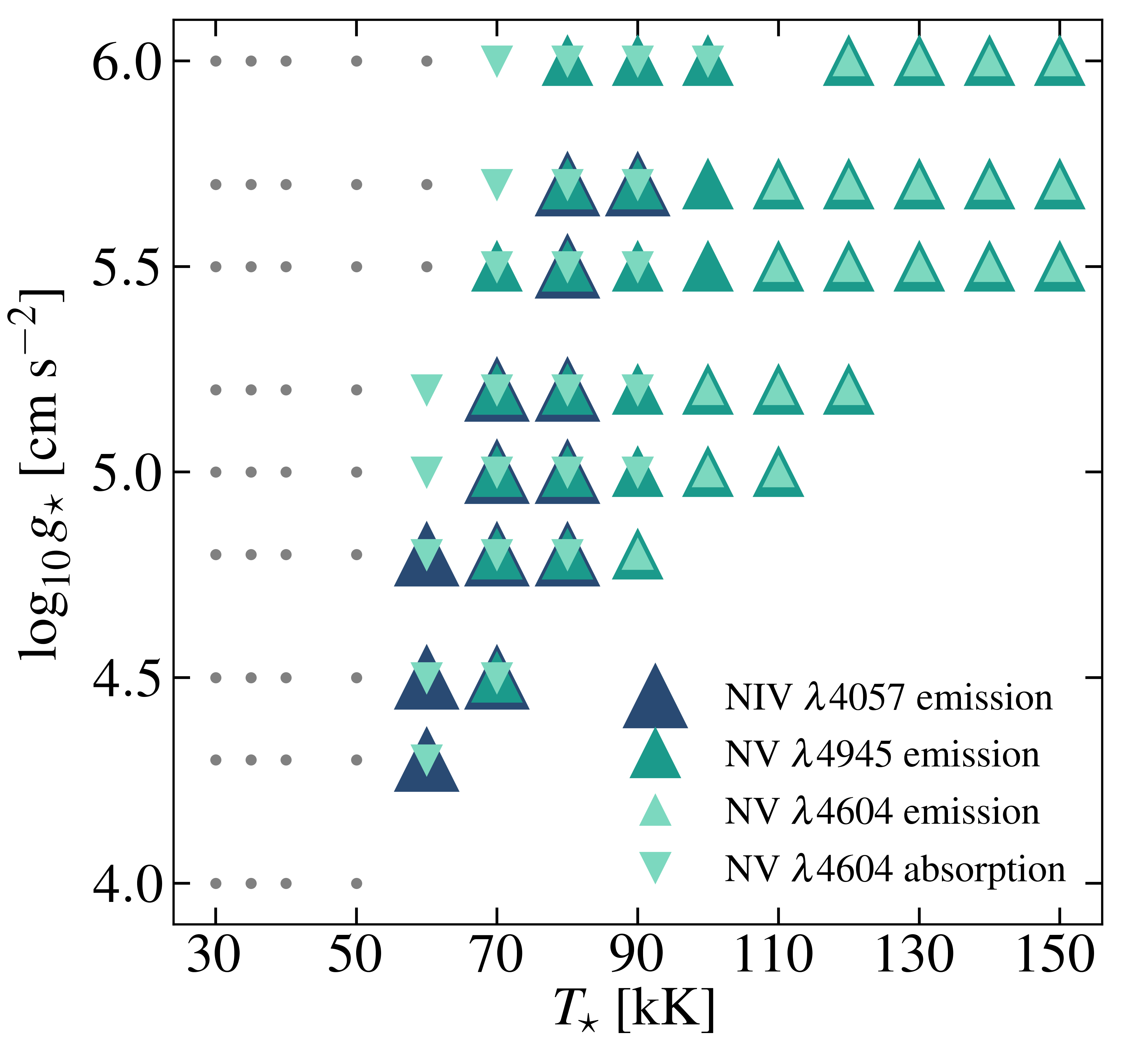}
\includegraphics[width=0.35\textwidth]{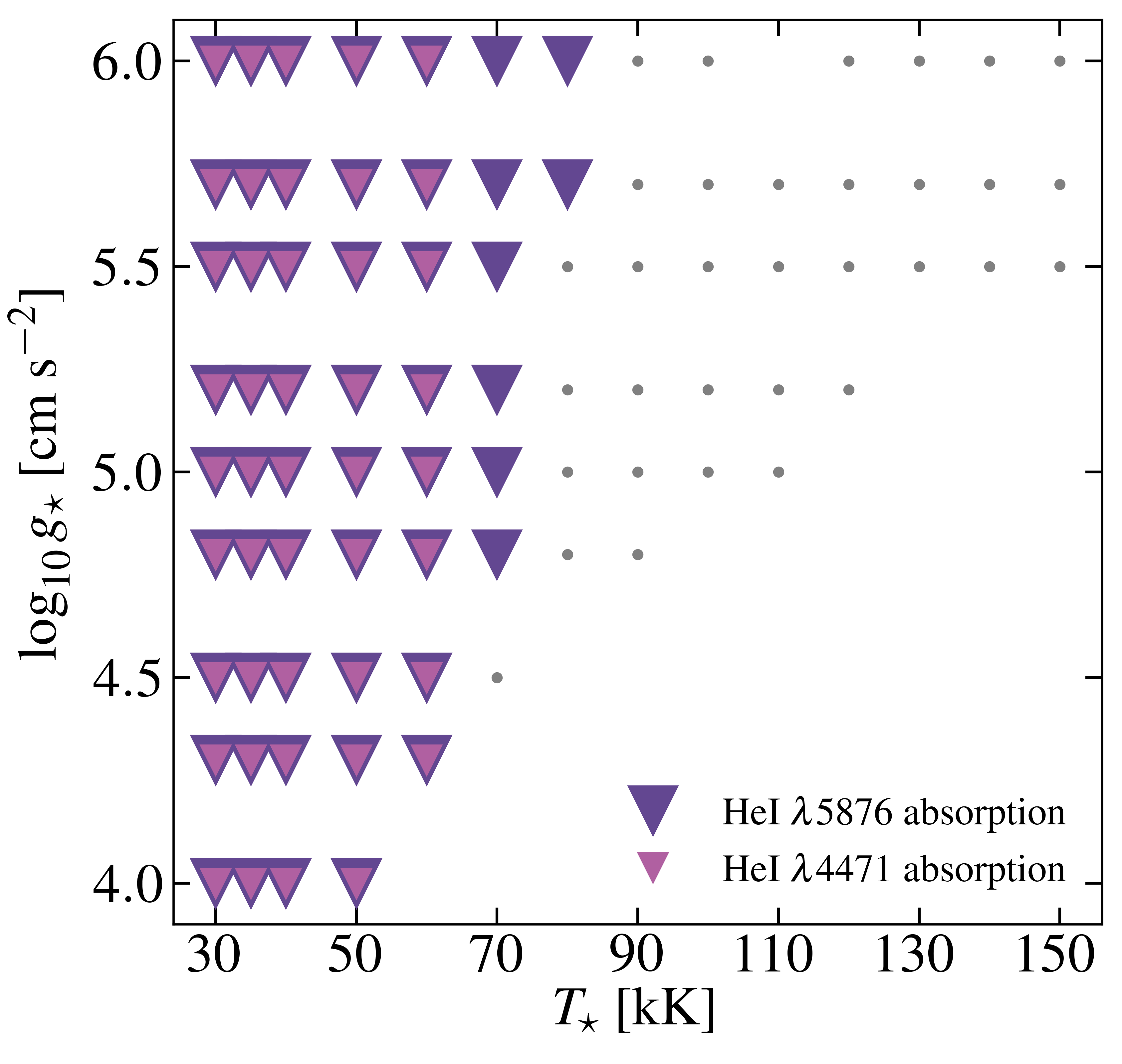}
\caption{We use the presence of a set of nitrogen lines (top) and helium lines (bottom) to constrain the model grid used when fitting the observed spectra. As an example, we show here the models with surface hydrogen mass fraction $X_{\rm H, surf} = 0.3$, spread out in the \rev{temperature - surface gravity} plane. Triangular markers show the presence of nitrogen or helium lines that are used to constrain the grid (see also \tabref{tab:fit_lines}). Gray circles indicate models in which none of the lines specified by the legend are present.}
\label{fig:N_He_structure}
\end{figure}

In addition, we use the presence of various nitrogen and \HeI\ lines to  help constrain the the temperature range to from the full model grid to consider when fitting each individual star. While \HeII and H lines are present throughout the entire model grid, the same is not the case for nitrogen and \HeI. Specifically, although the strength and detailed line profile of the nitrogen features are dependent on the abundance of nitrogen (which we do not vary in our grid) their presence can provide a sensitive temperature diagnostic at \rev{$T_{\star} > 60$} kK. As demonstrated in the top panel of \figref{fig:N_He_structure}, \NIV $\lambda$ 4057 is present in emission roughly at \rev{$T_{\star} \sim 60-80$ kK} (dark blue triangles), \NV $\lambda$ 4604 appears in absorption for \rev{$T_{\star} \sim 60-90$ kK} (downward cyan triangles), while it flips into emission for \rev{$T_{\star} \gtrsim 100$ kK} (upward cyan triangles), and \NV $\lambda$ 4945 appears in emission for \rev{$T_{\star} \gtrsim 70$ kK} (teal triangles). On the low temperature end, \HeI\ can provide a similar discriminant. As the bottom panel of \figref{fig:N_He_structure} shows, \HeIl 5876 is present for \rev{$T_{\star} \lesssim 70$ kK} (purple triangles) and \HeIl 4471 for \rev{$T_{\star} \lesssim 60$ kK} (pink triangles).  We note that \figref{fig:N_He_structure} only shows the part of the grid with surface hydrogen mass fraction $X_{\rm H, surf} = 0.3$ for illustration purposes, but the line presence only varies slightly with different surface hydrogen mass fraction. 

When one or more of the described lines are present in an observed spectrum, we use it to constrain the model grid used in our fitting procedure. The constraints we use for each star are given in \tabref{tab:fit_lines} in Appendix~\ref{app:details_spectral_fitting}. We do not use the absence of lines to constrain the model grid since poor signal-to-noise ratio or exact nitrogen abundance can affect whether the line is visible. 


\subsubsection{Spectral fitting to obtain \rev{$T_{\star}$}, $T_{\rm eff}$, \rev{$\log_{10} g_{\star}$, $\log_{10} g_{\rm eff}$}, $X_{\rm H, surf}$, $X_{\rm He, surf}$, and $\mathcal{L}$}\label{sec:spectroscopic_fit}

For each star, we calculate the $\chi^2$ of all models in the interpolated and constrained grid and determine which one is the best-fit model by finding the one with smallest $\chi^2$ (designated $\chi^2_{\rm min}$). 
The models with $\chi^2 < \chi^2_{\rm min} + \Delta \chi^2$ are regarded as acceptable models and their properties are used to determine the errors on the fitted parameters. We determine $\Delta \chi^2$ by calculating the 68.27\% confidence interval based on the number of degrees of freedom. 
The calculation of $\Delta \chi^2$ is done using the \texttt{python} function \texttt{scipy.stats.chi2.ppf} (see however \citealt{1992nrfa.book.....P}).

We use \rev{the temperature and surface gravity at $\tau=20$ and $\tau=2/3$, along with the} surface hydrogen mass fraction of the best-fit model as the best-fit values for these parameters \rev{($T_{\star}$, $T_{\rm eff}$, $\log_{10} g_{\star}$, $\log_{10} g_{\rm eff}$, and $X_{\rm H, surf}$)}. For the 1$\sigma$ errors on these parameters, we use the maximum and minimum values among the models that fulfil $\chi^2 < \chi^2_{\rm min} + \Delta \chi^2$.

Two more stellar parameters can be derived directly from these model fits. First, the surface helium mass fraction, which is simply $X_{\rm He, surf} = 1-X_{\rm H, surf} - Z$. This corresponds to $X_{\rm He, surf} = 0.98547$, 0.89547, 0.69547, 0.49547, and 0.29547 for $X_{\rm H, surf}=$ 0.01, 0.1, 0.3, 0.5, and 0.7 (see \secref{sec:spectral_model_grid} for information on $Z$).
Second, the inverse of the flux-weighted gravity, $\mathcal{L} \equiv T_{\rm eff}^4/g$ \citep{2003ApJ...582L..83K, 2014A&A...564A..52L}, can be calculated for each model and thus also determined using the $\chi^2$ method outlined above. We present $\mathcal{L}$ in solar units, $\mathcal{L}_{\odot}$, calculated assuming $T_{\rm eff, \odot} = 5,777$ K and $g_{\odot} = 27,400$ cm~s$^{-2}$. Note that the inverse of the flux-weighted gravity is very sensitive to uncertainties in the effective temperature, due to the fourth power in its definition.

\subsubsection{Obtaining $L_{\rm bol}$, $A_V$, $R_{\rm eff}$, $M_{\rm spec}$, and $\Gamma _{\rm e}$} \label{sec:photometric_fit}

In order to determine bolometric luminosities, we fit the spectral energy distributions (SEDs) of the acceptable models to the observed photometry of each star, including extinction as a free parameter. 
For each spectral model, we scale the spectrum to produce a range of bolometric luminosities between roughly 1 and $10^6\Lsun$. We then apply a range of extinction values between $A_V = 0-1.5$ mag separated in steps of 0.01 mag, adopting the extinction curves from \cite{2003ApJ...594..279G}\footnote{We employ the functions \texttt{averages.G03\_LMCAvg} and \texttt{averages.G03\_SMCBar} of the \texttt{python} package \texttt{dust\_extinction} for this calculation (\url{https://dust-extinction.readthedocs.io/en/stable/}).}. \rev{For simplicity, we only adopt the average extinction curve for each of the Magellanic clouds, and do not explicitly include a separate Milky Way foreground component in the fitting. While the LMC and Milky Way extinction curves are comparable in the wavelength regions of interest, we discuss any impact of differences in the shape of the SMC and Milky Way curves in the ultraviolet in \secref{sec:stellar_properties}. The exception for this approach is star 26 evaluated at 10 kpc distance, where we only adopt the Milky Way extinction curve.} 
We calculate the AB magnitudes of each resulting model in the \emph{Swift} UVW2, UVM2, UVW1, and optical UBVI bands using the filter functions from the SVO filter service\footnote{\url{http://svo2.cab.inta-csic.es/theory/fps/}} \citep{2012ivoa.rept.1015R, 2020sea..confE.182R}. We then calculate the chi-square statistic for the resulting modeled magnitudes compared to the observed photometric data, adopting distances of 50 kpc to the LMC \citep{2013Natur.495...76P} and 62 kpc to the SMC \citep{2020ApJ...904...13G}\footnote{We consider both a foreground, 10 kpc distance and the LMC distance for star 26 when preparing the parameter fit.}. 
\rev{Because extinction has larger influence in the UV compared to the optical, we prefer to use the described method fitting to photometry, rather than for example assessing flux calibrated optical spectra, which furthermore often have larger systematic uncertainties in absolute calibration.}

We apply the above procedure to all models that fall within the $\chi^2 < \chi^2_{\rm min} + \Delta \chi^2$ threshold from the spectral fitting (Section~\ref{sec:spectroscopic_fit}), resulting in a range of $L_{\rm{bol}}$ and $A_V$ values for each star. (Because the photometric errors are small, we simply find a single best-fit value of these parameters for each spectral model.) For each star, we adopt the $L_{\rm{bol}}$ and $A_V$ found for the best-fit spectral model from Section~\ref{sec:spectroscopic_fit} as our baseline values. Errors are determined based on the minimum and maximum values found from fitting the larger sample of models accepted within $1\sigma$ from the spectral fitting.

For each model, we compute the effective radius using the bolometric luminosity and effective temperature following the Stefan-Boltzmann's law ($L_{\rm bol} = 4\pi R_{\rm eff}^2 \sigma T_{\rm eff}^4$) and the spectroscopic mass by combining the surface gravity and effective radius (\rev{$g_{\rm eff} = GM_{\rm spec}/R_{\rm eff}^2$)}. As with extinction and bolometric luminosity, for each star we adopt the \rev{effective} radius and spectroscopic mass found from the best-fit spectral model as our baseline values. Quoted errors similarly correspond to minimum and maximum values found from all models within 1$\sigma$ based on the spectroscopic fit.

With the bolometric luminosity and spectroscopic mass, we can also estimate the Eddington factor for Thomson scattering, $\Gamma _{\rm e}$, which describes how close the star is to the Eddington limit \citep{2011A&A...535A..56G}. The Eddington factor is defined as follows

\begin{equation}
\Gamma _{\rm e} = \dfrac{\kappa _{\rm e} L_{\rm bol}}{4\pi c G M_{\rm spec}} = \dfrac{\kappa_{\rm e} \sigma T_{\rm eff}^4}{cg_{\rm eff}}, 
\end{equation}
where $c$ is the speed of light, $G$ is the gravitational constant, and $\kappa _{\rm e}$ is the electron scattering opacity, defined as $\kappa_{\rm e} = 0.2(1+X_{\rm H, surf})$ cm$^2$ g$^{-1}$. 

\subsubsection{Estimating the evolutionary mass, $M_{\rm evol}$}\label{sec:Mevol}

\begin{figure}
\centering
\includegraphics[width=0.35\textwidth]{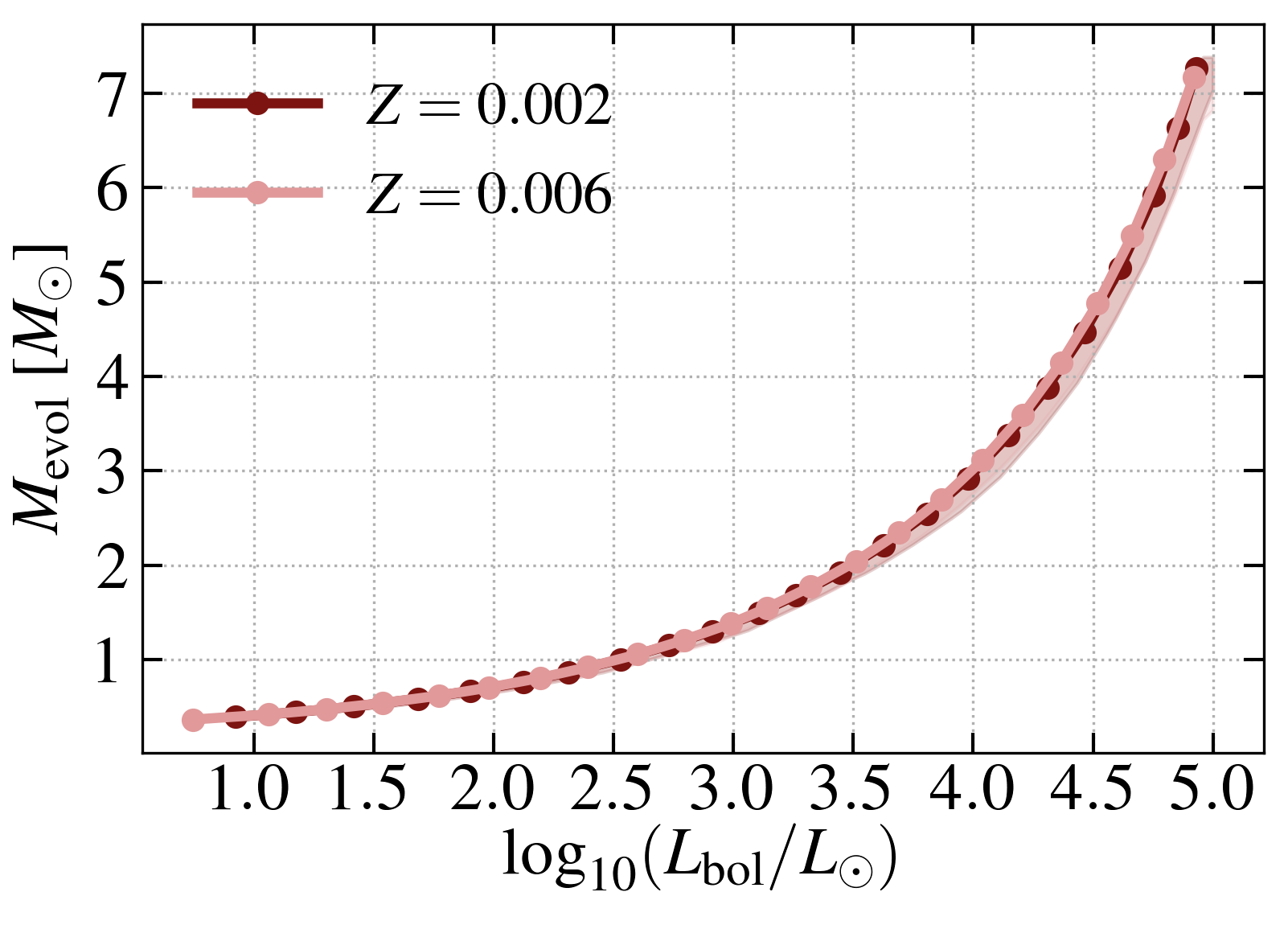}
\caption{The relation between mass and luminosity for stripped stars half-way through central helium burning ($X_{\rm He, center} \equiv$ 0.5) is used to estimate the evolutionary mass. Here, we show the relation for stars stripped via stable mass transfer using evolutionary models from \citet{2018A&A...615A..78G} for $Z=0.006$ (pink line) and $Z=0.002$ (dark red line). The shaded background color demonstrates the variation throughout helium-core burning when the central helium mass fraction is between 0.8 and 0.1.
}
\label{fig:ML}
\end{figure}

Finally, we estimate the evolutionary masses for the stars in our sample using the relation between mass and luminosity for stripped stars that have reached half-way through central helium burning, defined as when $X_{\rm He, center} = 0.5$. To find this relation, we use the evolutionary models of \citet{2018A&A...615A..78G} and plot the stellar mass and bolometric luminosity in \figref{fig:ML}. In the figure, we show the relations for both the $Z=0.002$ and $Z=0.006$ grids, which closely overlap. 
The mass-luminosity relation shown in \figref{fig:ML} roughly follows: 

\begin{equation}
\log_{10} (M_{\rm evol}/M_{\odot}) \approx 0.32 \log_{10} (L_{\rm bol}/L_{\odot}) -0.80.
\end{equation}

This mass-luminosity relation should be a decent approximation for the mass-luminosity relation throughout helium-core burning, since it does not significantly change during this phase. This is demonstrated in \figref{fig:ML} where we use shaded background to show the variation in these parameters for central helium mass fractions between 0.8 and 0.1. 
However, we note that this definition of the evolutionary mass assumes that the stripped stars are in the phase of helium-core burning and are not currently contracting or expanding \citep[cf.][]{2020A&A...637A...6L}. We emphasize that using this relation to estimate the mass for stripped stars that are inflated may lead to an overestimated evolutionary mass. We will directly assess this for stars in our sample (e.g., for star 16) in  \secref{sec:evol_stage}. 

The models of \citet{2018A&A...615A..78G} reach stripped star masses of $\sim 7.2 \Msun$ and bolometric luminosities up to $\sim 10^5 L_{\odot}$. As, in particular, star 1 could reach higher values, we allow for extrapolation of the mass-luminosity relation.

\section{Stellar properties}\label{sec:stellar_properties}

\begin{sidewaystable}
\begin{center}
\caption{Stellar properties of the stripped stars in our spectroscopic sample.}
\label{tab:stellar_properties}
{\scriptsize
\begin{tabular}{lcccccccccccccccccc}
\toprule\midrule
Star & $T_{\mathrm{eff}}$ & $T_{\star}$ & $\log_{10} g_{\mathrm{eff}}$ & $\log_{10} g_{\star}$ & $X_{\mathrm{H,surf}}$ & $X_{\mathrm{He,surf}}$ & $A_{\mathrm{V}}$ & $\log_{10} L_{\mathrm{bol}}$ & $\log_{10} \mathcal{L}$ & $R_{\mathrm{eff}}$ & $M_{\mathrm{spec}}$ & $M_{\mathrm{evol}}$ & $\log_{10} Q_0$ & $\log_{10} Q_1$ & $\log_{10} Q_2$ & $\Gamma _{\rm e}$ & $v_{\rm esc}$ & $\dot{M}_{\rm wind}$\\ 
 & [kK] & [kK] & [cm s$^{-2}$] & [cm s$^{-2}$] & & & [AB mag] & [$L_{\odot}$] & [$\mathcal{L}_{\odot}$] & [$R_{\odot}$] & [$M_{\odot}$] & [$M_{\odot}$] & [s$^{-1}$] & [s$^{-1}$] & [s$^{-1}$] & & [km s$^{-1}$] & [$M_{\odot}$/yr] \\ 
\midrule
Star 1 & $90^{+10}_{-4}$ & $92^{+12}_{-4}$ & $4.96^{+0.1}_{-0.1}$ & $5.0^{+0.1}_{-0.1}$ & $0.40^{+0.10}_{-0.10}$ & $0.59^{+0.10}_{-0.10}$  & $0.22^{+0.02}_{-0.01}$ & $5.09^{+0.14}_{-0.08}$ & $4.25^{+0.15}_{-0.06}$ & $1.44^{+0.05}_{-0.09}$ & $6.97^{+1.84}_{-1.70}$  & $8.45^{+1.04}_{-0.58}$ & $48.93^{+0.11}_{-0.07}$ & $48.72^{+0.14}_{-0.09}$ & $47.19^{+0.71}_{-0.33}$ & $0.379^{+0.113}_{-0.051}$ & $1358^{+ 173}_{- 179}$ & $\sim 10^{-7}$ \\ 
Star 2 & $63^{+28}_{-2}$ & $64^{+28}_{-2}$ & $4.99^{+0.5}_{-0.4}$ & $5.0^{+0.5}_{-0.4}$ & $0.35^{+0.25}_{-0.25}$ & $0.65^{+0.25}_{-0.25}$  & $0.30^{+0.06}_{-0.01}$ & $4.12^{+0.48}_{-0.07}$ & $3.62^{+0.69}_{-0.45}$ & $0.95^{+0.02}_{-0.17}$ & $3.18^{+6.96}_{-2.00}$  & $3.31^{+1.82}_{-0.18}$ & $47.94^{+0.50}_{-0.07}$ & $47.61^{+0.63}_{-0.10}$ & $43.55^{+2.82}_{-0.19}$ & $0.086^{+0.290}_{-0.053}$ & $1133^{+ 894}_{- 441}$ & $< 10^{-9}$ \\ 
Star 3 & $71^{+74}_{-7}$ & $72^{+76}_{-8}$ & $5.4^{+0.6}_{-0.5}$ & $5.4^{+0.6}_{-0.5}$ & $0.10^{+0.25}_{-0.09}$ & $0.90^{+0.09}_{-0.25}$  & $0.21^{+0.06}_{-0.04}$ & $4.15^{+0.84}_{-0.17}$ & $3.42^{+1.15}_{-0.60}$ & $0.77^{+0.05}_{-0.27}$ & $5.32^{+15.84}_{-3.96}$  & $3.38^{+4.34}_{-0.42}$ & $47.98^{+0.71}_{-0.18}$ & $47.71^{+0.87}_{-0.24}$ & $43.86^{+4.28}_{-0.60}$ & $0.044^{+0.557}_{-0.033}$ & $1629^{+1625}_{- 739}$ & $< 10^{-9}$ \\ 
Star 4 & $67^{+79}_{-11}$ & $68^{+82}_{-12}$ & $5.09^{+0.7}_{-0.4}$ & $5.1^{+0.7}_{-0.4}$ & $0.30^{+0.35}_{-0.20}$ & $0.69^{+0.20}_{-0.35}$  & $0.14^{+0.06}_{-0.04}$ & $4.01^{+0.95}_{-0.26}$ & $3.62^{+0.97}_{-0.86}$ & $0.73^{+0.08}_{-0.27}$ & $2.43^{+11.17}_{-1.59}$  & $3.04^{+4.43}_{-0.59}$ & $47.84^{+0.80}_{-0.31}$ & $47.54^{+1.00}_{-0.42}$ & $43.72^{+4.43}_{-1.37}$ & $0.084^{+0.612}_{-0.071}$ & $1123^{+1472}_{- 450}$ & $< 10^{-9}$ \\ 
\midrule 
Star 5 & $66^{+18}_{-6}$ & $68^{+18}_{-6}$ & $4.46^{+0.3}_{-0.3}$ & $4.5^{+0.3}_{-0.2}$ & $0.01^{+0.04}_{-0.00}$ & $0.98^{+0.00}_{-0.04}$  & $0.63^{+0.06}_{-0.05}$ & $4.34^{+0.33}_{-0.17}$ & $4.22^{+0.19}_{-0.30}$ & $1.12^{+0.05}_{-0.11}$ & $1.31^{+1.42}_{-0.59}$  & $4.06^{+1.45}_{-0.54}$ & $48.19^{+0.33}_{-0.19}$ & $47.86^{+0.42}_{-0.24}$ & $44.02^{+1.66}_{-0.63}$ & $0.258^{+0.143}_{-0.130}$ & $ 668^{+ 302}_{- 175}$ & $\sim 10^{-8}$ \\ 
Star 6 & $73^{+14}_{-15}$ & $74^{+14}_{-16}$ & $4.99^{+0.3}_{-0.3}$ & $5.0^{+0.3}_{-0.3}$ & $0.35^{+0.20}_{-0.15}$ & $0.65^{+0.15}_{-0.20}$  & $0.32^{+0.04}_{-0.06}$ & $4.24^{+0.24}_{-0.34}$ & $3.87^{+0.36}_{-0.50}$ & $0.81^{+0.09}_{-0.06}$ & $2.34^{+3.11}_{-1.04}$  & $3.74^{+0.91}_{-0.94}$ & $48.08^{+0.25}_{-0.37}$ & $47.80^{+0.31}_{-0.48}$ & $44.41^{+1.55}_{-1.45}$ & $0.153^{+0.185}_{-0.106}$ & $1046^{+ 500}_{- 292}$ & $< 10^{-9}$ \\ 
Star 7 & $61^{+8}_{-8}$ & $62^{+8}_{-8}$ & $4.79^{+1.1}_{-0.6}$ & $4.8^{+1.1}_{-0.5}$ & $0.01^{+0.29}_{-0.00}$ & $0.98^{+0.00}_{-0.29}$  & $0.41^{+0.05}_{-0.05}$ & $3.95^{+0.19}_{-0.21}$ & $3.76^{+0.55}_{-1.16}$ & $0.82^{+0.07}_{-0.05}$ & $1.53^{+18.77}_{-1.13}$  & $2.91^{+0.51}_{-0.46}$ & $47.77^{+0.21}_{-0.25}$ & $47.42^{+0.28}_{-0.35}$ & $43.17^{+0.57}_{-0.99}$ & $0.089^{+0.230}_{-0.083}$ & $ 840^{+2201}_{- 410}$ & $< 10^{-9}$ \\ 
Star 8 & $57^{+24}_{-14}$ & $58^{+24}_{-15}$ & $4.99^{+1.0}_{-0.8}$ & $5.0^{+1.0}_{-0.7}$ & $0.05^{+0.45}_{-0.04}$ & $0.94^{+0.04}_{-0.45}$  & $0.22^{+0.09}_{-0.09}$ & $3.55^{+0.49}_{-0.42}$ & $3.44^{+0.88}_{-1.40}$ & $0.60^{+0.09}_{-0.08}$ & $1.28^{+14.95}_{-1.09}$  & $2.14^{+1.00}_{-0.58}$ & $47.37^{+0.52}_{-0.58}$ & $46.98^{+0.68}_{-1.65}$ & $42.50^{+1.78}_{-4.52}$ & $0.045^{+0.293}_{-0.043}$ & $ 905^{+2142}_{- 547}$ & $< 10^{-9}$ \\ 
Star 16 & $33^{+4}_{-2}$ & $34^{+4}_{-3}$ & $4.16^{+0.5}_{-0.2}$ & $4.2^{+0.5}_{-0.2}$ & $0.35^{+0.20}_{-0.15}$ & $0.65^{+0.15}_{-0.20}$  & $0.35^{+0.08}_{-0.08}$ & $3.20^{+0.17}_{-0.13}$ & $3.32^{+0.25}_{-0.40}$ & $1.20^{+0.06}_{-0.08}$ & $0.76^{+1.73}_{-0.35}$  & $1.63^{+0.21}_{-0.15}$ & $46.37^{+0.45}_{-0.51}$ & $43.89^{+1.51}_{-1.00}$ & $36.51^{+2.45}_{-1.22}$ & $0.043^{+0.036}_{-0.026}$ & $ 493^{+ 404}_{- 125}$ & $< 10^{-9}$ \\ 
Star 26 & $51^{+4}_{-3}$ & $52^{+4}_{-4}$ & $5.7^{+0.3}_{-0.6}$ & $5.7^{+0.3}_{-0.6}$ & $0.01^{+0.14}_{-0.00}$ & $0.98^{+0.00}_{-0.14}$  & $0.24^{+0.01}_{-0.04}$ & $4.14^{+0.11}_{-0.13}$ & $2.56^{+0.50}_{-0.37}$ & $1.44^{+0.04}_{-0.05}$ & $37.95^{+36.10}_{-28.20}$  & $3.42^{+0.34}_{-0.38}$ & $47.91^{+0.12}_{-0.16}$ & $47.43^{+0.18}_{-0.58}$ & $41.84^{+1.05}_{-1.37}$ & $0.006^{+0.012}_{-0.003}$ & $3168^{+1285}_{-1574}$ & $< 10^{-8}$ \\ 
\midrule
Star 26 & $51^{+4}_{-3}$ & $52^{+4}_{-4}$ & $5.7^{+0.3}_{-0.6}$ & $5.7^{+0.3}_{-0.6}$ & $0.01^{+0.14}_{-0.00}$ & $0.98^{+0.00}_{-0.14}$  & $0.24^{+0.01}_{-0.04}$ & $2.73^{+0.11}_{-0.12}$ & $2.56^{+0.50}_{-0.37}$ & $0.29^{+0.01}_{-0.01}$ & $1.51^{+1.47}_{-1.11}$  & $1.17^{+0.09}_{-0.10}$ & $46.51^{+0.12}_{-0.16}$ & $46.03^{+0.18}_{-0.57}$ & $40.44^{+1.05}_{-1.37}$ & $0.006^{+0.012}_{-0.003}$ & $1414^{+ 579}_{- 700}$ & - \\ 
\bottomrule
\end{tabular}

}
\newline
{Notes. For Star 26, we present two sets of values designated by $^a$ for assuming the LMC distance (50 kpc), and $^b$ for assuming a 10 kpc distance. \rev{The parameters are presented for the photosphere ($\tau=2/3$) apart from $T_{\star}$ and $\log_{10} g_{\star}$, which we display for comparison and that correspond to the temperature and surface gravity at $\tau=20$.}}
\end{center}
\end{sidewaystable}

\begin{figure*}
\centering
\includegraphics[width=0.9\textwidth]{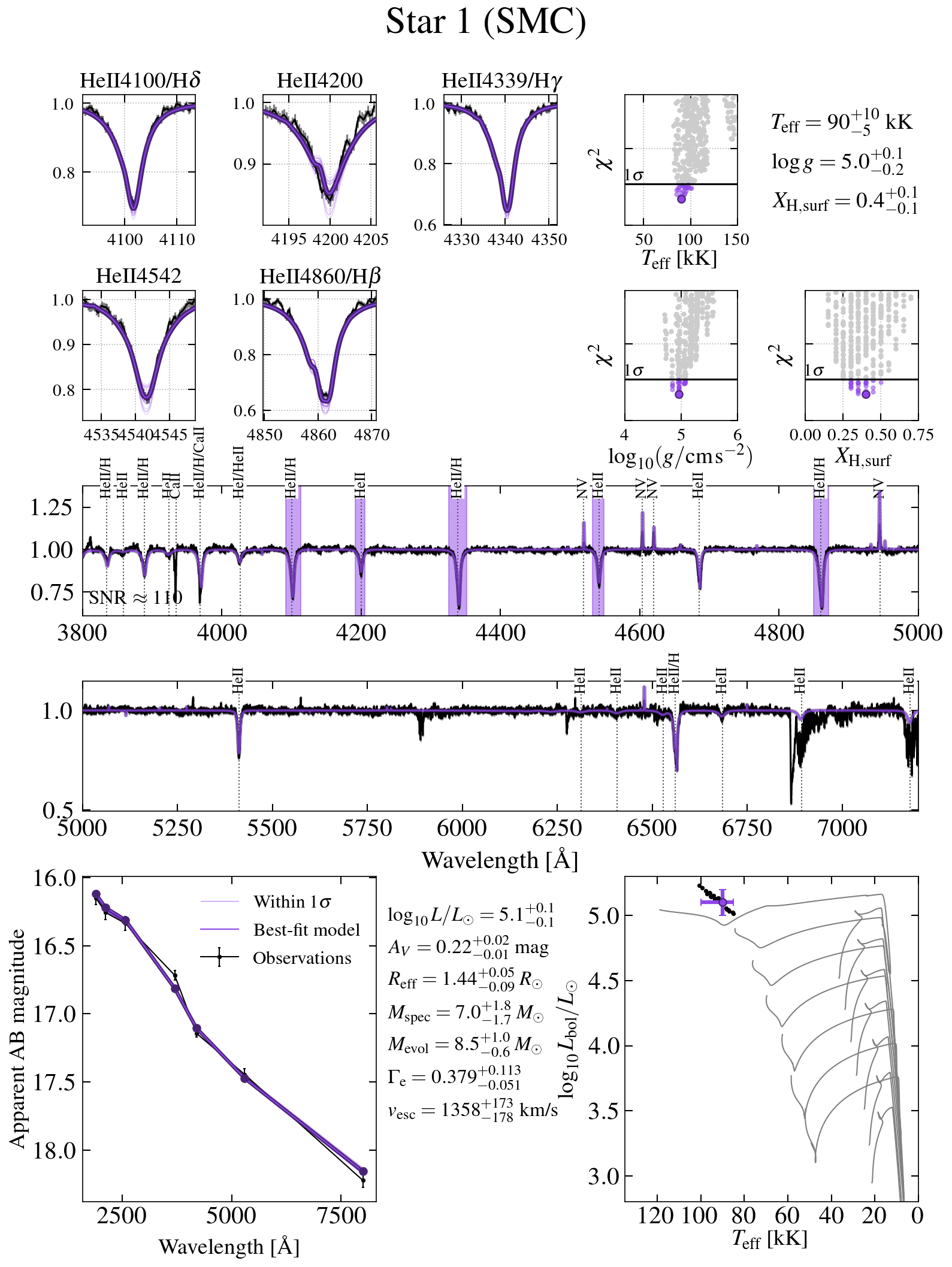}
\caption{Fit for star 1. See \appref{app:details_spectral_fitting} for the fits of the remaining stars. \rev{See \tabref{tab:stellar_properties} for the derived temperature and surface gravity at optical depth $\tau =20$.}}
\label{fig:fit_star1}
\end{figure*}

In \tabref{tab:stellar_properties}, we present the stellar properties that we obtain following the method described in \secref{sec:spectral_fitting}. We show the fit for star 1 as an example in \figref{fig:fit_star1}, while the fits for the other stars are presented in \appref{app:details_spectral_fitting}. The top left panels of the figure shows the spectral lines used for the spectral fit. The observed spectrum is shown in black, while the thick colored lines indicate the best-fit spectral model. Other models acceptable within 1$\sigma$ are shown as thin colored lines. 

The top right panels show $\chi^2$ as function of the effective temperature, surface gravity, and surface hydrogen mass fraction. The best-fit model (with the minimum $\chi^2$) is shown as a big colored circle, while the models acceptable within 1$\sigma$ are marked with smaller colored circles below the black line labeled $1\sigma$. The models marked with gray dots are not acceptable within $1\sigma$. \rev{As seen in these panels, none of the stars exhibit any ambiguity regarding where the true minimum and thus best-fit model lies.}

The two middle panels show the normalized observed spectrum in black and the best-fit model overplotted in a thick colored line. The spectral lines used for the spectral fit are marked by shaded background. The bottom left panel shows, in black, the observed photometric data in AB magnitudes and centered on the central wavelengths of each filter. The best-fit model is shown in a thick colored line and large colored circles, while the models allowed within 1$\sigma$ are plotted with thin lines. 

Finally, the derived best-fit effective temperature and bolometric luminosity with associated errors are plotted using color in a Hertzsprung-Russell diagram at the bottom right. The models allowed within $1\sigma$ are shown using black dots. For reference, we also plot evolutionary tracks for a sequence of stripped star models from \citet{2018A&A...615A..78G} using gray lines. These evolutionary models are for stripped stars with masses 1.5, 1.9, 2.5, 3.4, 4.5, 5.9, and 7.3\Msun, corresponding to initial masses of 5.5, 6.7, 8.2, 10, 12.2, 14.9, and 18.2\Msun. 

In the remainder of this section, we summarize and discuss the stellar parameters found for the 10 stars in our spectroscopic sample. In several instances, we compare with the evolutionary models from \citet{2018A&A...615A..78G}. Work presented in this manuscript suggests that the observed wind mass loss rate (see \secref{sec:wind}) is lower compared to what we assumed for the evolutionary models. 
However, although winds are important for the spectral morphology and future evolution of stripped stars, winds only mildly affect their broad surface properties \citep{2019MNRAS.486.4451G}. 


\paragraph{Effective temperature}
We measure effective temperatures above 50 kK for all but one star. The best-fit effective temperatures are in the range $50-95$ kK for stars 1, 2, 3, 4, 5, 6, 7, 8, and 26.
Star 16 is somewhat cooler, with about 35 kK. The tightest constraints on the effective temperature can be made when both \HeI and \HeII lines can be included in the spectral fit (see \secref{sec:fitting_routine}). However, for the hottest star (star 1) that does not display \HeI lines, the effective temperature can be well-constrained using the H and \HeII lines alone, because of the high signal-to-noise ratio. \rev{In other cases where \HeI lines are not present (stars 2, 3, and 6) and/or when the signal-to-noise ratio is lower (stars 3, 4, 7, and 8), we obtain large, sometimes asymmetrical errors for the effective temperature. This occurs because the \HeII lines have poor constraining power at high temperatures.}

\paragraph{Surface gravity}
We find typical surface gravities of \rev{$\log_{10} g_{\rm eff} \sim 5$}\footnote{We adopt cgs units when no units are given.} -- well above those of regular main-sequence stars, which are \rev{$\log_{10} g_{\rm eff} \sim 3.5-4.5$}, but below values for white dwarfs (\rev{$\log_{10} g_{\rm eff} \sim 6-9$}). Stars 5 and 16 have somewhat lower surface gravities, with \rev{$\log_{10} g_{\rm eff}$} of about 4.5 and 4.2 respectively. The derived surface gravities for stars 3 and 26 are somewhat higher, with \rev{$\log_{10} g_{\rm eff}$} of 5.4 and 5.7 respectively. 
\rev{We note that our obtained errors for surface gravity may be somewhat underestimated since it is challenging to identify the precise continuum adjacent to the broad Balmer and Pickering lines}

\begin{figure}
\centering
\includegraphics[width=\columnwidth]{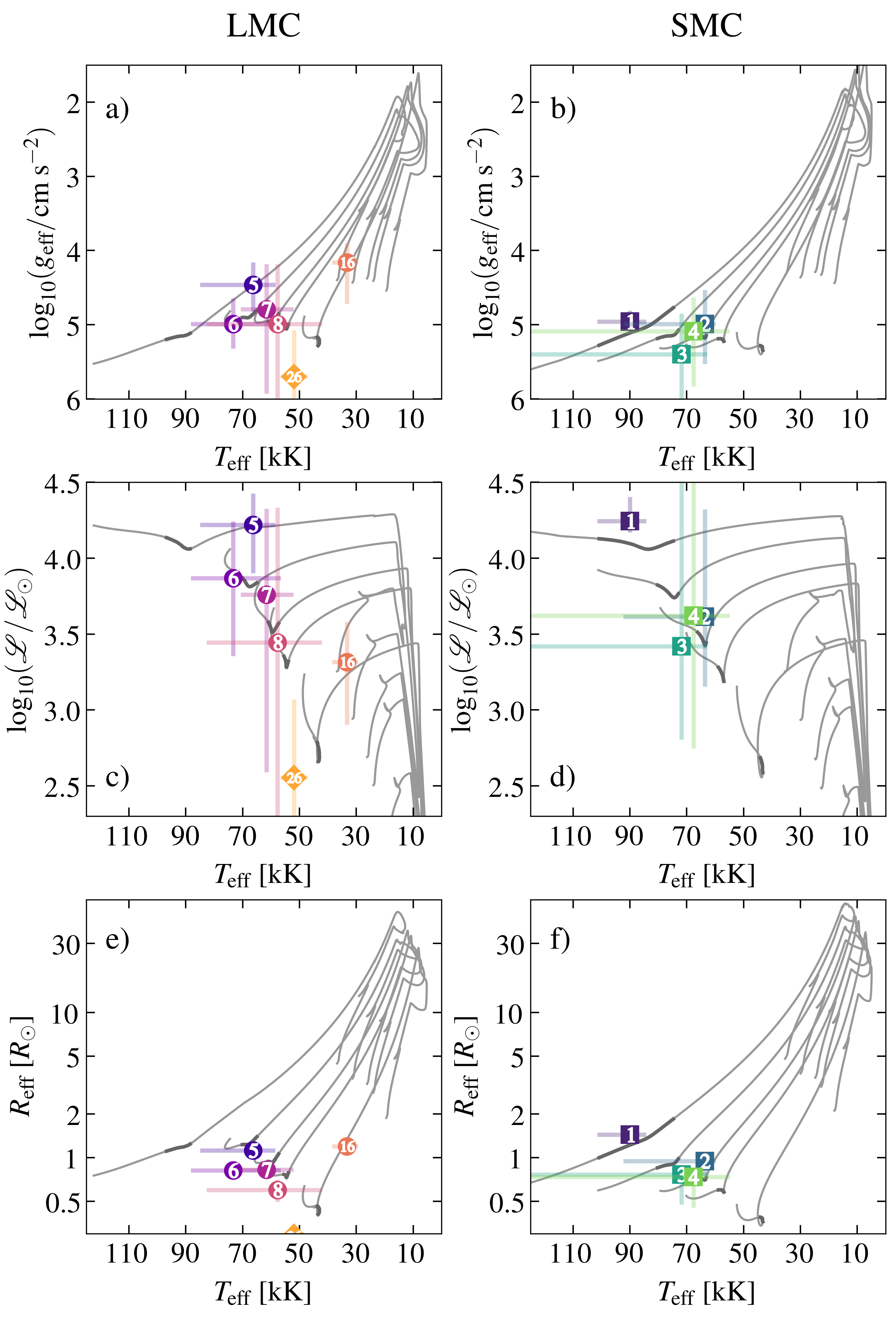}
\caption{The derived properties with associated errors for the spectroscopic sample shown with numbered markers plotted together with binary evolutionary models for donor stars in binary systems \citep{2018A&A...615A..78G}. Stars in the LMC are marked using circles, stars in the SMC with squares, and the foreground object with a diamond. From top to bottom we show the Kiel diagram, the spectroscopic Hertzsprung-Russell diagram, and effective radius as function of effective temperature. The left panels are for the Large Magellanic Cloud and the right panels for the Small Magellanic Cloud. The evolutionary models are for stars with initial masses of 4.5, 7.4, 9.0, 12.2, and 18.2\Msun, with corresponding stripped star masses of 1.1(1.2), 2.0(2.2), 2.7(2.9), 4.1(4.5) and 7.2(7.3)\Msun\ for the Large(Small) Magellanic Cloud. The central helium burning is marked with a thicker and darker line and the evolutionary tracks are cut at central helium depletion.}
\label{fig:grid}
\end{figure}

With constraints on effective temperature and surface gravity, the stars can be placed in Kiel diagrams, as shown in panels a) and b) of \figref{fig:grid}. Comparing to the Kiel diagram presented in \citetalias{DroutGotberg23} based on estimates of effective temperature and surface gravity using equivalent width diagnostics, this updated version is similar, illustrating the power of equivalent width analysis.  
In all panels of \figref{fig:grid}, we show the evolutionary tracks of donor stars in binary systems presented by \citet{2018A&A...615A..78G}. These models have initial masses of 4.5, 7.4, 9.0, 12.2, and 18.2\Msun, which results in masses of the stripped stars of 1.1(1.2), 2.0(2.2), 2.7(2.9), 4.1(4.5), and 7.2(7.3) \Msun\ for the LMC(SMC). We use the models with $Z=0.006$ and $Z=0.002$ to represent the LMC and SMC, respectively. We display the stars in the LMC using circles and the stars in the SMC with squares. Star 26 is displayed using a diamond.
The figures show that stars 1-8 and 26 agree well with being helium-core burning stars stripped of their hydrogen-rich envelopes through mass transfer in binary systems. This can be seen by comparing their locations in the Kiel diagram to the binary evolution tracks that we have displayed for reference. Star 16 appears to be more inflated than typical helium-core burning stripped stars. 

\paragraph{Inverse of flux-weighted gravity}
For the inverse of the flux-weighted gravity, we obtain values of $\log_{10} (\mathcal{L}/\mathcal{L}_{\odot}) \sim 2.5 - 4.5$. Since the inverse of the flux-weighted gravity behaves as a luminosity, we create spectroscopic Hertzsprung-Russell diagrams in panels c) and d) of \figref{fig:grid} using this quantity and the effective temperature. 
In this diagram, we see that all stars agree well with being donor stars stripped of their hydrogen-rich envelopes since they overlap with the expected location for stripped stars from the evolutionary models. Also in the spectroscopic Hertzsprung-Russell diagrams, the stars agree well with being central-helium burning stars, apart from star 16, which appears to be somewhat cooler than typical helium-core burning stripped stars. 

\paragraph{Surface hydrogen and helium mass fraction}
The best-fit surface mass fraction of hydrogen is well below what is expected for stars with hydrogen-rich envelopes, such as main-sequence stars. Five stars (star 1, 2, 4, 6, and 16) have surface hydrogen mass fractions between 0.3 and 0.4, while the remaining five stars (star 3, 5, 7, 8, and 26) have surface hydrogen mass fractions between 0 and 0.1. Conversely, the surface helium mass fraction for these two groups correspond roughly to between 0.6 and 0.7 and between 0.9 and 1. It is likely that three stars (stars 5, 7, and 26) are completely hydrogen free. These values are broadly consistent with the estimates presented in \citetalias{DroutGotberg23} based on equivalent width diagnostics.

\paragraph{Extinction}
We find small values for the extinction, between $A_V =$ 0.1 and \rev{0.7} mag. \rev{Generally, we find lower extinction values for the stars located in the SMC ($A_V\sim 0.1-0.4$ mag) compared to those located in the LMC ($A_V \sim 0.2-0.7$ mag). These values agree with the low end of the distributions found for stars in the Magellanic Clouds by \citet{Zaritsky2002,Zaritsky2004}.} This is expected since the stars were identified through their UV excess, meaning that our spectroscopic sample would be biased against stars whose sight-lines are strongly affected by dust extinction. 

\rev{Indeed, for a few stars (e.g. star 4 and star 8) the extinction values values are consistent with the expectation for foreground Milky Way extinction \citep{2011ApJ...737..103S}, implying negligible internal extinction in the SMC/LMC, respectively. On this point, we note that }
the extinction curves we employ \citep{2003ApJ...594..279G} are averages over the Magellanic Clouds. They do well in representing the extinction curves for our observed sample as seen from the photometric fits, \rev{although the foreground should be better represented by a Milky Way average extinction curve. While the LMC and Milky Way extinction curves are similar over the wavelength regions we consider \citep{2003ApJ...594..279G}, differences exist in the UV for the SMC.  To ensure that the stellar parameters that depend on the extinction estimate are robustly estimated, we run the spectral fitting routine on the SMC star 4 using an average extinction curve for the Milky Way \citep{2009ApJ...705.1320G}, which, in contrary to the SMC curve, contains the bump around 2175\AA. Despite this significant difference, we obtain estimates for the stellar parameters that are negligibly different from those obtained when using the SMC extinction curve. }


\paragraph{Bolometric luminosity}
The bolometric luminosities that we infer from the model fits are between $10^3$ and $10^5$~$L_{\odot}$. This range is typical, for example, for main-sequence stars with masses between $\sim$5 and $\sim$30\Msun \citep{2013A&A...558A.103G}. The bolometric luminosity determination is sensitive to how well the effective temperature is determined since the peak of the spectral energy distribution is located in the un-observable ionizing regime and needs to be inferred from the shape of the modeled spectral energy distribution. This dependency is reflected in the larger errors on bolometric luminosity when the effective temperature also has larger errors (for example, see star 4, \figref{fig:fit_star4}).  
The bolometric luminosity is also dependent on the distance. This is not an issue for stars 1-8 and 16, which are members of the Magellanic Clouds, but affects star 26, which has a more uncertain distance.

\begin{figure*}
\centering
\includegraphics[width=0.85\textwidth]{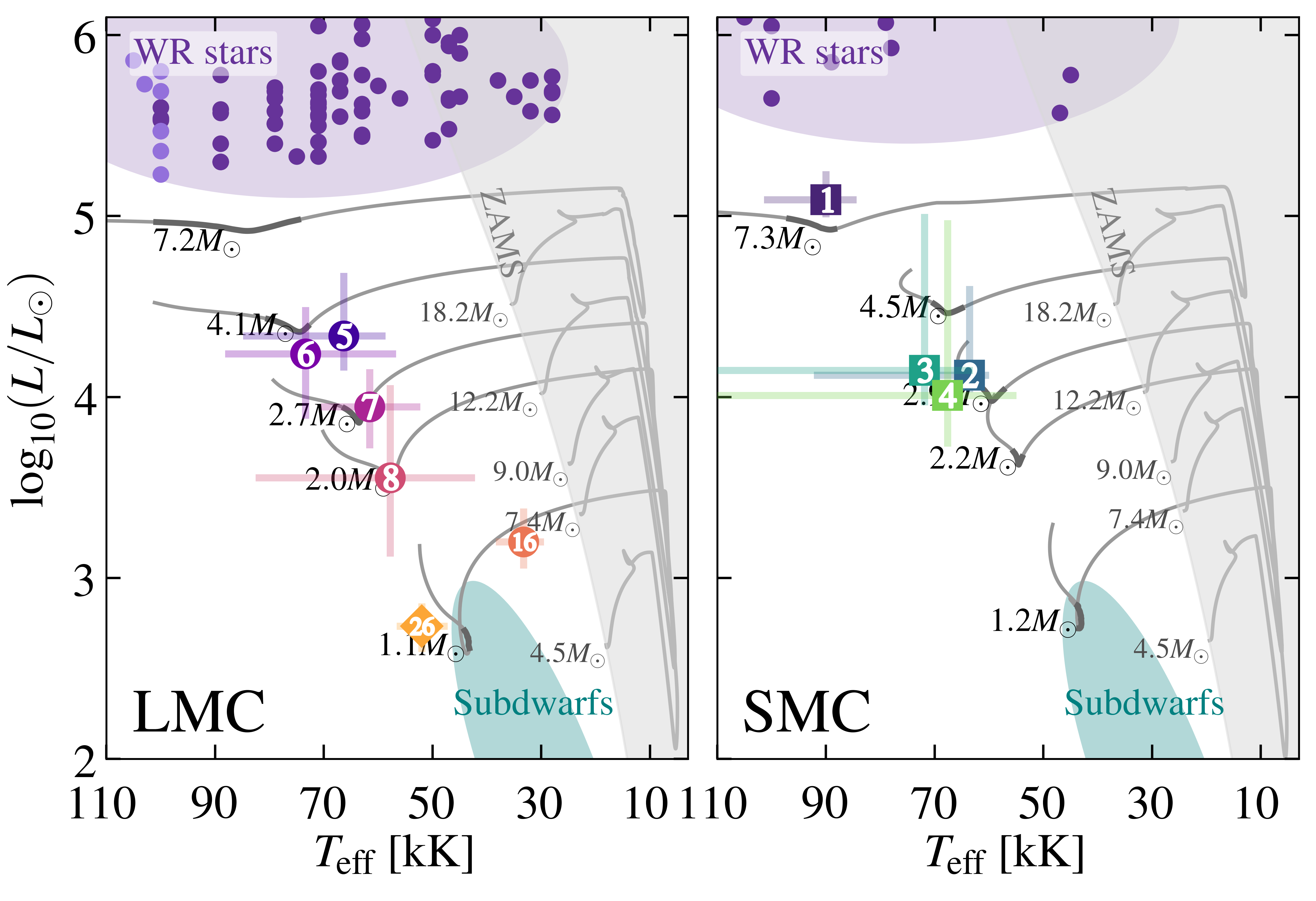}
\caption{The stars in our spectroscopic sample, shown with numbered markers, match well with models of stars stripped in binaries (gray lines) in the Hertzsprung-Russell diagram. The left panel shows the stars from the LMC plotted together with models of $Z=0.006$ and the right panel shows stars in the SMC plotted together with models of $Z=0.002$. Star 26, which likely is a foreground object, is plotted using an assumed distance of 10 kpc and diamond-shaped marker. We label the zero-age main-sequences and gray-shade the parts of the diagrams with cooler temperatures. Wolf-Rayet stars in each of the clouds are shown using purple circles and a shaded region, while the expected locations of bright subdwarfs are marked with a green-shaded ellipse. The weak-wind WN3/O3 stars in the LMC are indicated using lighter purple color.}
\label{fig:HRD}
\end{figure*}

When placed in the Hertzsprung-Russell diagram in \figref{fig:HRD}, it is again clear that the stars in our spectroscopic sample are poorly matched with main-sequence stars. Instead, they overlap with the helium main-sequence. The exception is again star 16, which instead appears to overlap with an inflated phase. The assumed 10 kpc distance of star 26 as displayed in \figref{fig:HRD} matches well with the expected location for helium-core burning, massive subdwarfs.
Compared to the set of Wolf-Rayet stars \citep[dark purple circles,][]{2014A&A...565A..27H,2015A&A...581A..21H,2016A&A...591A..22S}, WN3/O3 stars \citep[lighter purple circles in the LMC plot,][]{2017ApJ...841...20N}, and the expected location of subdwarfs in the two clouds \citep[teal shaded regions, cf.][]{2016PASP..128h2001H}, it is clear that the stars in our spectroscopic sample create a connecting bridge between faint subdwarfs and bright Wolf-Rayet stars. 


\paragraph{Effective radius}
The effective radii we derive are well constrained and all close to $1R_{\odot}$, spanning a range from $0.3 \Rsun$ to $1.4 \Rsun$. Within the uncertainties, none of the stars exceed $1.6\Rsun$, suggesting that they are indeed much smaller than typical main-sequence stars with the same temperatures -- the massive O-stars having radii $\gtrsim 10\Rsun$. The measured radii agree well with predictions from binary stellar evolution models ($0.6-1.4\Rsun$ for stripped stars with masses between $2$ and $7.2\Msun$, \citealt{2018A&A...615A..78G}). This can also be seen from panels e) and f) of \figref{fig:grid}. 
As shown in \tabref{tab:stellar_properties}, star 26 has an estimated radius of 1.4\Rsun\ when assumed to reside in the LMC, compared to 0.3\Rsun\ when assumed at a distance of 10 kpc. Given its high surface gravity, the smaller size is more compelling, and in agreement with the star being located in the foreground. 

\paragraph{Spectroscopic mass}
We find spectroscopic mass estimates between 0.8 and 6.9\Msun\ for stars 1-8 and 16. 
For stars where we have very good model fits, such as for star 1, the errors in the spectroscopic mass are only $\sim 20$\%. For fits with larger uncertainties, such as for star 8, the errors are very large, reaching a factor of 10. 
Star 26 has an estimated spectroscopic mass of 38\Msun\ when assumed to reside in the LMC, but instead the more realistic 1.5\Msun\ when placed at 10 kpc distance. 


\begin{figure}
\centering
\includegraphics[width=0.9\columnwidth]{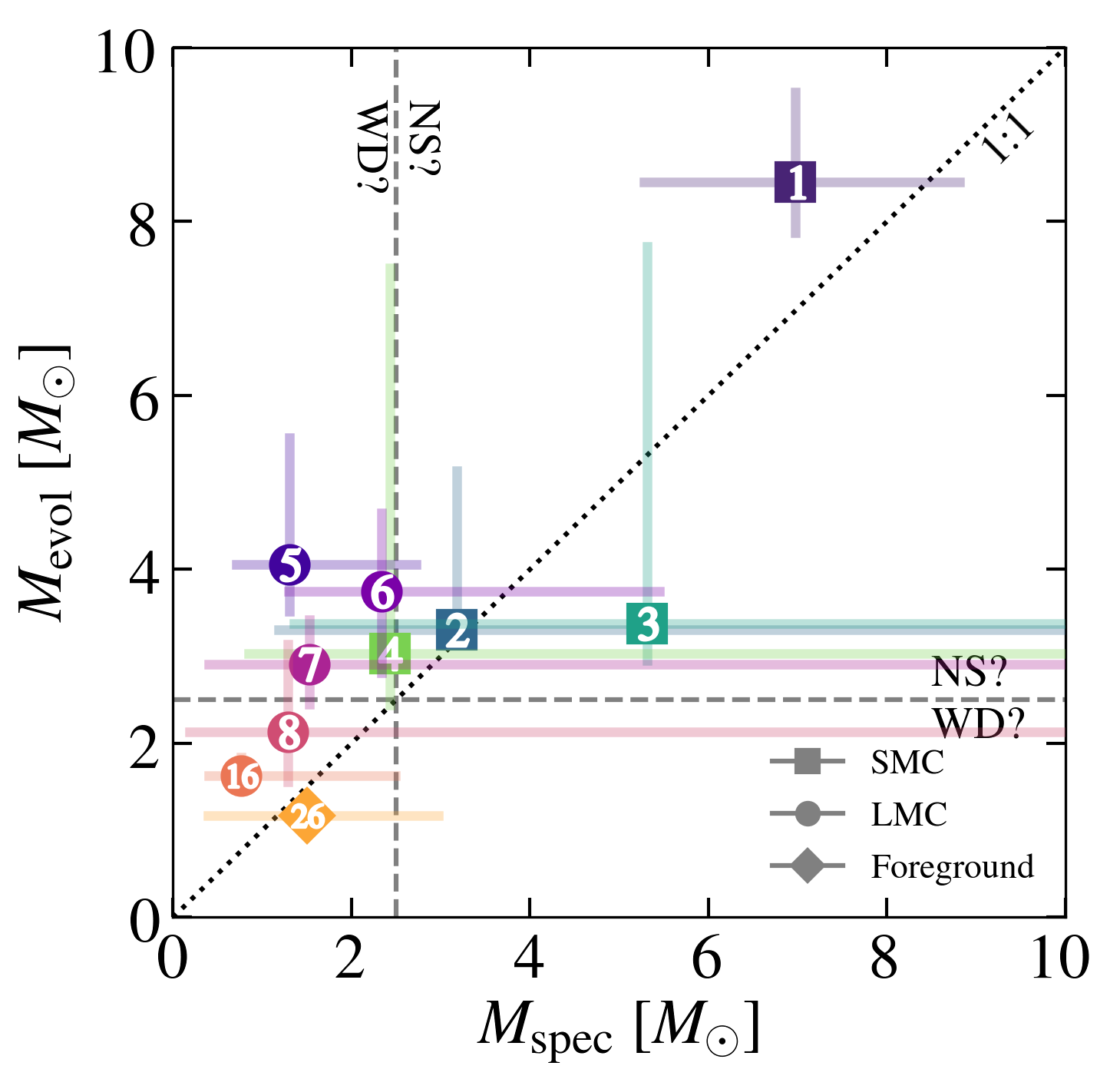}
\caption{Comparison of the spectroscopic and evolutionary masses for the stars in the spectroscopic sample. 
The lines at 2.5\Msun\ are meant as approximations for the limit for stripped stars that reach core collapse vs evolve to white dwarfs. }
\label{fig:Ms_Me}
\end{figure}

\paragraph{Evolutionary mass}
The evolutionary mass provides an additional handle on the stellar mass. On average, we find somewhat higher evolutionary masses than spectroscopic masses, stretching from 1.2 to 8.4 \Msun. Among the sample, all but stars 8, 16 and 26 have evolutionary masses above 2.5\Msun, which can be used as an approximation for the boundary for what stars will undergo core collapse \citep{2015MNRAS.451.2123T}. 

We plot the evolutionary mass versus the spectroscopic mass found from our analysis in \figref{fig:Ms_Me}. The figure shows that the best constrained spectroscopic masses belong to stars with either high SNR (star 1) or spectra with both \HeI and \HeII lines present (stars 5, 16, and 26, however not stars 7, or 8, likely because of their low SNR). We note that star 16 appears inflated (see above) and its mass may be poorly represented by the mass-luminosity relation we adopt when calcuating evolutionary mass (see \secref{sec:Mevol}). 
Dynamically inferred masses would be ideal to use for resolving what the true stellar masses are.

\paragraph{Eddington factor} 
We estimate that the stars in the spectroscopic sample have bolometric luminosities that mostly are far from their Eddington limits. Star 1 and star 5 are the closest to their Eddington limits, with Eddington factors of $\sim 0.4$ and $\sim 0.25$, respectively. The other stars all have Eddington factors of $\Gamma_e \sim 0.006$-$0.15$. The Eddington factors we find are quite similar to those of O-type stars \citep{1993ApJ...412..771L}.

\section{Evolutionary stage: contracting, helium-core burning, or expanding?} \label{sec:evol_stage}

\begin{figure*}
\centering
\includegraphics[width=0.65\textwidth]{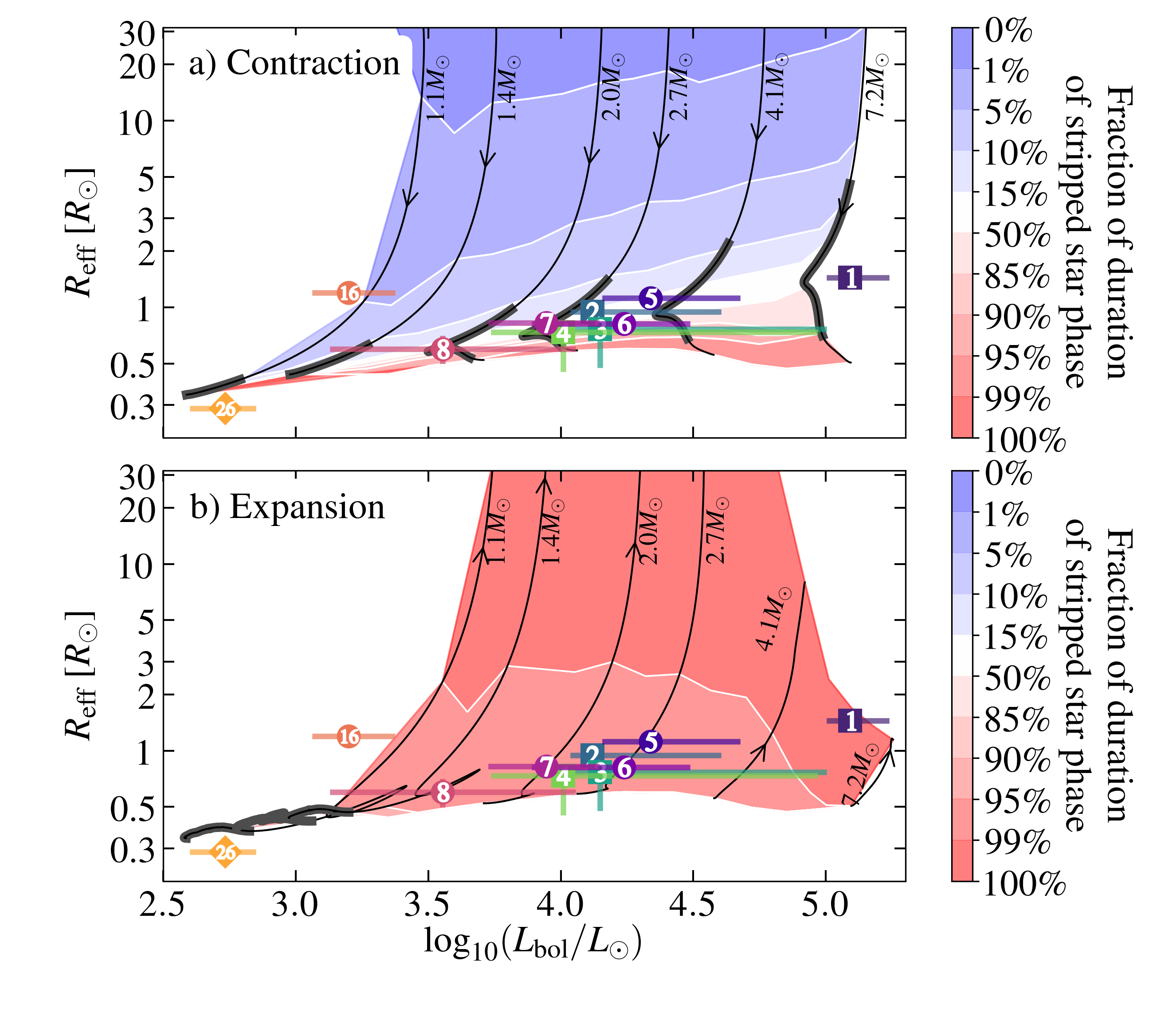}
\vspace{-7mm}
\caption{Contraction (top) and expansion (bottom) phases for stripped stars demonstrated using the $Z=0.006$ stripped star models of \citet{2018A&A...615A..78G}, labeled by stripped star mass. We show the fraction of the stripped star duration using blue and pink shades and the central helium burning phase when $0.9 > X_{\mathrm{He, c}} > 0.01$ using dark gray background for the evolutionary tracks. The stars in the spectroscopic sample are plotted using their effective radii and bolometric luminosities with numbered markers (see \tabref{tab:stellar_properties}). 
The top panel shows that contraction lasts $\sim 10\% $ of the stripped star duration, while the bottom panel shows the expansion phase lasts $\sim 1-5\%$. All stars but star 16 agree with the helium-core burning phase and the expansion phase, while star 16 could either be contracting or expanding.}
\label{fig:evolutionary_stage}
\end{figure*}

Stripped stars burn helium in their centers during the large majority of the remaining stellar lifetimes after envelope-stripping is complete. Unlike the central hydrogen burning during the main-sequence, the radii of stripped stars only moderately change during the central helium burning phase \citep[e.g.,][]{2019A&A...629A.134G}. There are, however, two shorter-lasting inflated stages predicted for stripped stars. First, the contraction phase after envelope-stripping is complete, and, second, the expansion phase initiated after helium-core depletion \citep{2020A&A...637A...6L}. 

We show these evolutionary phases in \figref{fig:evolutionary_stage}, using the binary evolution models of \citet{2018A&A...615A..78G}. In the figure, we plot the radii of  models of stripped stars with masses $\sim1-7\Msun$ (corresponding to initial masses $\sim 4.5-18.2\Msun$) as function of their bolometric luminosity. The models are represented by solid black lines and arrows that demonstrate the evolutionary direction. In the top panel, we plot the contraction phase followed by the helium-core burning phase until the star reaches its minimum radius, while in the bottom panel we show the expansion phase during helium-shell burning, from the point where the star has reached its minimum radius, until death or the model evolves off the plot. We use dark gray background for the tracks to mark the central helium burning, which here is defined as when the central mass fraction of helium is between 0.9 and 0.01. The blue and red shading is used to show what fraction of the temporal duration of the stripped star phase has passed. Comparing the color shading with the dark gray background of the tracks, it is clear that central helium burning indeed coincides with the majority of the stripped star duration, while contraction and expansion correspond to about 10\% and 1-5\% of the stripped star phase, respectively. Thus, we expect that most stripped stars should be helium-core burning. 

\figref{fig:evolutionary_stage} also shows that the radius change during central helium burning is somewhat mass dependent, with a larger change for the more luminous, higher-mass stripped stars. For example, we expect that a 7\Msun\ stripped star with $L_{\rm bol} \sim 10^5 L_{\odot}$ can have radii between $\sim$0.7 and 5\Rsun\ during central helium burning, while a 3\Msun\ stripped star, with $L_{\rm bol} \sim 10^4 L_{\odot}$, should be limited to radii between $\sim$0.6 and 1.5\Rsun\ in the same evolutionary phase. The reason is twofold: first because more massive stars ignite helium in their cores earlier during the evolution, and second because of wind mass-loss, which allows deeper, more compact layers of the stellar models to be revealed \citep[cf.][]{2019MNRAS.486.4451G}. We note that the binary evolution models we use were created for stars stripped via stable mass transfer, which leaves a layer containing hydrogen on the stellar surface \citep{2017A&A...608A..11G,2020A&A...637A...6L}. 
Stripped stars with no hydrogen layer are expected to be more compact and smaller than stripped stars that retain hydrogen \citep{2017ApJ...840...10Y}.

We overplot the stars in our spectroscopic sample in both panels of \figref{fig:evolutionary_stage}. 
All stars overlap with expectations for the central helium burning stage, apart from star 16. While it is possible that the stars are during the early stages of expansion, the different timescales make the helium-core burning stage more likely. 
More precise measurements for the stellar masses than what we currently have could be used to determine the evolutionary stage more accurately. 
As an example, according to the models displayed in \figref{fig:evolutionary_stage}, star 1 could either match a helium-core burning star with mass $\sim 8\Msun$ or a $\sim 5\Msun$ expanding stripped star. Similarly, star 5, for example, matches either a $\sim 4\Msun$ helium-core burning stripped star or a $\sim 3\Msun$ expanding stripped star. 

Star 16 is about twice as large compared to what is expected for helium-core burning stripped stars with its determined bolometric luminosity. We, therefore, consider that star 16 likely is experiencing an inflated stage \citep[cf.][]{2018A&A...615A..30S}, which agrees with its lower surface gravity and lower effective temperature compared to the other stars in the sample (see \figref{fig:grid} and \secref{sec:stellar_properties}). Whether the star is in the contraction or expansion phase is not evident from current data: contraction stages should be slower and thus more common, but expansion phases should be brighter, favoring their detection \citep[see][]{2018A&A...615A..30S}. Again, more precise mass measurements will provide insight in what evolutionary stage star 16 is in. 

Even though we do not know the distance to star 26 very accurately, \figref{fig:evolutionary_stage} suggests that the star is likely a helium-core burning subdwarf with mass of $\sim 1\Msun$, demonstrated by the closeness to that evolutionary track. Especially its effective temperature also matches such a massive subdwarf scenario better than either that of a typical subdwarf B-star or a helium-core burning stripped star in the LMC (cf.\ \citealt{2018A&A...615A..78G}). 
If star 26 would have been located in the LMC (which would also require that it was a runaway star; Appendix~\ref{app:kinematic}), it would overlap with an inflated stage (see \tabref{tab:stellar_properties}), which does not match well with its high surface gravity. The 10 kpc distance we adopt here gives rise to a bolometric luminosity, stellar radius and spectroscopic mass that roughly match the expectations for a helium-core burning stripped star with the effective temperature of star 26 \citep{2018A&A...615A..78G}, also accounting for the complete loss of hydrogen, which likely results in the slightly higher surface gravity and effective temperature. 
It is worth to note that star 26 has a significantly higher temperature ($T_{\rm eff}>50$kK) than typical subdwarf~B type stars ($T_{\rm eff}\sim 25$kK), and is in fact much more similar to the $\sim 1.5\Msun$ subdwarf in the Galactic binary HD~49798 \citep{2009Sci...325.1222M, 2017ApJ...847...78B}.

\section{Constraints on stellar wind mass-loss}\label{sec:wind}

In contrast to the original spectral models created for stripped stars by \citet{2018A&A...615A..78G}, the stars in our spectroscopic sample do not show any strong/broad emission lines indicative of mass loss through stellar winds. However, it is possible that some wind is driven off the surfaces, for example through metal line driving and radiation pressure. The somewhat higher Eddington factors for stars 1 and 5 (see \tabref{tab:stellar_properties}), for example, suggest some contribution from radiation pressure to the wind driving, and these stars could therefore perhaps have somewhat higher wind mass loss rates than the other stars. While ultraviolet spectroscopic will ultimately provide the most precise measurements of the wind properties from these stars, here we investigate what rough constraints can be placed from the optical spectra alone.

As seen in \figref{fig:optical_spectra}, the optical spectra contain only absorption features with the exception of weak \NIV and \NV emission lines. While these nitrogen lines may occur in emission, they are, in these cases, not signs of a stellar wind, instead the result of photospheric level inversion \citep[cf.][]{2011A&A...536A..58R, 2012A&A...537A..79R}. This is also clear from their narrow widths, which are not expected for the fast speed that is necessary for stellar winds to escape the surface of the compact stripped stars ($\gtrsim 1,000\kms$). In fact, for example, when the \NV $\lambda\lambda$ 4604/20 doublet appears in emission, it is most likely because of high surface temperature causing the upper level to be pumped ($\gtrsim 90$kK, see \figreftwo{fig:grid_exploration}{fig:N_He_structure}). 

The lines that are most sensitive to wind mass-loss in the optical spectrum are H$\alpha$ and \HeIIl 4686, since they are both $\alpha$-lines \citep[cf.\ e.g., the WN3/O3 stars discovered by ][which show moderate wind mass-loss]{2014ApJ...788...83M, 2017ApJ...841...20N}. 
Because H$\alpha$ is very sensitive to contributions from surrounding H~\textsc{ii} regions, we choose to focus on the effect of winds on \HeIIl 4686 to very roughly estimate the wind mass-loss rate of the observed sample of stars. 

\begin{figure}
\centering
\includegraphics[width=\columnwidth]{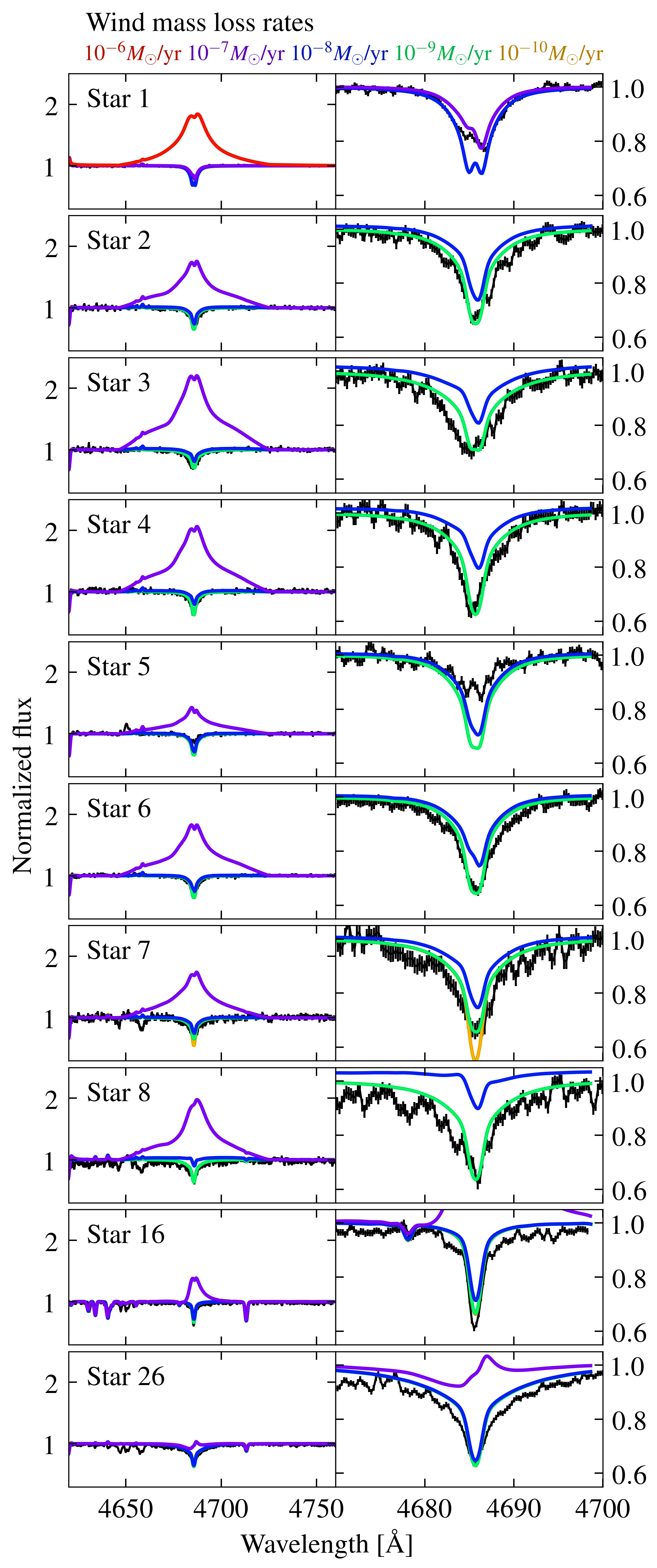}
\caption{The shape of the \HeIIl 4686 spectral line is very sensitive to surrounding gas and we, therefore, use it to estimate wind mass-loss rates. The observed \HeIIl 4686 lines are shown from top to bottom for each star along with models for a range of wind mass-loss rates ($\dot{M}_{\rm wind} = 10^{-6}$ - red, $10^{-7}$ - purple, $10^{-8}$ - blue, $10^{-9}$ - green, and $10^{-10} \Msunyr$ - yellow). Right panels show zoom-ins of the observed spectral lines, while left panels show zoom-outs that also include the expectations for wind emission.}
\label{fig:mdot_plot}
\end{figure}

To estimate wind mass-loss rates, we take the best-fit spectral models for each star following the parameters presented in \tabref{tab:stellar_properties}, and then compute new versions of these models assuming a range of wind mass-loss rates ($\dot{M}_{\rm wind} = 10^{-10}$, $10^{-9}$, $10^{-8}$, $10^{-7}$, and $10^{-6} \Msunyr$), while fixing the terminal wind speed ($v_{\infty} = 2500 \kms$), the amount of wind clumping ($f_{\rm vol} = 0.5$), and the wind velocity profile ($\beta = 1$). While the wind speed is uncertain,
we adopt $2500\kms$ because it matches reasonably well with the ratio between terminal wind speed and surface escape speed, $v_{\rm esc}$, for massive O-stars, which is $v_{\infty}/v_{\rm esc} \sim 2.5$ \citep{1995ApJ...455..269L}. This ratio also matches reasonably well with the expectations for subdwarfs that was computed by \citet{2016A&A...593A.101K} and the computed values for a range of helium star masses of \citet{2017A&A...607L...8V}. We estimate the surface escape speeds for the stars using the derived parameters ($v_{\rm esc} = \sqrt{2GM_{\rm spec}/R_{\rm eff}}$) and present the values in \tabref{tab:stellar_properties}.

After computing the spectral models with varying wind mass-loss rate, we find the upper limit for wind mass-loss rate acceptable for each star by identifying, by eye, the model with the highest wind mass-loss rate that still matches the line shape of \HeIIl 4686. This comparison is plotted in \figref{fig:mdot_plot}, where we show the observed spectra in black and the models with mass-loss rates $10^{-10}$, $10^{-9}$, $10^{-8}$, $10^{-7}$, and $10^{-6}\Msunyr$ in yellow, green, blue, purple, and red, respectively. The left panels show a zoomed-out version displaying the development of wind emission, while the right panels show the detailed comparison between the models and the data. All wind mass-loss rates were not computed for all models. The $10^{-10}\Msunyr$ models exist for stars 7 and 16, and the $10^{-6}\Msunyr$ model exists for star 1. The reason is that the lowest wind mass-loss rate models are cumbersome to converge numerically and the highest wind mass-loss rate model was not necessary for other stars than star 1.

We find that stars 1 and 5 have some in-filling in \HeIIl 4686, suggesting there could be a stellar wind affecting the optical spectra. This aligns well with their somewhat higher Eddington factors of $\Gamma_e \sim 0.38$ and $\sim0.26$, respectively (see \tabref{tab:stellar_properties}). The model with mass-loss rate $10^{-7}\Msunyr$ and $10^{-8} \Msunyr$ match best the \HeIIl 4686 line for star 1 and star 5, respectively. We, therefore, adopt these values as a rough mass-loss rate estimate for stars 1 and 5. 
For the remaining stars, no line-infilling is evident and all spectral line shapes are well-matched by the wind mass-loss rate models with $\dot{M}_{\rm wind} = 10^{-9}\Msunyr$. We therefore adopt $10^{-9}\Msunyr$ as the upper limit for the wind mass-loss rate for the remaining stars. In the case of star 7, it appears that the $10^{-10}\Msunyr$ model produces a too deep spectral feature, therefore we do not consider the $10^{-9}\Msunyr$ an upper limit for star 7, but a rough estimate. These low mass-loss rates match well given the lower Eddington factors of $\Gamma_e \sim 0.04-0.15$ for stars 2, 3, 4, 6, 7, 8, and 16, suggesting that wind driving from radiation pressure is small. 
Star 26 may be an exception, because we cannot distinguish between the $10^{-9}$ and $10^{-8}\Msunyr$ models and therefore adopt $10^{-8}\Msunyr$ as an upper limit. However, we note that for this analysis, we adopted the stellar properties that correspond to membership of the LMC for star 26. 
We provide these rough estimates for the wind mass-loss rates in \tabref{tab:stellar_properties}. 
We emphasize that the method we employ is approximate since the fixed wind parameters also influence the line shapes, although perhaps less than the wind mass-loss rates, within reasonable ranges. 

The wind mass-loss rate of stripped stars is thought not only to change the spectral morphology, but primarily to affect the properties and future evolution of the stripped star \citep{2017ApJ...840...10Y, 2017A&A...608A..11G, 2019MNRAS.486.4451G, 2020A&A...637A...6L}. 
Because of the lack of observed stripped stars, it has been difficult to construct a suitable wind mass-loss prescription. From the analysis of the Galactic quasi Wolf-Rayet star in HD~45166 \citep{2008A&A...485..245G}, it previously appeared as if an extension of the empirical Wolf-Rayet wind mass-loss scheme of \citet{2000A&A...360..227N} was appropriate. However, a weaker wind prescription, for example, the one made for subdwarfs by \citet{2016A&A...593A.101K} could also be accurate. Recently, efforts have been made to improve our understanding of wind mass loss from helium stars, in particular with the single-temperature models from \citet{2017A&A...607L...8V} and the high-mass helium star models from \citet{2020MNRAS.499..873S}. Interestingly, these studies predict lower wind mass-loss rates than what is expected from extrapolated Wolf-Rayet wind mass-loss schemes. Anticipating the results from these teams' ongoing theoretical efforts, we hope to provide a tentative, yet useful, comparison.

\begin{figure}
\centering
\includegraphics[width=\columnwidth]{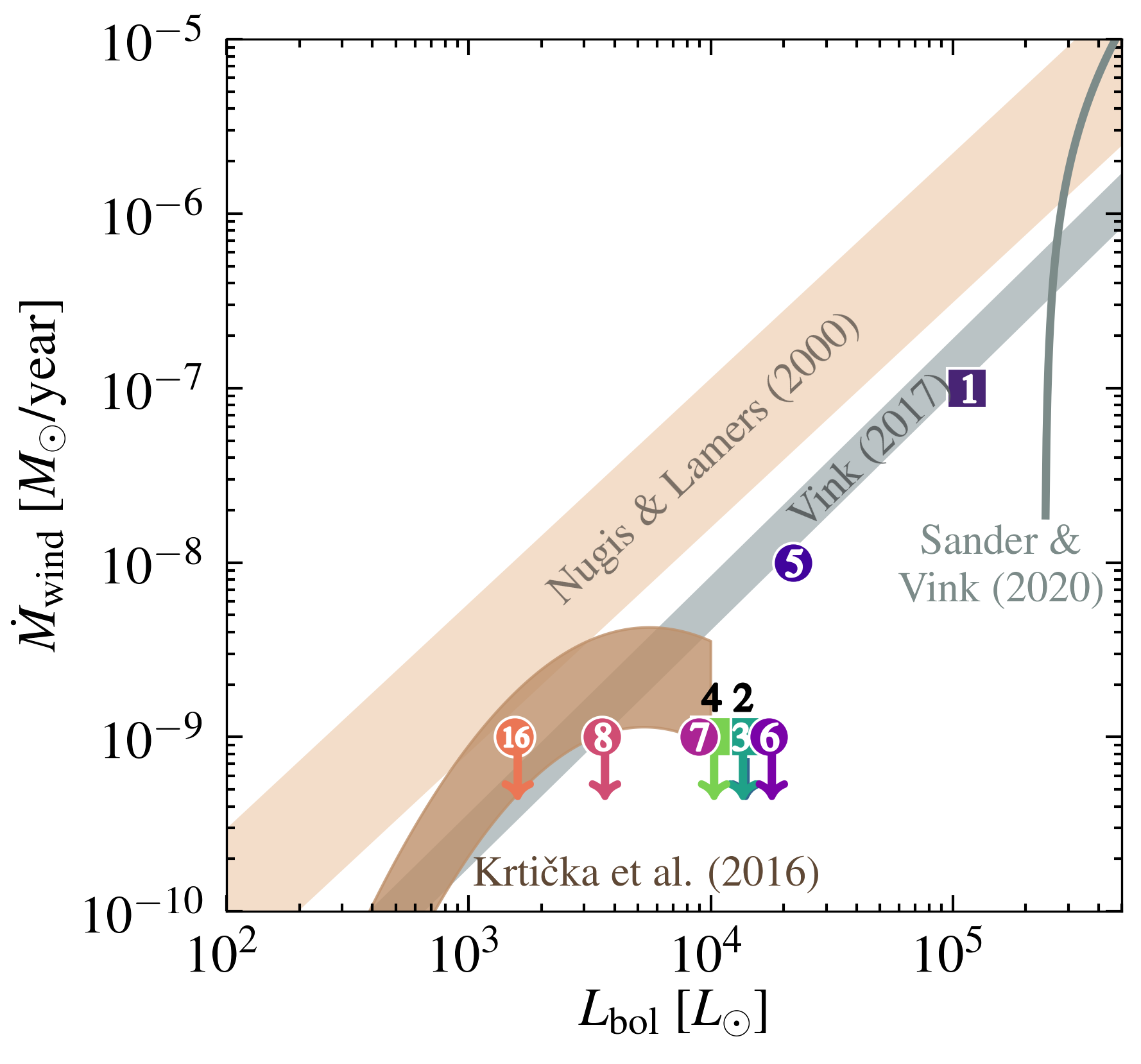}
\caption{Rough estimates for the mass-loss rate upper limits (and tentative number in the case of stars 1, 5 and 7) plotted as function of bolometric luminosity for the stars in the sample using colored and numbered symbols (because the symbols for stars 2 and 4 are behind other markers, we label them above). We also plot the mass-loss rate prescriptions from \citet{2000A&A...360..227N}, \citet{2016A&A...593A.101K},  \citet{2017A&A...607L...8V}, and \citet{2020MNRAS.499..873S} in beige, brown, light gray, and dark gray. 
We do not extrapolate the \citet{2016A&A...593A.101K} scheme above $10^4 L_{\odot}$ since these models were created for subdwarfs. 
For the \citet{2020MNRAS.499..873S} scheme, we only show $Z=0.006$ since the lower metallicity predictions are beyond the parameter space of the plot. 
}
\label{fig:Mdot_L}
\end{figure}

For radiation driven winds, mass-loss rate prescriptions are often described as luminosity dependent (see for example the review by \citealt{2014ARA&A..52..487S}). We, therefore, plot the estimates for wind mass-loss rates as function of the bolometric luminosity for the observed sample in \figref{fig:Mdot_L}. To compare, we also display the predictions from \citet{2000A&A...360..227N}, \citet{2016A&A...593A.101K}, \citet{2017A&A...607L...8V}, and \citet{2020MNRAS.499..873S}. For these, we adopt, when possible, surface helium mass fractions between 0.4 and 1, metallicity between 0.002 and 0.006, and effective temperature between 50 and 100 kK. These ranges result in the broad, colored bands that we display in \figref{fig:Mdot_L}.  

\figref{fig:Mdot_L} shows that the mass-loss rate estimates from our observations are low compared to most schemes. None of the stars match the extrapolation of the Wolf-Rayet scheme from \citet{2000A&A...360..227N}, and the massive helium star scheme from \citet{2020MNRAS.499..873S} does, understandably, not extend to sufficiently low luminosities. Stars 1, 5, 8, and 16 appear to agree with the predictions from the \citet{2017A&A...607L...8V} scheme, but stars 2, 3, 4, 6, and 7 appear to have significantly lower mass-loss rates, resulting in a poor match. The flattening of the subdwarf prescription from \citet{2016A&A...593A.101K} appears to better represent the low mass-loss rates of stars 2, 3, 4, 6, 7, 8, and 16, but it could be that the actual wind mass-loss rates are even lower than the expectations from this prescription. We also note that the prescription of \citet{2016A&A...593A.101K} was fitted to data with $L_{\rm bol} < 10^4 L_\odot$ and their models were tailored for cooler stars ($T_{\rm eff} \sim 15-55$kK). 
We emphasize that, to obtain an accurate comparison, it is necessary to also allow other wind parameters than mass-loss rate to vary. If, for example, the winds were faster than the fixed $v_{\infty} = $ 2500\kms, higher mass-loss rates compared to our estimates would be allowed. 

We note that the optical spectral lines that are sensitive to circumstellar gas cannot be used to determine the exact origin of this moving material. While stellar winds are expected for hot and helium-rich stars, these stars are binaries and gas could originate from disks, outflows, or ejecta \citep[e.g.,][]{1998ApJ...493..440G,2011MNRAS.418.1959S,2015MNRAS.450.2551M}. Such gas could, potentially, have an impact on these optical spectral lines that could be confused with stellar winds. To measure direction, speed, and better constrain the amount of circumstellar material -- thus also its origin -- UV spectroscopy is needed. This is the focus of an upcoming study in our series (HST/COS cycle 29 PI: Drout, HST/COS cycle 30 PI: G\"{o}tberg). 

\section{Emission rates of ionizing photons}\label{sec:ionizing_flux}

The emission rates of ionizing photons cannot be directly measured. But, they can be inferred from the shapes of the modeled spectral energy distributions. 
We estimate the emission rates of H, He, and He$^+$ ionizing photons, referred to as $Q_0$, $Q_1$, and $Q_2$, by integrating the spectral energy distributions of the best-fit model and the models within 1$\sigma$ error, following:
\begin{equation}\label{eq:Q}
Q = \int_{50} ^{\lambda_{\rm lim}} \dfrac{L_{\lambda}}{hc/\lambda} d\lambda,
\end{equation}
where we integrate from 50\AA, which is the shortest wavelength included in the spectral models, until $\lambda_{\rm lim}$, which is the ionization edge for the given atom or ion (912\AA, 504\AA, and 228\AA\ for H, He, and He$^+$, respectively) and thus sets whether $Q$ refers to $Q_0$, $Q_1$, or $Q_2$. In \eqref{eq:Q}, $h$ is Planck's constant, $c$ is the speed of light, $\lambda$ is the wavelength, and $L_{\lambda}$ is the wavelength dependent luminosity. 
We also do not account for the effect of wind mass loss when estimating the ionizing emission rates. However, within the expected regime of weak winds (see \secref{sec:wind}), we do not expect large variations in either of the ionizing emission rates \citep[cf.][]{1992PASP..104.1164S}. 

\begin{figure*}
\centering
\includegraphics[width=0.8\textwidth]{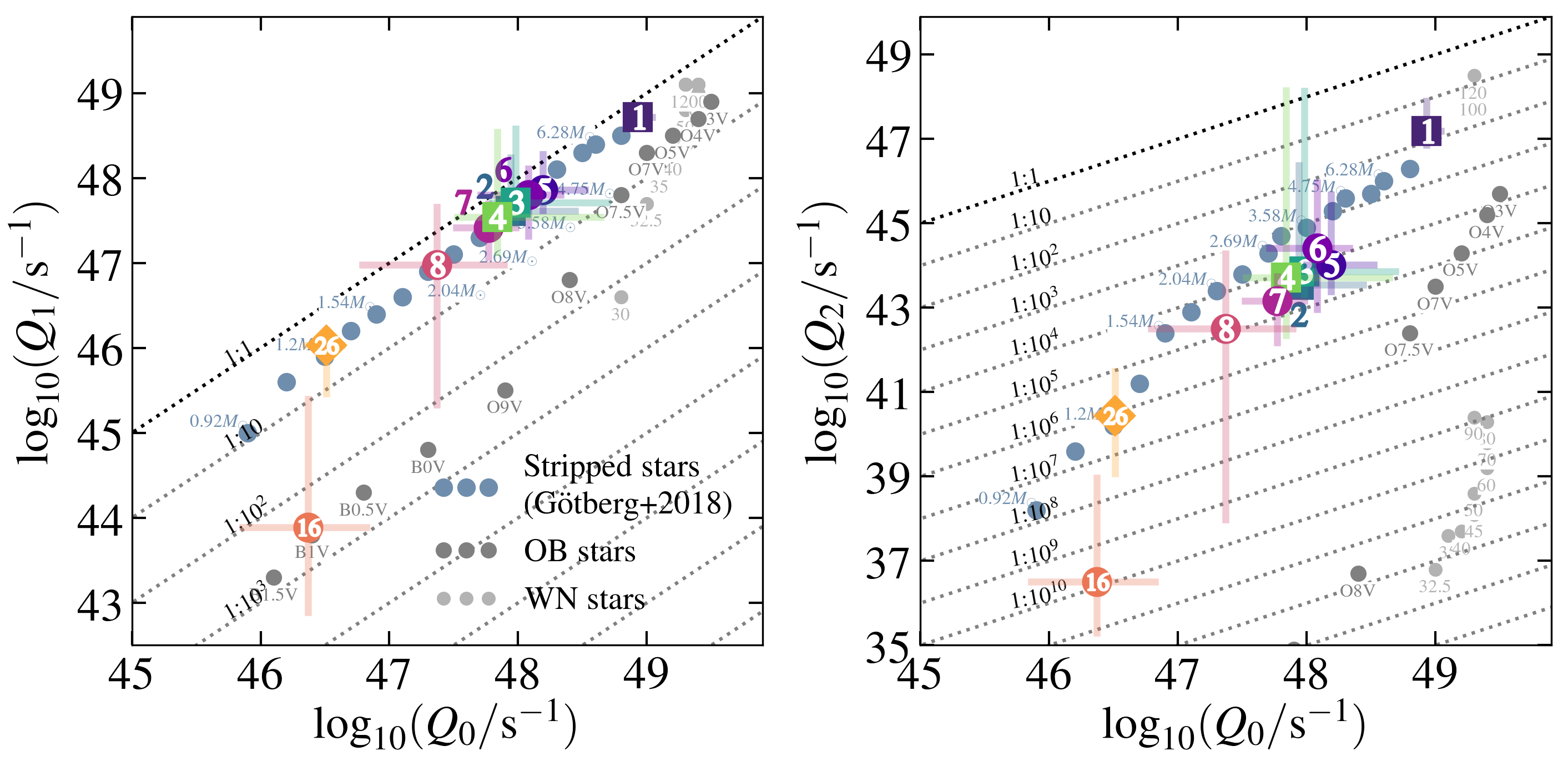}
\caption{Inferred emission rates of H-, He-, and He$^+$-ionizing photons ($Q_0$, $Q_1$, and $Q_2$, respectively), plotted against each other to explore ionizing hardness for the stars in the spectroscopic sample and using numbered colored symbols. A large fraction ($\sim50\%$) of the H-ionizing photons are He-ionizing, but only a small fraction ($\sim0.001-1\%$) are He$^+$-ionizing. This shape of the spectral energy distribution is expected for stars with temperatures $\sim 50-100$kK, but remains to be observationally confirmed. 
For comparison, we also display \rev{models with $Z=0.006$ for stripped stars by \citet{2018A&A...615A..78G} using pale blue and labeled with the stripped star mass, along with} models with $Z=0.4Z_{\odot}$ from \citet{2002MNRAS.337.1309S} for OB-type main-sequence stars in dark gray, labeled by spectral types, and for WN-type WR stars in light gray, labeled by temperature in kK.  
}
\label{fig:Qs}
\end{figure*}

We present the emission rates of ionizing photons in \tabref{tab:stellar_properties} and plot them in \figref{fig:Qs}. The figure shows hardness diagrams, where we plot $Q_1$ as function of $Q_0$ in the left panel, and $Q_2$ as function of $Q_0$ in the right panel. The dotted lines show the ratio between the helium to hydrogen ionizing emission rates as labeled. The figures show that, while roughly half of the hydrogen-ionizing photons are also helium-ionizing photons (for all stars but star 16), only a small fraction of them are also He$^+$-ionizing (typically $\sim 0.001-0.1\%$). 

We expect that stars 1-8 have $Q_0 \sim 10^{47.5}-10^{49}$ s$^{-1}$, $Q_1 \sim 10^{47}-10^{49}$ s$^{-1}$, and $Q_2 \sim 10^{43}-10^{47}$ s$^{-1}$. We compare these to the expected emission rates of ionizing photons from \rev{models of stripped stars with $Z=0.006$ \citep{2018A&A...615A..78G} and models of} OB main-sequence stars and WN-type WR stars from the 0.4 $Z_{\odot}$ models from \citet{2002MNRAS.337.1309S} in \figref{fig:Qs}. As the figure shows, the H-ionizing emission rates of stars 1-8 are similar to mid-late O-type main sequence stars, but lower by a factor of a few compared to WN-stars. Compared to OB-stars, stars 1-8 and 26 have harder ionizing emission, with typically more than an order of magnitude higher He$^0$-ionizing emission rates compared to OB stars of the same $Q_0$. Main-sequence stars with similar $Q_0$ as stars 2-8 are expected to emit many orders of magnitude lower rates of $Q_2$. In fact, WN stars with similar temperatures as stars 2-8 also are expected to emit He$^+$-ionizing photons at substantially lower rates, because of their opaque stellar winds. 

\figref{fig:Qs} demonstrates the important role the effective temperature plays for the emission rate of ionizing photons. Star 1 is the hottest star in the sample, and also the star with the hardest ionizing spectrum, where more than 1\% of the hydrogen-ionizing photons also are He$^+$-ionizing. In fact, star 1 is expected to have a similar emission rate of hydrogen-ionizing photons as an O7V-type star, but a three orders of magnitude higher emission rate of He$^+$-ionizing photons \citep{2002MNRAS.337.1309S}. 

\citet{2018A&A...615A..78G} predicted that stripped stars with masses $\sim 3-4\Msun$ should have $Q_0\sim10^{48}$~s$^{-1}$, $Q_1 \sim 10^{47.5}$~s$^{-1}$ and $Q_2 \sim 10^{44}-10^{45}$~s$^{-1}$. As seen from \tabref{tab:stellar_properties} \rev{and \figref{fig:Qs}}, stars 2-7 agree well with these predictions. We note that large variations in $Q_2$ were already predicted by \citet{2018A&A...615A..78G} (see also \citealt{2017A&A...608A..11G}) as a result of both metallicity variations and wind mass-loss rates. While the right panel of \figref{fig:Qs} exhibits an apparently smooth trend for $Q_2$ with $Q_0$, we note that further observational explorations are needed to accurately determine the emission rates of ionizing photons from stripped stars. Such observational explorations could include for example nebular ionization studies.

\section{Implications for binary evolution}\label{sec:implications}

With the parameter determinations described in this paper, there are several topics interesting to discuss in the context of interacting massive binary stars. We choose a subset here. 

\subsection{Resulting surface composition from envelope-stripping}\label{sec:XHs_envstrip}

The stripped stars in our sample have a range of surface hydrogen mass fractions, from about 0.4 down to negligible amounts (see \secref{sec:stellar_properties} and \tabref{tab:stellar_properties}; and also Appendix~\ref{app:impact_companion}). This suggests that envelope-stripping results in both hydrogen-poor and hydrogen-free stars. 
Because leftover hydrogen can affect both the effective temperature, ionizing emission rates, future expansion and thus binary interaction, and supernova type, this result suggests that approximating stripped stars with pure helium stars may lead to a poor representation. 

A range of surface hydrogen mass fractions has been predicted from models \citep[e.g.,][]{2017ApJ...840...10Y} and is thought to arise from how deeply the stars are stripped into the chemical gradient that results from the receding main-sequence core. The depth of stripping could depend on how large the Roche lobe was at detachment (for the case of stable mass transfer), the metallicity and thus opacity of the stellar envelope \citep[e.g.,][]{2019ApJ...885..130S}, and perhaps also whether the envelope was stripped via common envelope ejection or stable mass transfer \citep[e.g.,][]{2011ApJ...730...76I}. Given the weak stellar winds, we consider it unlikely that wind mass loss after envelope-stripping significantly affects the surface hydrogen content of these stars. Because, with a typical wind mass-loss rate of  $10^{-9}\Msunyr$ and typical stripped star durations of 1 Myr, only about 0.001\Msun\ of material can be removed during the stripped star phase. The total mass of hydrogen expected for stripped stars with surface hydrogen mass fraction of 0.3 and stellar masses  2-7 \Msun\ is 0.03-0.06\Msun\ \citep{2018A&A...615A..78G}. 

To establish the relation between the amount of left-over hydrogen and the envelope-stripping mechanism, orbital monitoring is needed. 
If stripped stars with hydrogen-depleted surfaces predominantly have short ($\lesssim 1$ day) orbital periods, this would suggest that common envelope ejection removes more hydrogen. The surface hydrogen content could thus provide an easy way to determine the envelope-stripping mechanism and identify different types of binary systems.

\subsection{Companion types}\label{sec:companion_type}

In this paper, we have chosen to analyze stripped stars whose flux dominates the optical spectrum and for which no evident sign of a bright companion is present (see also \appref{app:impact_companion}). Despite this apparent lack of a companion star, the stripped stars exhibit radial velocity variations consistent with orbital motion. 
This suggests that optically faint companion stars are present. Such companions can only be lower-mass main sequence companions or compact objects. 

In \citetalias{DroutGotberg23}, we found that stripped star + main-sequence star systems will appear as ``Helium-star-type'' if the main-sequence star is (1)  $\lesssim 0.6$ times as massive as the stripped star, and (2) early on its main-sequence evolution (which is expected from binary evolution if the companion is that much less massive). Assuming that stripped stars typically are about a third as massive as their progenitors,
this critical mass ratio of $q_{\rm crit} = 0.6$  translates to a critical initial mass ratio of $q_{\rm crit, init} = 0.6 \times 1/3 = 0.2$. If interaction is initiated in a system with $q_{\rm init} < 0.4$, it is thought that a common envelope should develop \citep{2002MNRAS.329..897H}. We have therefore reason to believe that the stripped stars of ``Helium-star-type'' are the result of common envelope ejection when orbiting MS stars or stable mass transfer/common envelope ejection when orbiting compact objects. 

To better explore what kinds of objects have stripped these stars, orbital monitoring, lightcurve studies, and X-ray observations will be important. The ``composite-type'' and ``B-type stars'' with UV excess presented by \citetalias{DroutGotberg23} provide an opportunity to study companion stars and assess how they were affected by the previous envelope-stripping phase, which could have led to mass gain and spin-up for the accretor stars. 
To further explore the masses and types of accretor stars, methods such as those of  \citet{2018ApJ...853..156W,2021AJ....161..248W}, who used cross-correlation of spectra in the ultraviolet regime to search for subdwarf companions to rapidly rotating Be stars, could be of interest, since it successfully reaches the part of the population of stripped star systems that do not exhibit UV excess.

\subsection{Future evolution to supernovae and compact objects}

According to our evolutionary mass estimates, seven stars are more massive than 2.5\Msun, meaning that they most likely will reach core collapse \citep[cf.][]{2015MNRAS.451.2123T}, and thus explode as stripped-envelope supernovae \citep[e.g.,][]{2011ApJ...741...97D, 2016MNRAS.457..328L, 2017ApJ...840...10Y}.
With some that have leftover hydrogen and others that are consistent with no leftover hydrogen (\secref{sec:XHs_envstrip}), in conjunction with low wind mass-loss rates (\secref{sec:wind}), these stars likely will result in both type Ib (hydrogen-free) and type IIb (hydrogen-poor) supernovae. 

The structure models of stripped stars with mass $>2.5\Msun$ from \citet{2018A&A...615A..78G} have surface hydrogen mass fractions of $X_{\rm H, surf} \sim 0.25-0.30$ and corresponding total hydrogen masses of $0.04-0.06 \Msun$. According to computations from \citet{2012MNRAS.422...70H}, such hydrogen masses should result in type IIb supernovae. If the stellar structure of these models is representative of stripped stars, this should mean that stars 1, 2, 4, and 6 should result in IIb supernovae. Stars 3, 5, and 7 have substantially lower or negligible surface hydrogen mass fractions (see \tabref{tab:stellar_properties}). The type of their resulting stripped-envelope supernovae is less evident, and they could result in either IIb \citep{2011MNRAS.414.2985D} or Ib \citep{2012MNRAS.422...70H}.


It is possible (likely for short-period systems) that the stripped star will fill its Roche-lobe anew after central helium depletion, during helium-shell burning \citep{2020A&A...637A...6L}. This interaction stage should remove some or all leftover hydrogen, depending on when the interaction is initiated and how much hydrogen is left. The helium can only be removed for extremely short period systems \citep[$P_{\rm orb} \lesssim 0.5$ days, cf.][]{2013ApJ...778L..23T,2015MNRAS.451.2123T}, thus limiting the evolutionary pathways leading to type Ic supernovae, unless any leftover helium remains hidden during the explosion \citep[e.g.][]{Piro2014}. 


Assuming core-collapse will lead to the creation of a $1.4\Msun$ neutron star, we expect that the stripped stars in our sample should produce ejecta masses of $\sim 1.5-2.7\Msun$ for all stars with masses $>2.5\Msun$ apart from star 1, which could have as much as $\sim7\Msun$ ejecta. These numbers agree with the obsessionally constrained ejecta masses for most stripped-envelope supernovae \citep[e.g.][]{2011ApJ...741...97D,2016MNRAS.457..328L}. 

Because of its higher mass, it is possible that star 1 will create a black hole. While it is difficult to know what mass such a black hole would have, it could be similar to the mass of the carbon/oxygen core. \citet{2021A&A...656A..58L} estimate the carbon/oxygen core mass to be 6.2\Msun for a 8.2\Msun helium-core mass, which is similar to the evolutionary mass of star 1. In conjunction with its low metallicity, this could make star 1 a good calibrator for evolutionary pathways leading to merging black hole binaries. 

Stars 8, 16, and 26 (assuming it is residing in the foreground) have lower predicted masses compared to the rest of the sample, sand should lead to white dwarf creation. Stars 8 and 16 likely have current masses above the Chandrasekhar limit and therefore should lose some material before white dwarf creation. Assuming the mass lost will be the outermost layers, they should lose all of the remaining hydrogen and could thus result in DB type white dwarfs. Given that stars 8, 16, and 26 most likely are, or will be, helium-burning objects, they should evolve into C/O white dwarfs. 

Depending on the magnitudes of potential kicks present at compact object formation, the orbit of these binaries will be affected. Orbital solutions for the current systems will help constrain possible future evolutionary pathways, in some cases potentially leading to double compact object formation.

\section{Summary \& Conclusions}\label{sec:summary_conclusion}

We present a spectroscopic analysis to obtain the stellar properties for a set of 10 stars first presented in \citetalias{DroutGotberg23} that we argue are stripped of their hydrogen-rich envelopes via binary interaction. We measure directly from the spectral fitting, for all but one star, effective temperatures confidently above 50kK, surface gravities $\log g \sim 5$ and surface hydrogen (helium) mass fractions $\sim$0-0.4 ($\sim$1-0.6). By fitting the spectral energy distribution of the models to UV and optical photometry, we obtain low extinction values ($A_V \sim 0.1-0.65$) and bolometric luminosities of $\sim 3\times 10^3$-$10^5 L_{\odot}$. Combined with effective temperature and surface gravity, we then estimate stellar radii $\sim 0.6-1.5 R_{\odot}$ and spectroscopic masses $\sim 0.8-6.9 M_{\odot}$. Using a mass-luminosity relation from binary evolution models, we estimate the evolutionary masses to $\sim 1.2-8.4 M_{\odot}$. 

These properties agree well with the expectations from detailed binary evolution models for helium-core burning stars that have been stripped of their hydrogen-rich envelopes in binaries. This confirms the prediction that the large majority of hydrogen-rich envelopes can be stripped off during binary interaction, leaving the helium core exposed with no or only a thin layer of hydrogen-polluted material left on the surface \citep{2017A&A...608A..11G}. 

Our analysis of the observed properties of stripped stars helps to strengthen several expectations about envelope-stripping in binaries that have existed for several years, but which have remained untested:
\begin{enumerate}
\item Stars stripped in binaries can be sufficiently massive to reach core-collapse. Thus, they most likely can produce neutron stars and black holes. However, they can also be progenitors for white dwarfs.
\item Stars stripped in binaries can have some or no residual hydrogen left on their surfaces after envelope-stripping. This suggests that binary-stripped stars are progenitors of both Ib and IIb supernovae. 
\item Stars can be stripped by compact objects or low-mass stars. This must be true because the stripped stars we analyze here dominate the optical spectrum. 
\item The stellar properties expected from binary evolution models where stars are stripped via stable mass transfer reflect the observed stellar properties reasonably well. 
\item While detailed analysis of ultraviolet spectra is needed, the optical spectra indicate that the wind mass-loss rates from stripped stars are likely lower ($\dot{M}_{\rm wind} \lesssim 10^{-9}\Msunyr$) than expected from extrapolations of Wolf-Rayet wind mass-loss schemes, and possibly also single-temperature helium star schemes. 
These low mass-loss rates suggest that winds are unimportant in the removal of residual hydrogen or stripping of the helium layer, suggesting such removal only can happen through future binary interaction. 
\end{enumerate}


The derived stellar masses and general stellar properties of the stripped stars indicate that we have filled the gap in the helium-star mass range, creating a bridge between subdwarfs and Wolf-Rayet stars. This observed stellar sample offers opportunities to constrain uncertain physics, such as understanding wind mass loss from hot and helium-rich stars and the period evolution of interacting binaries. 

To explore the full parameter space of stripped star binaries, studies reaching systems with massive and exotic companions, along with a Galactic sample, will be needed. A more complete coverage over the binary parameter space will provide better constraints for binary evolution and population synthesis models. Larger samples will also provide the opportunity to study the effect of metallicity on massive binary interaction, which could lead to a better understanding of the distant, young Universe when metallicity was low. 
The research field of massive stars, and especially stripped helium stars, is and will be even more dependent on incoming ultraviolet data from the Hubble Space Telescope. These data are crucial for studying stellar winds, but also likely the vast majority of stripped stars, which are thought to orbit brighter and more massive main-sequence stars \citep{2021AJ....161..248W}. Conversely, identifying and studying the effects on the companion stars, affected by significant mass accretion and spin-up due to binary interaction, will require UV spectroscopy.

\acknowledgments
\revv{We are thankful to the anonymous referee for providing a constructive report that helped improve the manuscript. In addition, we acknowledge input from Peter Senchyna, Cole Johnston, JC Bouret, Anna O'Grady, Beryl Hovis-Afflerbach, Ashley Carpenter, Anaelle Roc, Alex Laroche, Thomas Kupfer, Katie Breivik, Katie Auchettl, Jielai Zhang, Gwen Rudie, and Stephen Justham.}

Support for this work was provided by NASA through the NASA Hubble Fellowship Program grant \#HST-HF2-51457.001-A and the HST grants GO-15824 and GO-16755 awarded by the Space Telescope Science Institute, which is operated by the Association of Universities for Research in Astronomy, Inc., for NASA, under contract NAS5-26555.
MRD acknowledges support from the NSERC through grant RGPIN-2019-06186, the Canada Research Chairs Program, the Canadian Institute for Advanced Research (CIFAR), and the Dunlap Institute at the University of Toronto. 
PAC is supported by the Science and Technology Facilities Council research grant ST/V000853/1. 
Computing resources used for this work were made possible by a grant from the Ahmanson Foundation. 
This research has made use of the SVO Filter Profile Service (\url{http://svo2.cab.inta-csic.es/theory/fps/}) supported from the Spanish MINECO through grant AYA2017-84089.

\appendix

\section{Details for spectral fitting}\label{app:details_spectral_fitting}

\begin{table}
\centering
\caption{Spectral lines used for the model fits.}
\label{tab:fit_lines}
\begin{tabular}{lll}
\toprule \midrule
Star & Spectral lines & Grid constraints \\
\midrule
Star 1 & H$\delta$/\HeIIl 4100, \HeIIl 4200, H$\gamma$/\HeIIl 4339, \HeIIl 4542, H$\beta$/\HeIIl 4859 & N\textsc{v} 4604 emission, N\textsc{v} 4945 emission \\  
Star 2 & H$\delta$/\HeIIl 4100, \HeIIl 4200, H$\gamma$/\HeIIl 4339, \HeIIl 4542, H$\beta$/\HeIIl 4859, \HeIIl 5412 & N\textsc{v} 4604 absorption, N\textsc{v} 4945 emission  \\
Star 3 & H$\delta$/\HeIIl 4100, \HeIIl 4200, H$\gamma$/\HeIIl 4339, \HeIIl 4542, H$\beta$/\HeIIl 4859, \HeIIl 5412  & N\textsc{v} 4945 emission \\
Star 4 & H$\delta$/\HeIIl 4100, \HeIIl 4200, H$\gamma$/\HeIIl 4339, \HeIIl 4542, H$\beta$/\HeIIl 4859, \HeIl 5876 & None  \\ 
\midrule
Star 5 & H$\delta$/\HeIIl 4100, \HeIIl 4200, H$\gamma$/\HeIIl 4339, \HeIIl 4542, H$\beta$/\HeIIl 4859, \HeIl 5876 &  N\textsc{v} 4604 absorption, N\textsc{iv} 4057 emission \\ 
Star 6 & H$\delta$/\HeIIl 4100, \HeIIl 4200, H$\gamma$/\HeIIl 4339, \HeIIl 4542, H$\beta$/\HeIIl 4859 & N\textsc{v} 4604 absorption, N\textsc{iv} 4057 emission  \\
Star 7 & H$\gamma$/\HeIIl 4339, \HeIl 4471, \HeIIl 4542, H$\beta$/\HeIIl 4859, \HeIIl 5412, \HeIl 5876 & \HeIl 4471 and \HeIl 5876 absorption \\
Star 8 & H$\delta$/\HeIIl 4100, \HeIIl 4200, H$\gamma$/\HeIIl 4339, \HeIIl 4542, H$\beta$/\HeIIl 4859, \HeIl 5876 & \HeIl 5876 absorption \\
Star 16 & \HeIIl 4200, H$\gamma$/\HeIIl 4339, \HeIl 4471, \HeIIl 4542, H$\beta$/\HeIIl 4859, \HeIl 5876 & \HeIl 4471 and \HeIl 5876 absorption \\
\midrule
Star 26 & \HeIIl 4200, H$\gamma$/\HeIIl 4339, \HeIl 4471, \HeIIl 4542, H$\beta$/\HeIIl 4859, \HeIl 5876 & \HeIl 4471 and \HeIl 5876 absorption \\
\bottomrule
\end{tabular}

\end{table}

In this appendix, we show, for each star, the detailed fits that give rise to the properties that we present in this paper (see \tabref{tab:stellar_properties}). A description for how these fits are performed, see \secref{sec:fitting_routine}.  

We use a set of the strongest and most robustly modeled spectral lines of hydrogen and helium for the spectral fitting. These usually include H$\delta$/\HeIIl 4100, \HeIIl 4200, H$\gamma$/\HeIIl 4339, \HeIIl 4542, and H$\beta$/\HeIIl 4859, and when present we also include \HeIl 5876. We avoid to use H$\alpha$ and \HeIIl 4686 for the fits since they are $\alpha$ lines and are therefore very sensitive to stellar wind and surrounding ionized gas, which can impact the determination of the stellar properties we focus on here (see \secref{sec:wind}). We also avoid using \HeIIl 5412 when possible because this spectral line sometimes has contributions from the outer parts of the stellar atmosphere, which is affected by the density and thus also the stellar wind. 

Which exact spectral lines that we use for the different stars are presented in \tabref{tab:fit_lines}. 
In the case of stars 2 and 3, \HeIIl 4200 is affected by noise and we therefore chose to include also \HeIIl 5412 in the fits. 
When observing star 7, we needed to rotate the telescope out of the parallactic angle to avoid including nearby stars in the slit, which led to poor signal-to-noise ratio in the blue part of the spectrum and we therefore chose to exclude H$\delta$/\HeIIl 4100 and \HeIIl 4200. 

\begin{figure}
\centering
\includegraphics[width=0.9\textwidth]{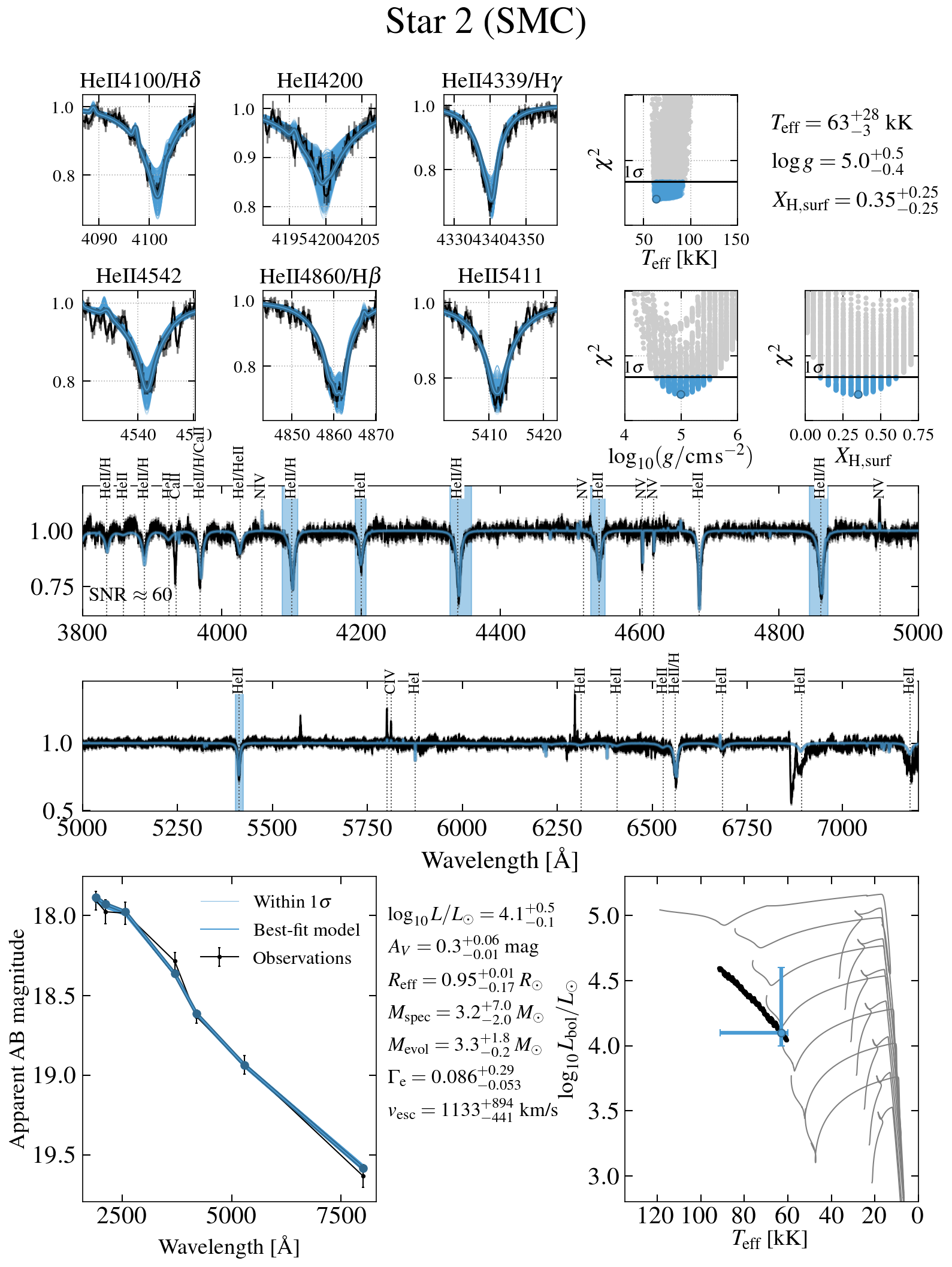}
\caption{Fit for star 2. }
\label{fig:fit_star2}
\end{figure}

\begin{figure}
\centering
\includegraphics[width=0.9\textwidth]{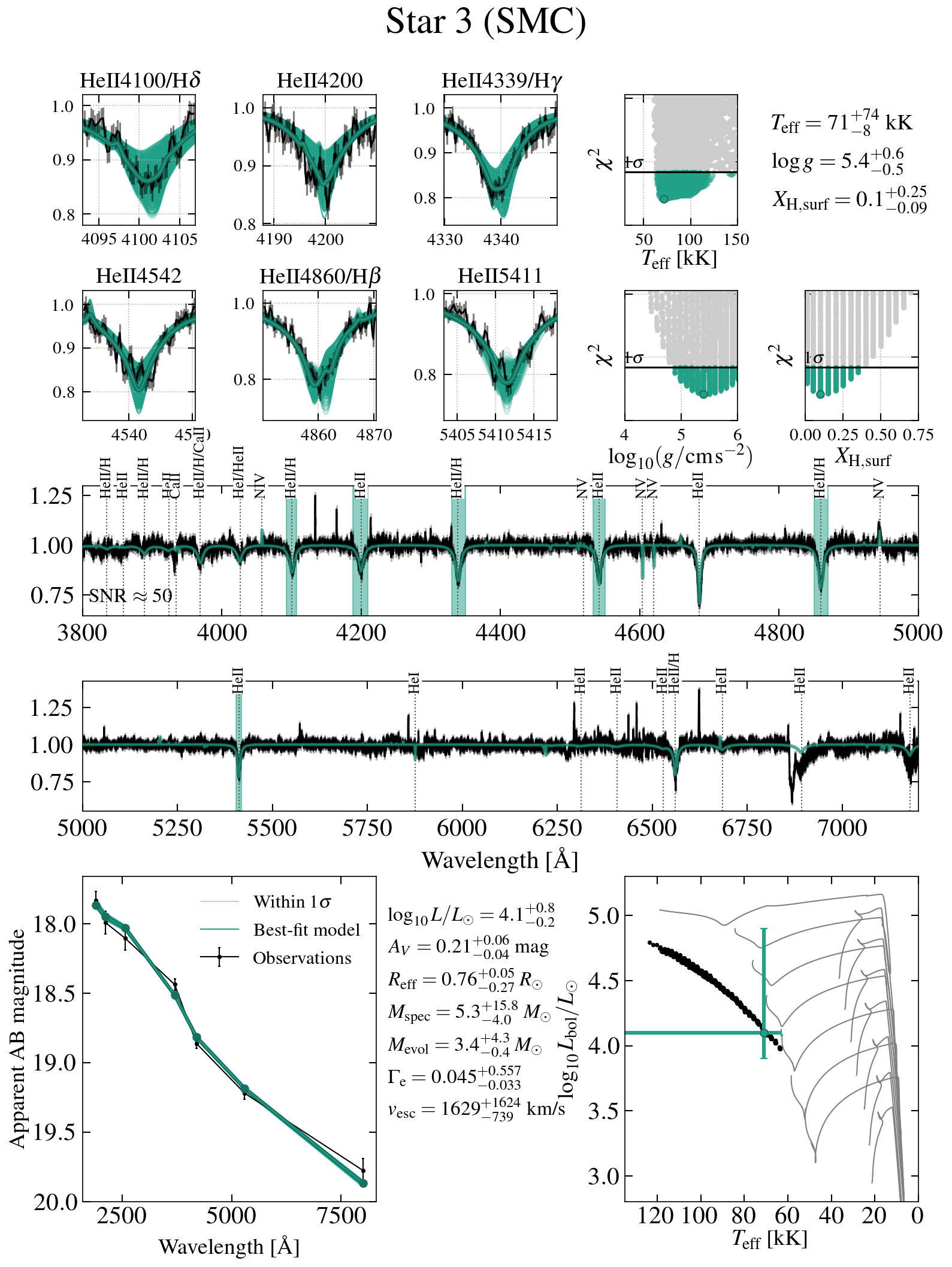}  
\caption{Fit for star 3. }
\label{fig:fit_star3}
\end{figure}

\begin{figure}
\centering
\includegraphics[width=0.9\textwidth]{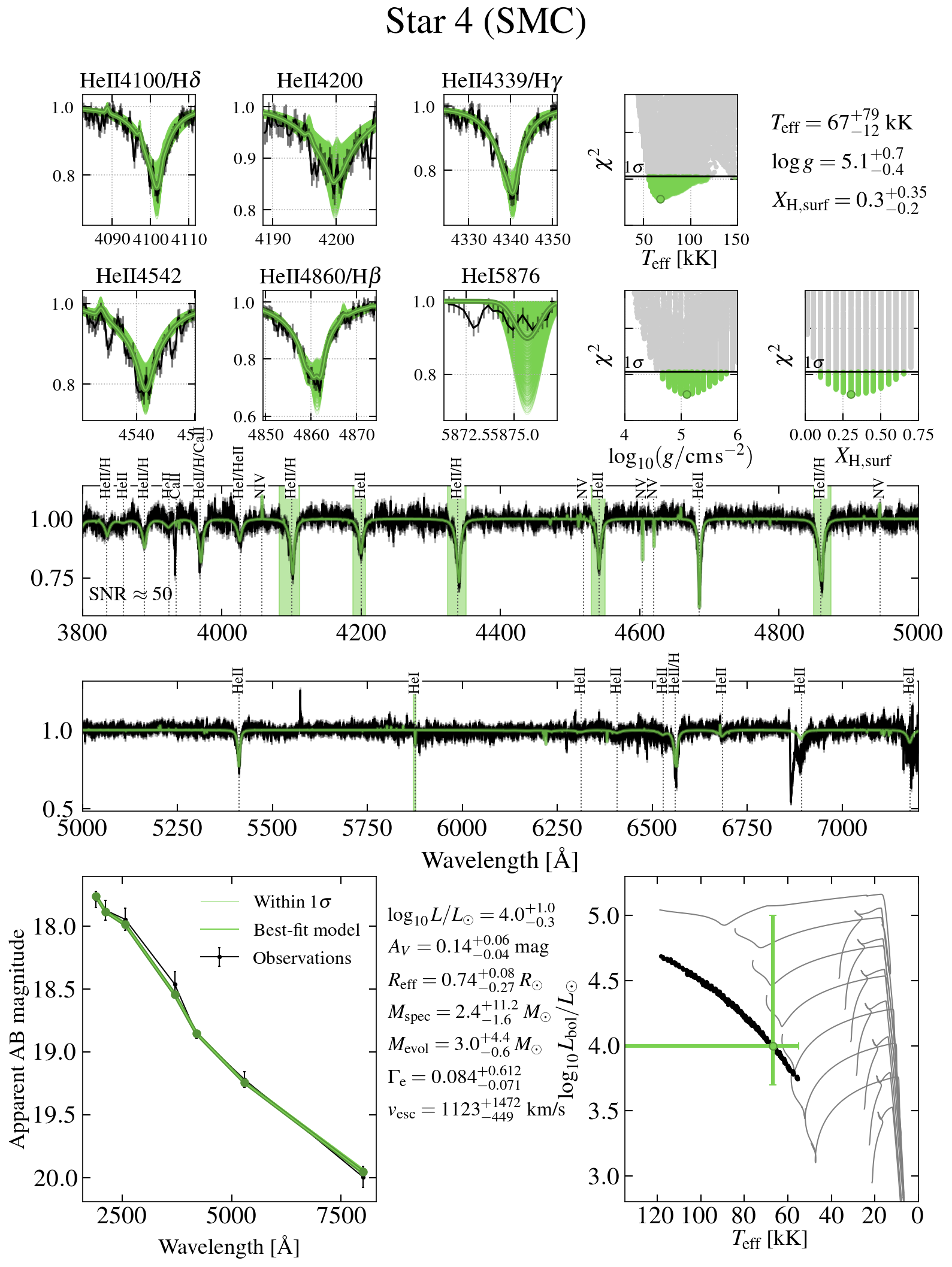}
\caption{Fit for star 4. }
\label{fig:fit_star4}
\end{figure}

\begin{figure}
\centering
\includegraphics[width=0.9\textwidth]{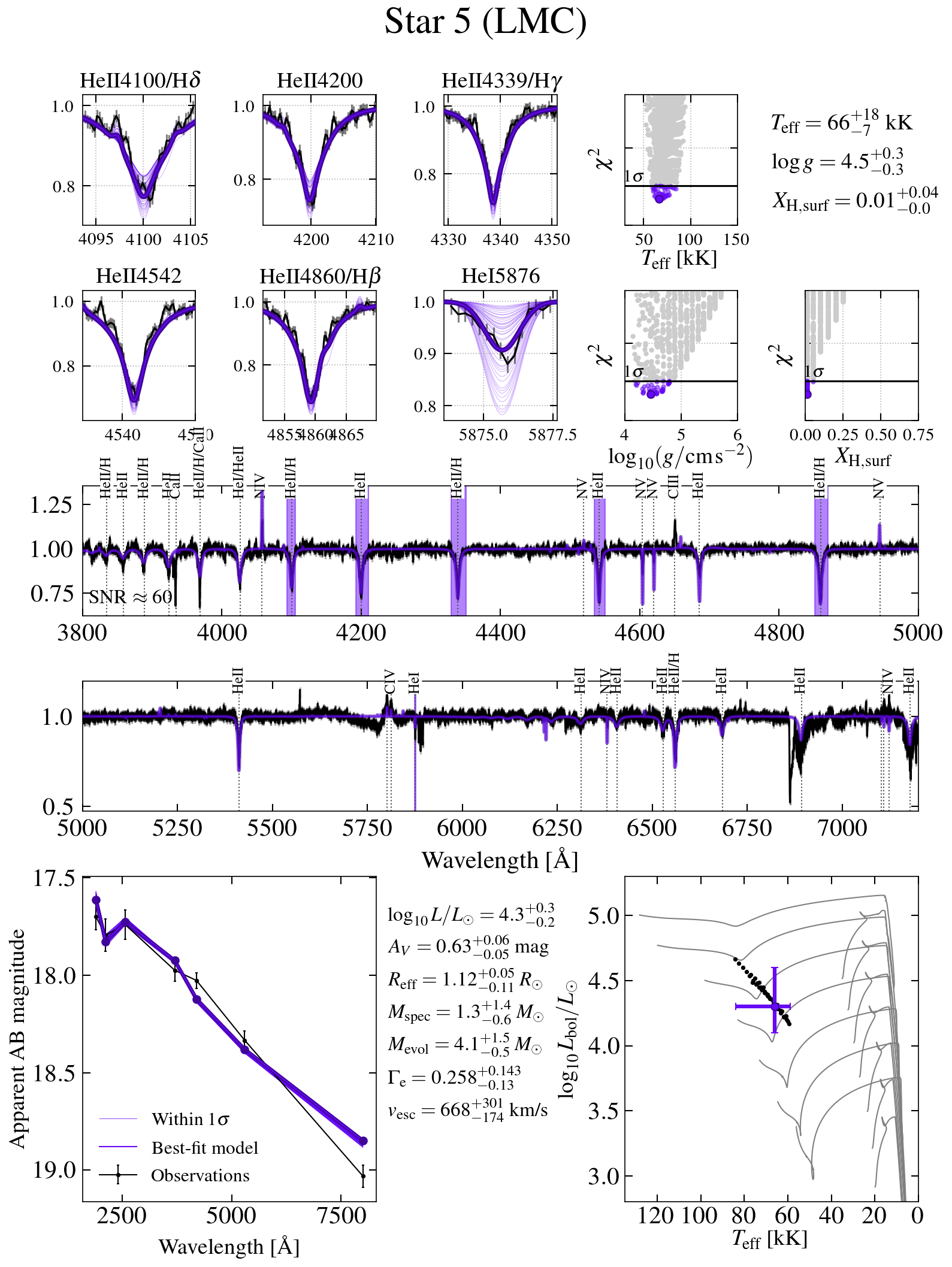}
\caption{Fit for star 5. }
\label{fig:fit_star5}
\end{figure}

\begin{figure}
\centering
\includegraphics[width=0.9\textwidth]{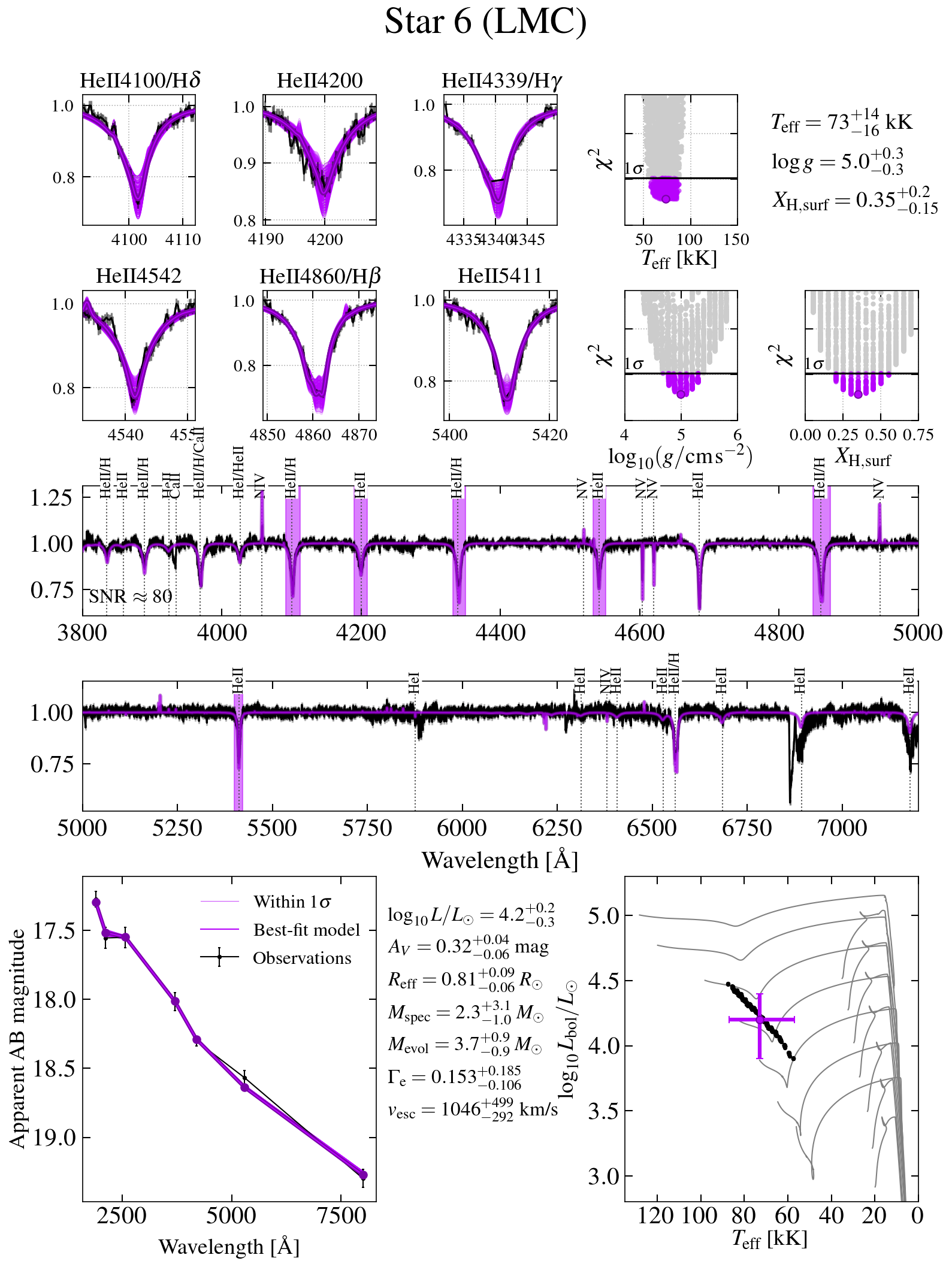}
\caption{Fit for star 6. We have clipped out the line cores of \HeIIl 4339/H$\gamma$, \HeIIl 4860/H$\beta$, \HeIIl 5412, and \HeIIl 6563/H$\alpha$.}
\label{fig:fit_star6}
\end{figure}

\begin{figure}
\centering
\includegraphics[width=0.9\textwidth]{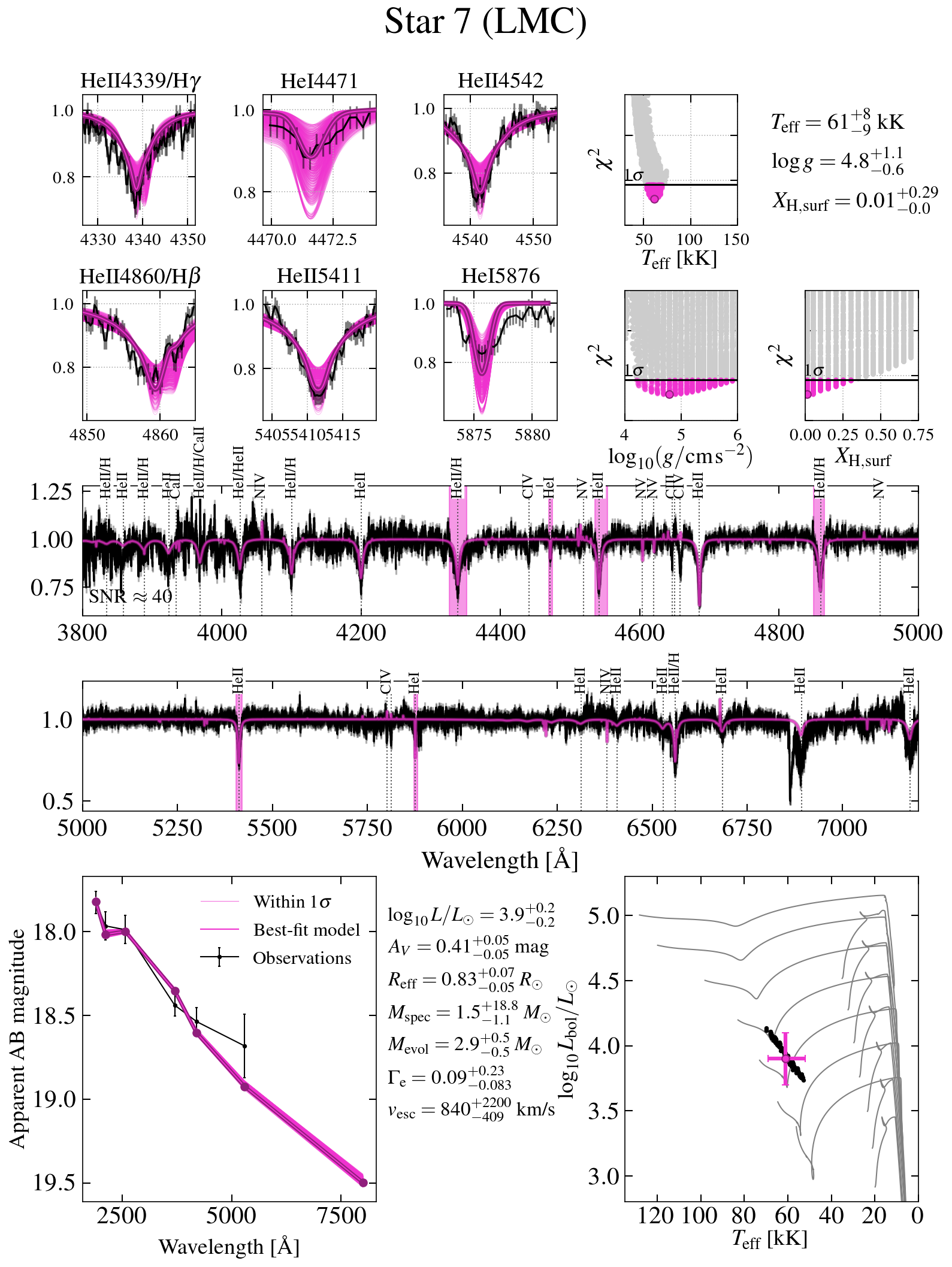}
\caption{Fit for star 7. }
\label{fig:fit_star7}
\end{figure}

\begin{figure}
\centering
\includegraphics[width=0.9\textwidth]{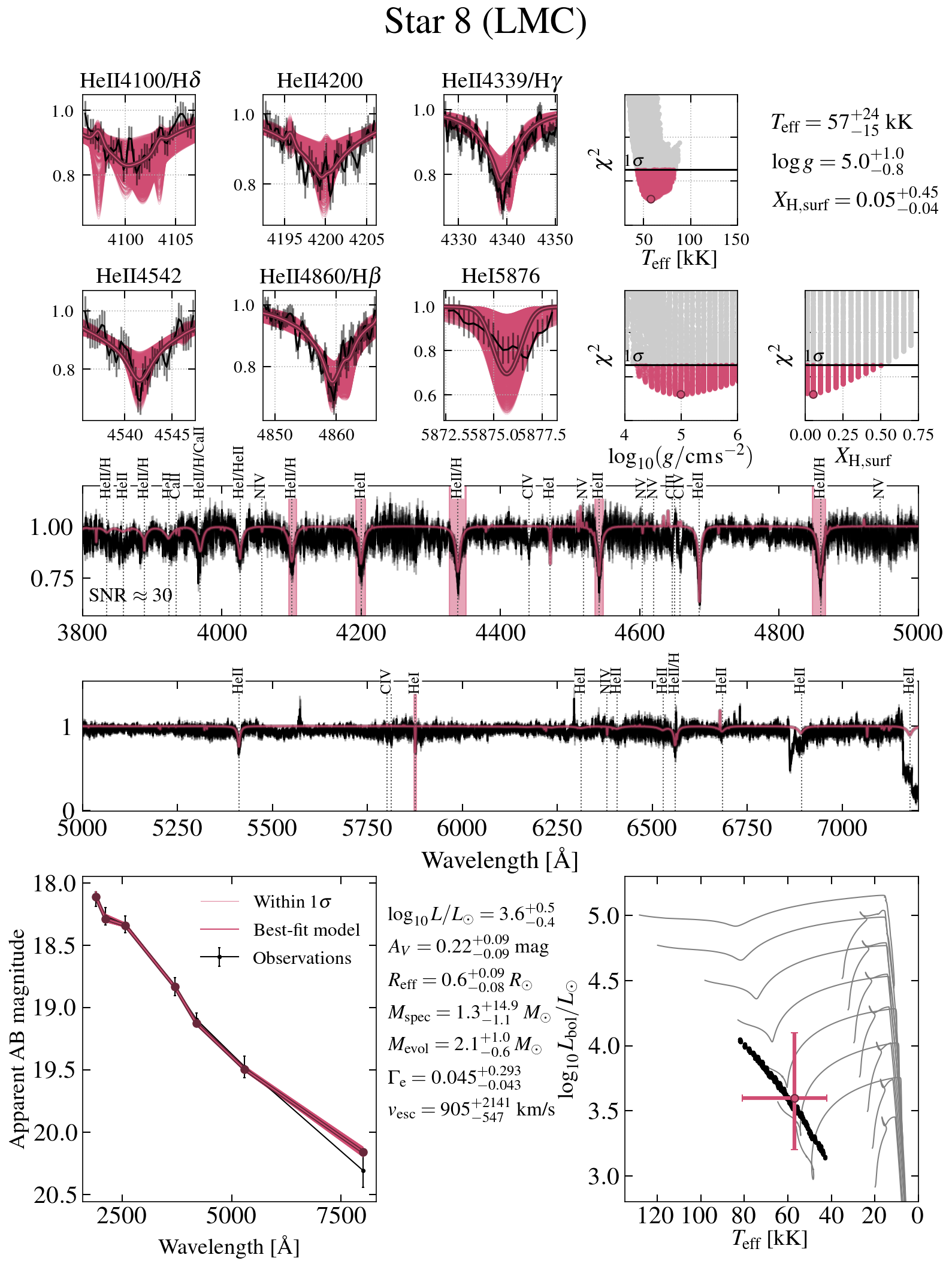}
\caption{Fit for star 8. }
\label{fig:fit_star8}
\end{figure}

\begin{figure}
\centering
\includegraphics[width=0.9\textwidth]{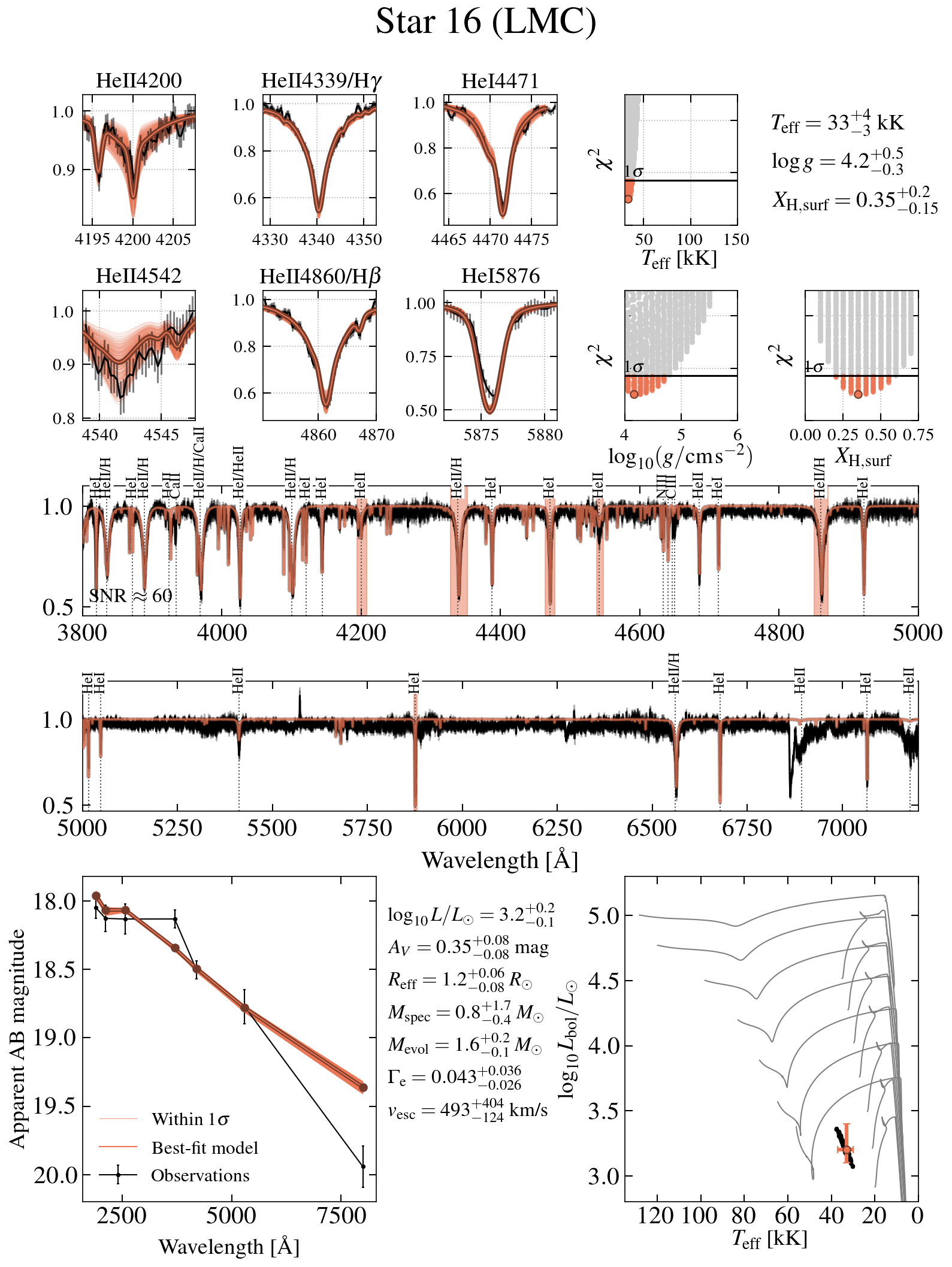}
\caption{Fit for star 16. }
\label{fig:fit_star16}
\end{figure}

\begin{figure}
\centering
 \includegraphics[width=0.9\textwidth]{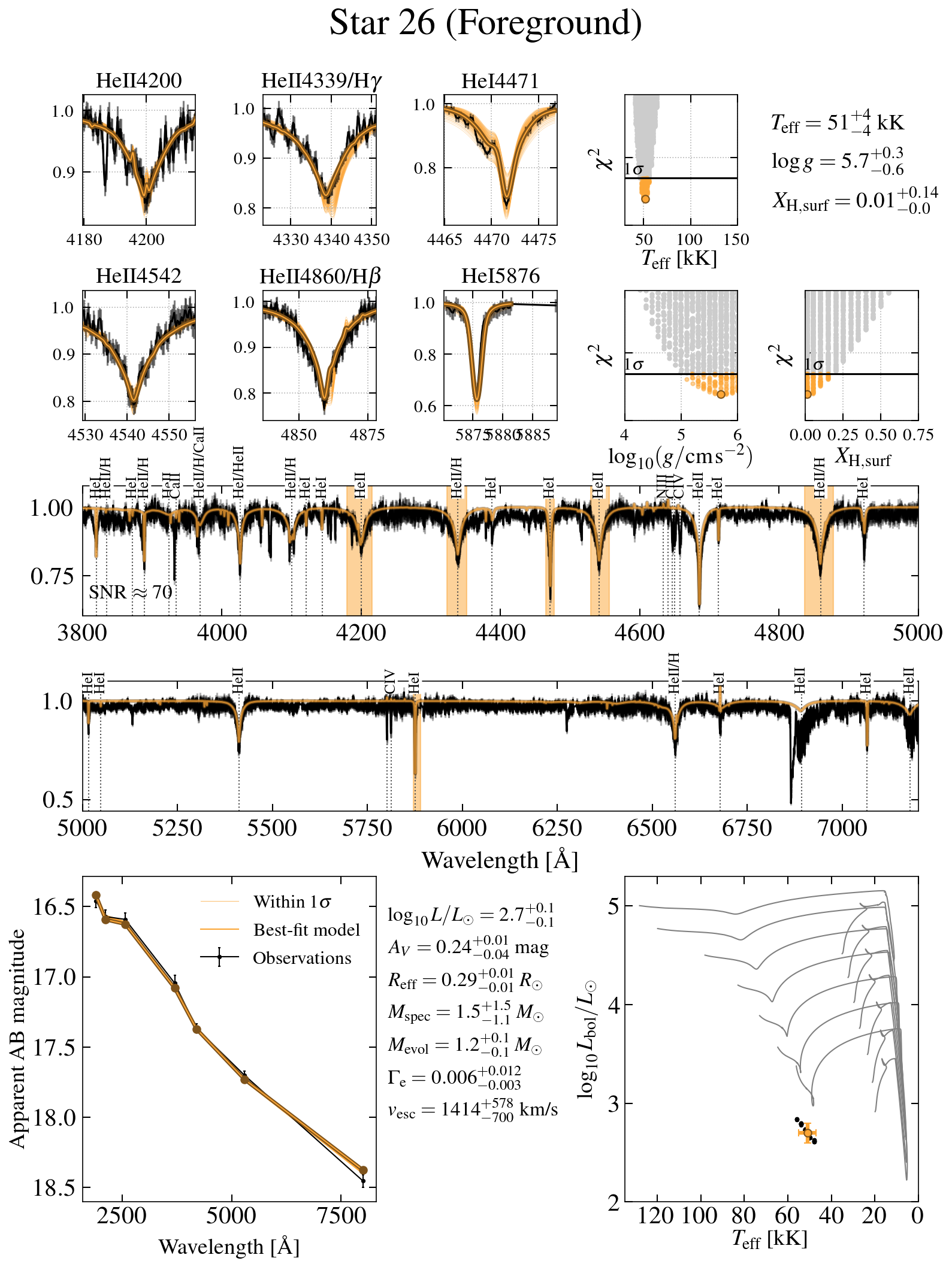}
\caption{Fit for star 26. Assuming a foreground distance of 10 kpc. }
\label{fig:fit_star26_fg}
\end{figure}

\begin{figure}
\centering
 \includegraphics[width=0.9\textwidth]{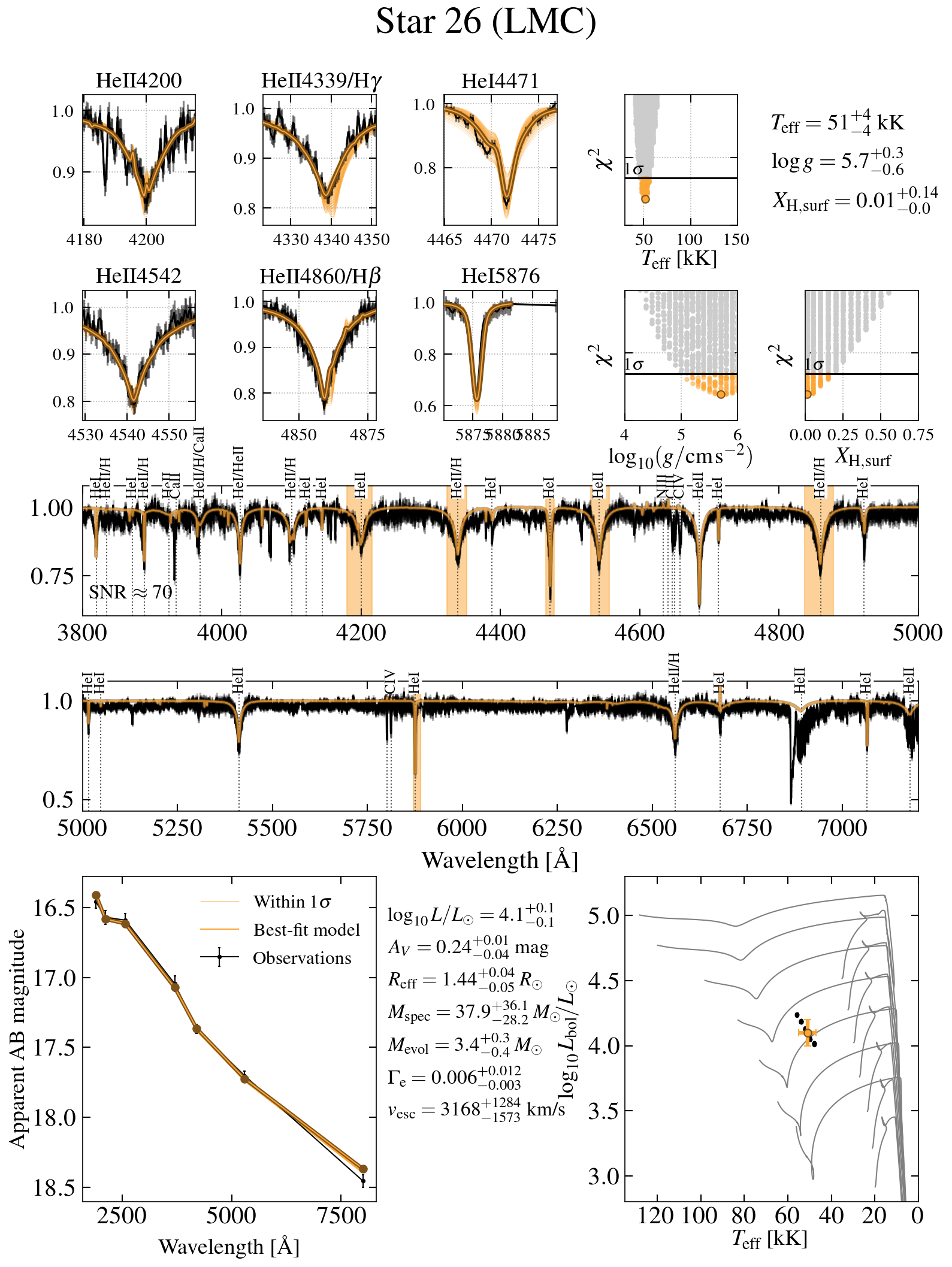}
\caption{Fit for star 26.  Assuming it is a member of the LMC.}
\label{fig:fit_star26_LMC}
\end{figure}

In Figs.\ \ref{fig:fit_star1}, \ref{fig:fit_star2}, \ref{fig:fit_star3}, \ref{fig:fit_star4}, \ref{fig:fit_star5}, \ref{fig:fit_star6}, \ref{fig:fit_star7}, \ref{fig:fit_star8}, \ref{fig:fit_star16}, \ref{fig:fit_star26_fg}, and \ref{fig:fit_star26_LMC}, we show the detailed fits to the spectroscopy and photometry of the stars. Each set of panels display the same things for each star and we describe them below. 

The top left panels show zoom-in panels for the wavelength range of each spectral line that is used for the spectral fit. The black line with errorbars show the observed spectrum, the colored thick line shows the best-fit model, and the colored thin lines show the models allowed within 1$\sigma$ errors. 

The 1$\sigma$ errors are determined using $\chi^2$ (see \secref{sec:fitting_routine}), and we therefore display the $\chi^2$ for each included model as function of the three parameters the model grid spans (effective temperature, surface gravity and surface hydrogen mass fraction) in the top right panels. The best-fit model, which has the minimum $\chi^2$, is marked with a large colored circle and the models allowed within 1$\sigma$ are shown with colored circles located below the black line marked $1\sigma$. Models that are not allowed within 1$\sigma$ are shown as gray circles. The properties resulting directly from the spectral fit are written at the very top right. 

To demonstrate that the best-fit model also matches other spectral features, we show a larger wavelength range together with the best-fit model in the two middle panels. For convenience, we mark the lines used for the spectral fit with colored background and we also give a rough estimate for the signal-to-noise ratio (SNR) of the observed spectrum. 

We show the fit to the photometry in the bottom left panel. The panel shows the observations with associated errors from Swift (the three bluest datapoints) and Swope (the four reddest datapoints) in black and located at the mid-wavelength of the filter function \citep{2012ivoa.rept.1015R, 2020sea..confE.182R}. All models allowed within 1$\sigma$ from the spectroscopic fit are shifted to their respective best-fit magnitude and extinction and shown in color. The best-fit model from the spectroscopic fit is shown with large colored circles and a thick line. The resulting bolometric luminosity and extinction are written in the middle at the bottom together with the estimates for stellar radius and spectroscopic mass that follows (see \secref{sec:fitting_routine}).  
The evolutionary mass is estimated from the mass-luminosity relation described in \secref{sec:Mevol}. 

In addition, we also show the models allowed within 1$\sigma$ in the Hertzsprung-Russell diagram and marked with black dots. The best-fit model is showed as a large colored circle and the errorbars indicate the extent of the models allowed within 1$\sigma$. For reference, we display detailed evolutionary models for donor stars in binary systems from \citet{2018A&A...615A..78G} and for initial masses of 5.5, 6.7, 8.2, 10, 12.2, 14.9, and 18.2\Msun, which correspond to stripped star masses of 1.5, 1.9, 2.5, 3.4, 4.5, 5.9, and 7.3\Msun. The evolutionary models are monotonically brighter with mass. For stars in the LMC and SMC, we show models from the $Z=0.006$ and $Z=0.002$, respectively.


\clearpage

\section{Impact of the companion star on fit}\label{app:impact_companion}

\begin{figure*}
\centering
\includegraphics[width=\textwidth]{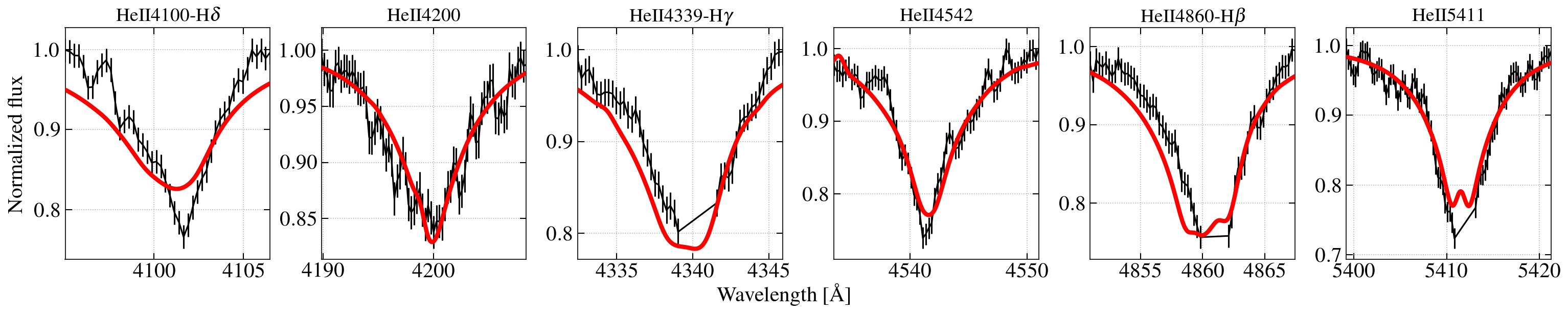} 
\includegraphics[width=\textwidth]{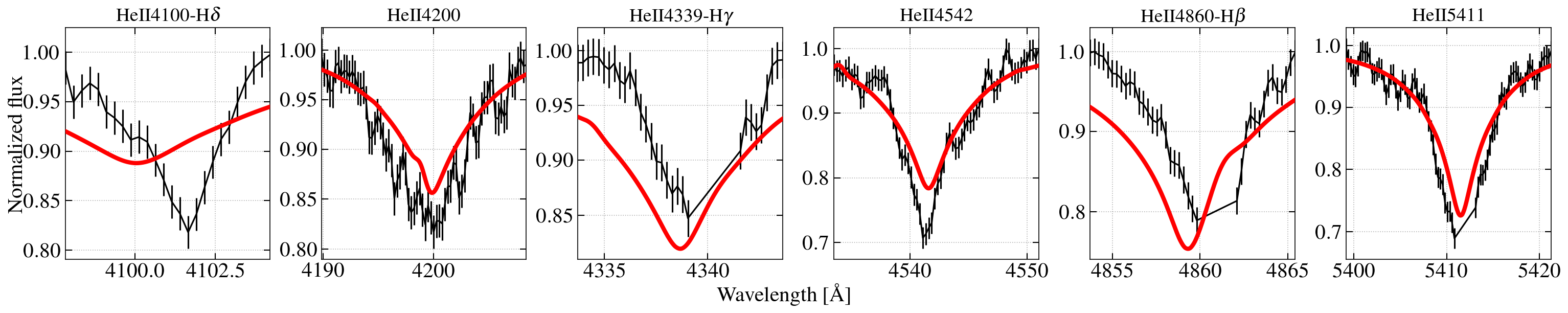} 
\caption{Best-fit models for star 6 after having removed the contribution from a 2.2\Msun\ late B-type companion star, assuming it contributed 10\% (top) and 20\% (bottom) of the optical flux. In both examples, the fits are poor.}
\label{fig:test_835}
\end{figure*}

\begin{figure*}
\centering
\includegraphics[width=\textwidth]{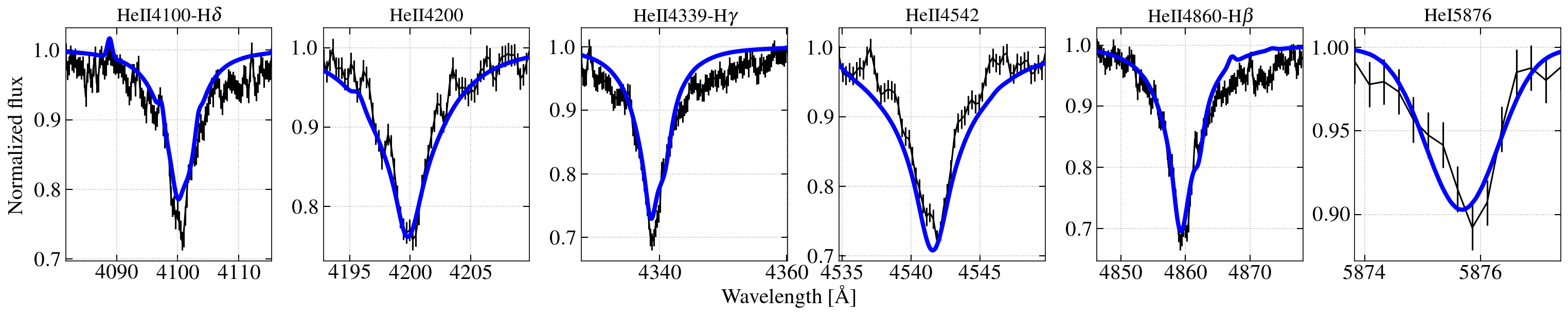} 
\includegraphics[width=\textwidth]{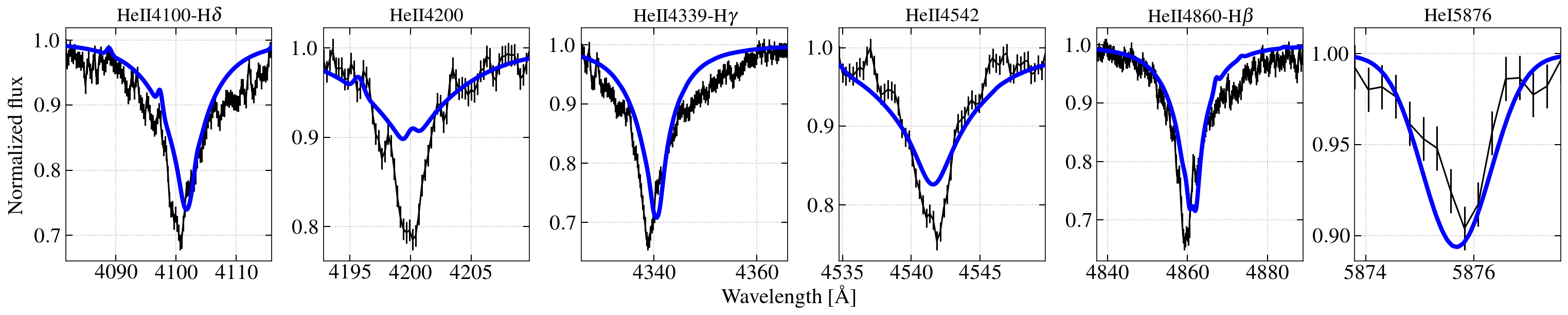} 
\caption{Best-fit models for star 5 after having added the contribution from a 2.2\Msun\ late B-type companion star, assuming it contributes 10\% (top) and 20\% (bottom) of the optical flux. While the 20\% contribution results in a poor fit, the 10\% contribution is acceptable and almost reproduces the effective temperature, surface gravity and surface hydrogen mass fraction derived for star 5.}
\label{fig:test_2273}
\end{figure*}

In this paper, we chose to fit the spectra of stars with ``Helium-star-type'' spectral morphology, approximating their spectra as single, although these stars exhibit binary motion. While these stars at maximum have a very minor contribution from a main-sequence companion, because their spectral morphologies do not show typical signs of main-sequence stars, it is valid to investigate whether a minor contribution can affect the derived stellar properties. 

Here, we test the performance of the spectral fitting routine when (1) removing the contribution from a main-sequence companion from the spectrum of star 6, and (2) adding the contribution from a main-sequence companion to the spectrum of star 5. 
Because we expect that a main-sequence companion should contribute with hydrogen lines, we choose star 6 for the first experiment, since it has measured surface hydrogen content. This experiment is meant to explore whether we could have mistaken the contribution from a main-sequence companion for surface hydrogen content of the stripped star. If true, fitting the spectrum after subtracting a companion star should result in a good fit as well. 
Similarly, for the second experiment, we choose star 5, because it does not show any signs of surface hydrogen content. If a main-sequence companion could be mistaken by surface hydrogen content, fitting the composite spectrum should result in good fits, but higher derived surface hydrogen content for star 5. 

For both tests, we use a spectral model of a late B-type star created using the modeled stellar properties from a 2.2\Msun\ evolutionary model, 20\% through the main-sequence evolution (see supplementary material of \citetalias{DroutGotberg23}). We scale the contribution of the B-star such that it contributes both 10\% and 20\% of the total optical flux in the binary composite. The B-type model does not show any \HeI\ lines and its spectrum is dominated by Balmer lines in the optical. We do not simulate smearing of its spectral features that should occur by stacking after correcting for radial velocity shifts of the stripped star in stars 5 and 6. However, we expect that the effect from such smearing on the spectral features is small. We also do not adapt the B-type model for stellar rotation, since it is likely such systems are created through common envelope ejection. 

We then fit the test spectra with the models as described in \secref{sec:fitting_routine}. When removing the contribution from the B-type star from star 6's spectrum, we find poor spectral fits both when assuming 10\% and 20\% contribution, as visualized in \figref{fig:test_835}. This illustrates that the Balmer lines from the B-type companions are so prominent that subtracting their contribution results in spectral features (in particular hydrogen lines) that are poorly fit by single stripped star models. 

When instead adding the B-type contribution to the spectrum of star 5, we find a poor fit when assuming 20\% contribution, but a realistic fit when assuming 10\% contribution with only slightly deep Balmer lines, as evidenced in \figref{fig:test_2273}. This suggests that the presence of a B-type companion that contributes 20\% of the flux should be detectable from the spectral morphology. It results in poor fits to the single stripped star models, requiring a fit to two components simultaneously. However, a 10\% flux contribution could potentially be missed. The derived stellar properties for the fit with 10\% contribution are very similar to those derived for star 5, but with a slightly higher hydrogen mass fraction ($X_{\rm H, surf} = 0.05$). 

Deeper investigation of the binary companions is needed, but requires several additional analyses and will be addressed in a future study. However, from the analysis presented in this appendix, we conclude that the optical contribution from a companion star must be small for the spectral model fits to be good. Therefore, if any, we expect small influence from the companion star on the derived stellar properties. 


\section{Kinematic Assessment of Star 26}\label{app:kinematic}

\begin{figure*}
\centering
\includegraphics[width=0.47\textwidth]{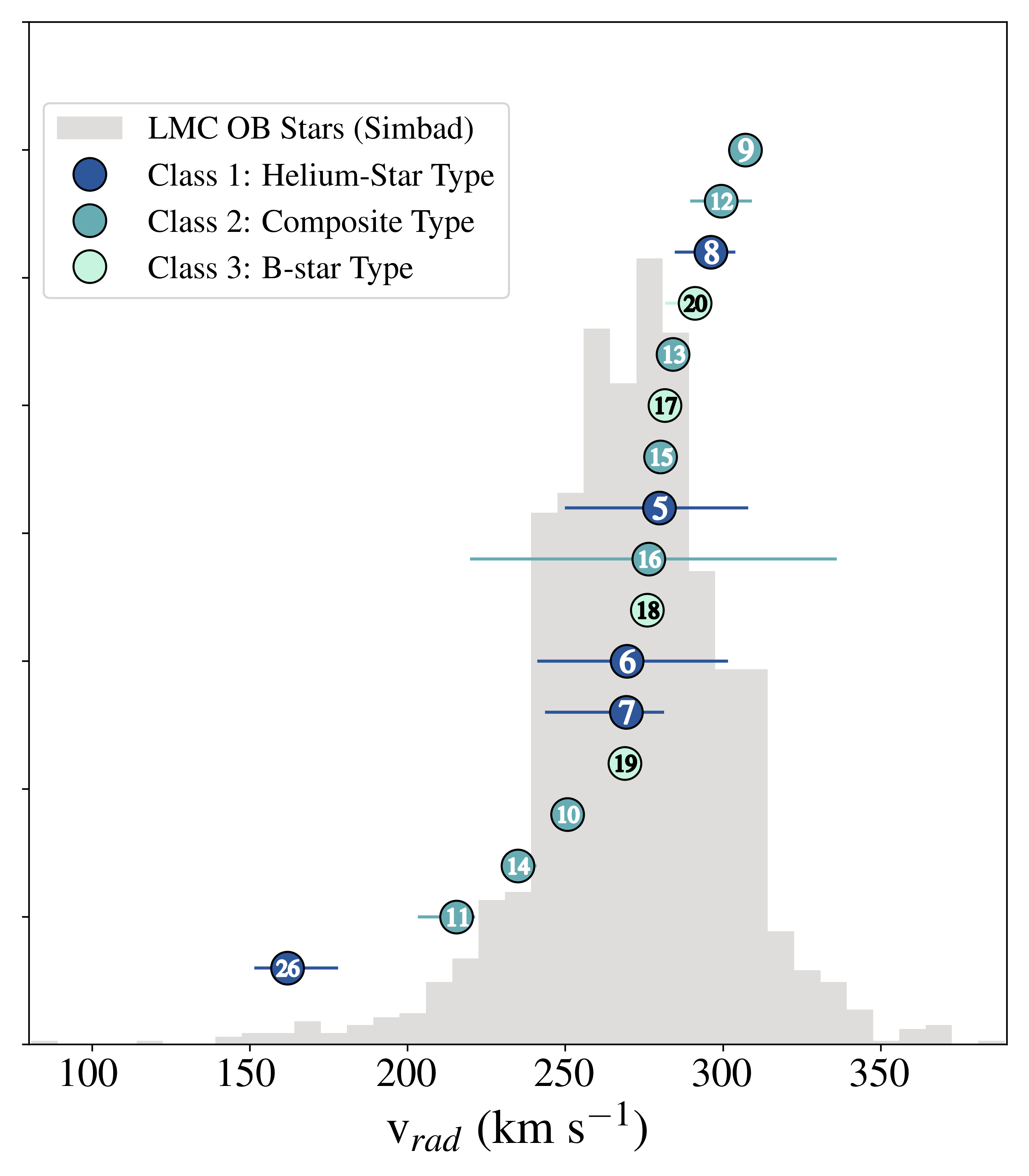}
\includegraphics[width=0.47\textwidth]{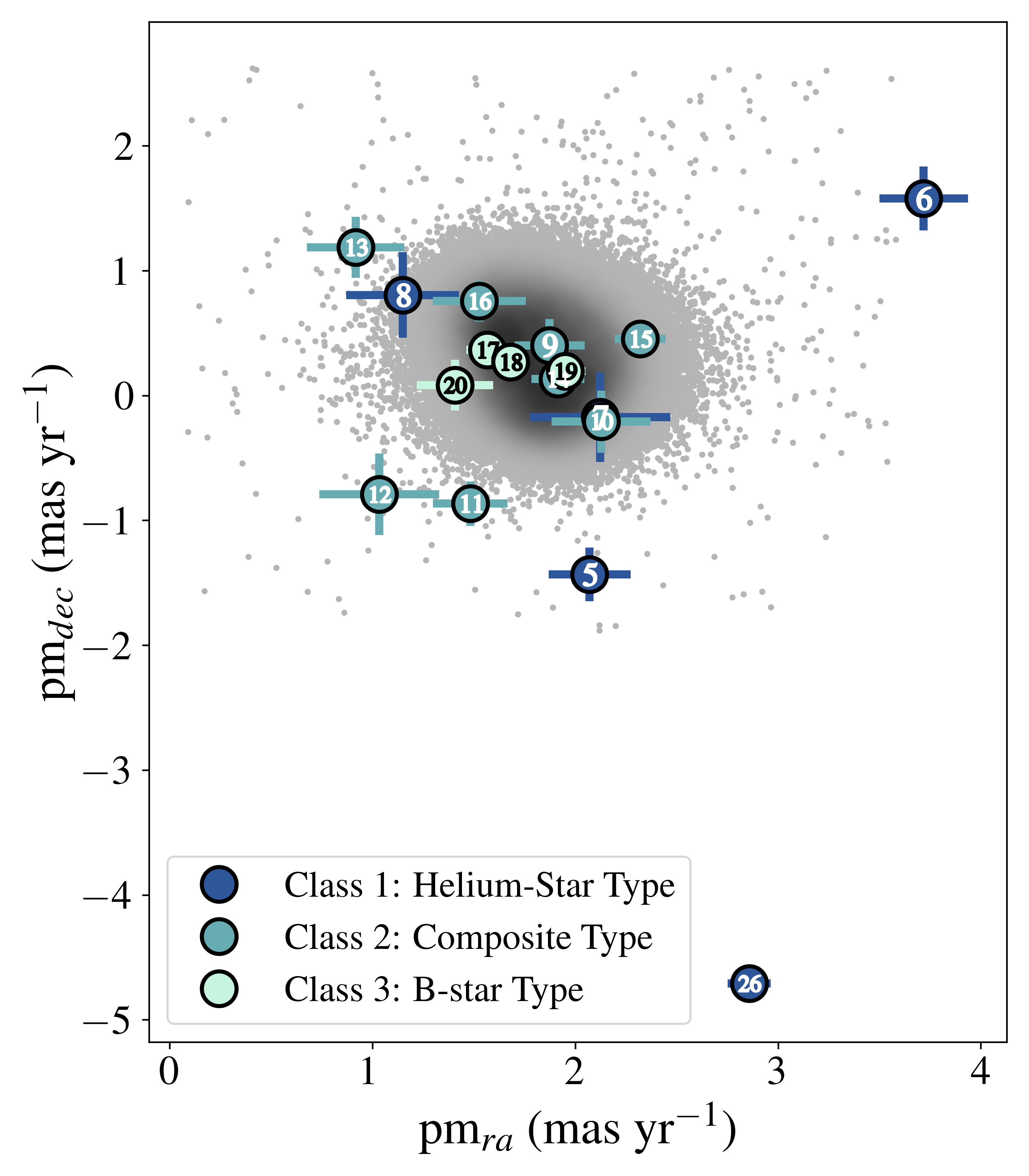}
\caption{\emph{Left:} Comparison of the mean radial velocity observed for Star 26 to known OB stars in the LMC (grey histogram). Horizontal ``errorbars'' designate the range of radial velocities observed at different epochs. \emph{Right:} Comparison of the \emph{Gaia} proper motions measured for Star 26 to likely LMC members (grey points). Other LMC stars presented in \citetalias{DroutGotberg23} are shown as colored/numbered circles in both panels. Figures adapted from \citetalias{DroutGotberg23}.}
\label{fig:kinematics}
\end{figure*}

Here we carry out a detailed kinematic assessment of star 26 compared to the bulk of objects in the LMC, following the same methodology outlined in \citetalias{DroutGotberg23}. In \figref{fig:kinematics} we show both the average radial velocity measured for star 26 (left panel; based on 10 epochs of observations between 2018 and 2022) and the proper motion in RA and DEC from \emph{Gaia EDR3} \cite{Gaia.Collaboration.2020.EDR3}. For comparison, we also show (i) the 16 LMC members presented in \citetalias{DroutGotberg23} (colored dots; both panels), (ii) a sample of OB stars pulled from Simbad that overlap with the LMC and have radial velocity measurements (grey dots; left panel), and (iii) a sample of bright likely LMC members pulled from \emph{Gaia EDR3} (grey dots; right panel; see \citetalias{DroutGotberg23} for details of sample selection).

From this, we see that the mean radial velocity of 162 km s$^{-1}$ is slightly low for the LMC. It overlaps with only the extreme tail of the full sample of OB stars listed on Simbad, and falls below the common threshold of 200 km s$^{-1}$ often adopted for membership (see e.g. \citealt{GF2015}, \citealt{Davies2018}). In addition, the proper motion values of ($\mu_{\alpha}$,$\mu_{\delta}$) $=$ $(2.86,-4.71)$ mas yr$^{-1}$ are significantly offset from the bulk of LMC stars, which have median values of ($\mu_{\alpha}$,$\mu_{\delta}$) $=$ $(1.83,0.30)$ mas yr$^{-1}$. Comparing these proper motion values with the distribution of likely LMC members, we find a $\chi^2$ value of $\sim$165. This indicates that star 26 is located \emph{significantly} outside the region that contains 99.7\% of likely LMC members (designated by  $\chi^2$ $<$ 11.6). In addition, \emph{Gaia} DR3 lists zero excess noise and an astrometric goodness-of-fit close to zero (\texttt{astrometric\_gof\_al} $=$ $-0.28$) for this object, indicating the the astrometric fit was high quality.

While it is possible for stripped helium stars can receive a kick upon the death of their companion stars, the proper motions observed for star 26, would imply a systematic velocity of $\sim$1200 km s$^{-1}$ relative to the mean values for the LMC (assuming a distance of 50 kpc). These values are \emph{significantly} larger the those predicted for runaway stripped stars of $\sim$100 km s$^{-1}$ by \citet{Renzo2019}. Thus, we consider it more likely that star 26 is a foreground halo object. This is supported by the fits presented above, which exhibit both a cooler temperature and higher surface gravity than other objects modeled here, consistent with a subdwarf interpretation. In \tabref{tab:kinematics26} we provide the same kinematic information presented for all objects in the sample of \citetalias{DroutGotberg23}.

\begin{table*}
\centering
\caption{Kinematic Information for Star 26. }
\label{tab:kinematics26}
\begin{tabular}{lcc ccc ccc cc}
\toprule\midrule 
 Star & Class & N$_{\rm spec}$ & $v_{\rm avg}$ &$v_{\rm min}$ & $v_{\rm max}$&  $\pi$ & $\mu_{\alpha}$ & $\mu_{\delta}$ & $\chi^2_{\mathrm{2D}}$ & $\chi^2_{\mathrm{3D}}$\\ 
&& & [km s$^{-1}$] & [km s$^{-1}$] & [km s$^{-1}$] & [mas] & [mas/yr] & [mas/yr] & & \\ 
\midrule 
26	& 1 & 	10	&	162	&	151	&	178	&	 $0.183\pm0.084$ 	&	 $2.857\pm0.106$ 	&	 $-4.709\pm0.099$ 	&	164.7	&	166.4 \\ 
\bottomrule
\end{tabular}
\end{table*}

\bibliographystyle{aasjournal}
\bibliography{references_bin.bib,add.bib}

\end{document}